\documentclass[twocolumn,amsmath,nofootinbib]{revtex4-1}

\usepackage{url}
\usepackage{amssymb}
\usepackage{amsfonts}
\usepackage{amsbsy}
\usepackage{graphicx}
\usepackage{epstopdf}
\usepackage{hyperref}
\usepackage{array} 
\usepackage{multirow}

\newcolumntype{x}[1]{%
>{\centering\hspace{0pt}}p{#1}}%
\providecommand{\openone}{\leavevmode\hbox{\small1\kern-3.8pt\normalsize1}}

\begin{document}
\input{epsf.sty}

% PAPER TEXT

\title{Testing Two-Field Inflation}

\def\harvard{1}
\def\mit{2}
\def\affilmrk#1{$^{#1}$}

\author{
Courtney M. Peterson\affilmrk{\harvard},
Max Tegmark\affilmrk{\mit}
}

\address{
\parshape 1 -3cm 24cm
\affilmrk{\harvard} Dept.~of Physics, Harvard University, Cambridge, MA 02138, USA \\
\affilmrk{\mit} Dept.~of Physics \& MIT Kavli Institute, Massachusetts Institute of Technology, Cambridge, MA 02139
}

\date{Submitted to Journal Month day year, revised Month day, accepted Month day}
%\date{\today}
\date{Submitted May 20, 2010, Revised October 29, 2010}

\vspace{10mm}

\begin{abstract}
We derive accurate semi-analytic formulae for the power spectra of two-field inflation assuming an arbitrary potential and arbitrary non-canonical kinetic terms, and we use them both to build phenomenological intuition and to constrain classes of two-field models using WMAP data.
% DERIVATIONS:
Using covariant formalism, we first develop a framework for understanding the background field kinematics and introduce a ``slow-turn'' approximation.  Next, we find covariant expressions for the evolution of the adiabatic/curvature and entropy/isocurvature modes, and we discuss how the evolution of modes can be inferred directly from the background kinematics and the geometry of the field manifold. From these expressions, we derive semi-analytic formulae for the curvature, isocurvature, and cross spectra, and the standard spectral observables, all to second-order in the slow-roll and slow-turn approximations.  
% INTUITION:
In tandem, we show how our covariant formalism provides useful intuition into how the general features of the inflationary Lagrangian translate into distinct features in the observable power spectra.  In particular, we find that key features of the power spectra can be directly read off from the nature of the roll path, the curve the field vector rolls along with respect to the two-dimensional field manifold.  For example, models whose roll path makes a sharp turn around 60 $e$-foldings before the end of inflation tend to be ruled out because they produce stronger departures from scale invariance than are allowed by the latest CMB observations.  
% DATA:
Finally, we apply our formalism to confront four classes of two-field models with WMAP data, including doubly quadratic and quartic potentials and non-standard kinetic terms, showing how whether a model is ruled out or not depends not only on certain features of the inflationary Lagrangian, but also on the initial conditions.  Ultimately, for a two-field model to be consistent with observations, we show that it must possess the right balance of certain kinematical and dynamical behaviors, which we reduce to a set of functions that represent the main characteristics of any two-field model of inflation. \end{abstract}

\maketitle

\section{Introduction}

Cosmic inflation is currently the leading model for generating the primordial density perturbations that seeded structure formation.  According to the inflationary paradigm, our universe experienced an early period of accelerated expansion, which solved the horizon, flatness, and relic problems (see, e.g., \cite{Guth-1981,Linde-1990,LythAndRiotto-1998,LiddleAndLyth-2000,BassettEtAl-2005}).  The accelerated expansion also stretched quantum fluctuations beyond the causal horizon, freezing them in.  Over time, these perturbations were gravitationally amplified, eventually initiating the formation of galaxies and large-scale structure \cite{MukhanovAndChibisov-1981,GuthAndPi-1982,MukhanovAndChibisov-1982,Hawking-1982,Starobinksy-1982,BardeenEtAl-1983}.
  
In the simplest models, inflation is driven by a single, slow-varying scalar field whose potential serves as an effective cosmological constant.  However, there are good reasons to believe that inflation might have been driven by more than one field.  First, many theories beyond the standard model of particle physics---such as string theory, grand unified theories, supersymmetry, and supergravity---involve multiple scalar fields.  Second, introducing one or more fields may provide attractive features.  For example, hybrid models of inflation involving two scalar fields are able to achieve sufficient inflationary expansion and match the observed power spectrum of density fluctuations, while possessing more natural values for their coupling constants and occurring at sub-Planckian field values \cite{Linde-1991,Linde-1993,CopelandEtAl-1994}.  

Despite the attractiveness of multi-field inflationary models, the task of analyzing them and comparing them against observations is considerably more complicated than the single-field case.  First of all, when there are two or more fields, perturbations in the relative contributions to the energy density (entropy/isocurvature perturbations) are possible, in addition to perturbations in the total energy density (adiabatic/curvature perturbations) \cite{KodamaAndSasaki-1984,Linde-1984}.  These isocurvature perturbations can source the curvature perturbations, causing them to evolve on super-horizon scales \cite{KodamaAndSasaki-1984,WandsEtAl-2000,Bardeen-1980,Mollerach-1990}, which complicates the calculation of the density power spectrum.  Moreover, the isocurvature perturbations themselves give rise to their own power spectrum and potentially also a correlated cross spectrum, as was first recognized by \cite{Langlois-1999}.  Finally, multi-field models are accompanied by an uncountable number of initial conditions.  Since initial conditions may affect the power spectra (e.g., \cite{SasakiAndStewart-1995,Wands-2007}), this complicates testing multi-field models against observational data.  

It is therefore important to develop a complete framework that takes these issues into account and that can be used to test multi-field models of inflation against observations.  In this paper, we focus on developing a theoretical framework to intuitively understand and to test two-field models of inflation with non-canonical kinetic terms.  Pioneering work enabling the calculation of the power spectra for general multi-field inflation was done in \cite{KodamaAndSasaki-1984,SalopekEtAl-1989,Salopek-1995,SasakiAndStewart-1995,NakamuraAndStewart-1996,SasakiAndTanaka-1998,HwangAndNoh-2000,NibbelinkAndVanTent-2000,HwangAndNoh-2001,NibbelinkAndVanTent-2001,GongAndStewart-2002,VanTent-2003,LeeEtAl-2005,LangloisAndRenaux-Petel-2008}.\footnote{An interesting complementary approach, which avoids 
discussing an inflaton potential at all, is to treat inflation as an effective field theory 
\cite{CheungEtAl-2007}.}   For two-field inflation, a myriad specific models have been investigated in the literature, e.g., \cite{MukhanovAndZelnikov-1991,HolmanEtAl-1991,PolarskiAndStarobinsky-1992,DeruelleEtAl-1992,PeterEtAl-1994,PolarskiAndStarobinsky-1994,StarobinskyAndYokoyama-1995,PolarskiAndStarobinsky-1995,Garcia-BellidoAndWands-1995a,Garcia-BellidoAndWands-1995b,Garcia-BellidoAndWands-1996,LesgourguesAndPolarski-1997,Chiba-1998,Langlois-1999,KanazawaEtAl-1999,KawasakiAndTakahashi-2001,StarobinskyEtAl-2001,AshcroftEtAl-2002,KadotaAndStewart-2003,BondEtAl-2007,ChoiEtAl-2007,LinAndMcDonald-2007,YangAndMa-2008,AshoorioonEtAl-2009b,AshoorioonEtAl-2009}. For the general case of two-field inflation, an approximate solution for the metric perturbations was found to lowest-order in the slow-roll approximation in \cite{MukhanovAndSteinhardt-1997}.  Later, evolution equations for the adiabatic and entropy perturbations were found for models with canonical kinetic terms \cite{GordonEtAl-2000} and for models with certain non-canonical kinetic terms \cite{DiMarcoEtAl-2002}.  The complete set of curvature, isocurvature, and cross spectra were first estimated for canonical kinetic terms by \cite{BartoloEtAl-2001a} and for some particular non-canonical kinetic terms by \cite{DiMarcoAndFinelli-2005} (and was later extended to include second-order terms by \cite{LalakEtAl-2007}).  However, these analytical results were all derived under the assumption that the slow-roll parameters and the effective entropy mass are approximately constant in the super-horizon limit. In exploring double inflation, Tsujikawa \textit{et al.} (2003) \cite{TsujikawaEtAl-2002} questioned this assumption by numerically illustrating that this assumption often did not hold for their specific inflationary model.  Indeed, Lalak \textit{et al.} (2007) \cite{LalakEtAl-2007} reported similar problems with inaccuracies when comparing these analytical estimates to numerical integration of the full equations of motion, which they used to follow the evolution of the power spectra to high accuracy.  Not using this assumption, a general formalism for parameterizing the evolution of adiabatic and entropy modes was developed in terms of a transfer matrix by \cite{AmendolaEtAl-2001}.  This formalism was used to estimate the power spectra in the slow-roll limit in the case of canonical kinetic terms to first-order in slow-roll in \cite{WandsEtAl-2002} and to second-order in \cite{ByrnesAndWands-2006}.  Also of note, there have been more in-depth investigations of the evolution of and the cross-correlations between curvature and isocurvature modes around horizon-crossing for general models with canonical kinetic terms \cite{BartoloEtAl-2001b} and for models with extremely general Lagrangians in which the entropy modes are not even assumed to propagate at the speed of light \cite{LangloisAndRenaux-Petel-2008,Gao-2009}.

In the first half of this paper (Section \ref{theoretical framework}), we build on these past results by deriving covariant expressions valid to second-order in the slow-roll and slow-turn limits for the unperturbed and perturbed fields; for the curvature, isocurvature, and cross spectra; and for the associated spectral observables, in the case of an arbitrary inflationary potential with completely arbitrary non-canonical terms.  In doing so, we extend the work done by  \cite{DiMarcoEtAl-2002,DiMarcoAndFinelli-2005,LalakEtAl-2007}, which assumes a particular form for the non-canonical kinetic terms, and we improve substantially on analytic estimates that assume that the slow-roll parameters and effective entropy mass can be treated as approximately constant in the super-horizon limit \cite{BartoloEtAl-2001a,DiMarcoAndFinelli-2005,LalakEtAl-2007}.  We also provide new intuition into two-field models, explaining in detail how the general features of the evolution of modes and of the power spectra can largely be inferred from the kinematics of the background field vector and from the curvature of the field manifold.  And lastly, we reduce all two-field models to a set of just a handful of parameters that determine all the inflationary dynamics; these parameters provide a foundation for comparing the general features of all two-field models against each other.  In the second half of this paper (Section \ref{Applications}), we illustrate how to apply our theoretical framework by analyzing four different classes of inflationary models.  For each type of model, we test more than 10,000 different combinations of the initial conditions and a characteristic Lagrangian parameter in order to understand the power spectra they produce.   This paper provides the first thorough investigation of the role of initial conditions in determining two-field power spectra, and demonstrates how to rigorously test and constrain two-field models of inflation, using only minimal assumptions about the end of inflation and reheating.  Though our paper focuses on the two-field case, many of our results are more widely applicable, as in many multi-field models, only two fields dominate during the last several $e$-folds of inflation \cite{NibbelinkAndVanTent-2001}. 

This paper is organized as follows.  In Section \ref{unperturbed equations}, we present exact and approximate expressions for the background equations of motion, develop a framework for understanding the background kinematics, and we introduce a new approximation that we call the ``slow-turn'' approximation.  In Section \ref{perturbed equations}, we derive evolution equations for the perturbations in both the given and kinematical bases, and we find super-horizon solutions for the adiabatic/curvature and entropic/isocurvature modes.  In tandem, we discuss how the evolution of modes can be inferred from the background field kinematics and the field manifold.  We also explain why previous approaches based on assuming the slow-roll parameters and effective entropy mass are approximately constant in the super-horizon limit often lead to substantial inaccuracies in estimating the power spectra.  Thereafter, we find simple expressions for the power spectra, spectral indices, and other observables in Section \ref{Power Spectra}.  We discuss how the general features and the relative sizes of these spectra can be inferred from the background field kinematics and the field manifold, and we discuss when two-field models are effectively equivalent to single-field models.  In Section \ref{Applications}, we apply our theoretical framework to four general classes of inflationary models.  We vary both the initial conditions and a characteristic parameter of the Lagrangian to understand what sorts of kinematical behaviors, power spectra, and spectral observables each class of models can produce.  We use these results to test these models against observations.  We conclude this paper by discussing the general implications for constraining two-field models using observational data.

\section{Theoretical Framework}
\label{theoretical framework}

\subsection{Unperturbed Equations}
\label{unperturbed equations}

In Section \ref{background equations}, we present the background equations of motion for an arbitrary two-field inflationary potential with arbitrary non-canonical kinetic terms.  To simplify the equations, we use covariant vector notation and use the number of $e$-folds, $N$, as our time variable.  In Section \ref{kinematics}, we present a framework for understanding the kinematics of the background field vector.  This framework is very powerful, as it will later allow us to formulate a measure of multi-field effects, to predict the behavior of the field perturbations, and to connect the power spectra to certain features of the inflationary Lagrangian.  In Section \ref{SRA and STA}, we use this kinematical framework to generalize the single-field slow-roll approximation to two-field inflation, dividing it into a slow-roll approximation and a separate ``slow-turn'' approximation.  This new distinction is important because, as we show later, the ``rolling'' and ``turning'' behavior of the background field vector have different effects on the field perturbations and hence the power spectra.  We conclude Section \ref{SRA and STA} by presenting covariant first- and second-order approximations to the background equations.

\subsubsection{Background Equations}
\label{background equations}

We assume the background spacetime is a flat (3+1)-dimensional homogeneous and isotropic spacetime and is described by the familiar Robertson-Walker metric,
\begin{align}
\label{metric}
ds^2 = -dt^2 + a(t)^2 \left[dx^2 + dy^2 + dz^2\right],
\end{align}
where $a(t)$ is the scale factor.  

We investigate inflationary scenarios driven by two scalar fields, $\phi_i$, where $i=1,2$.  We assume Einstein gravity and that the non-gravitational part of the inflationary action is of the form
\begin{align}
\label{action}
S = \int \left[-\frac{1}{2} g^{\mu \nu} G_{ij} \frac{\partial \phi^i}{\partial x^{\mu}} \frac{\partial \phi^j}{\partial x^{\nu}} - V(\phi_1,\phi_2)\right] \sqrt{-g} \, d^4x,
\end{align}
where $ V(\phi_1,\phi_2)$ is the inflationary potential, $g_{\mu \nu}$ is the spacetime metric, and $G_{ij} \equiv G_{ij}(\phi_1,\phi_2)$ determines the form of the kinetic terms in the Lagrangian.\footnote{We assume that the non-canonical kinetic terms can be expressed in the form shown in equation (\ref{action}).  For an even more general inflationary action, see references \cite{LangloisAndRenaux-Petel-2008} and \cite{Gao-2009}, for example.}  We call $G_{ij}$ the field metric, and it can be viewed as inducing a field manifold.  If the kinetic terms are canonical, then $G_{ij} = \delta_{ij}$, and the field manifold reduces to Euclidean space.  In this paper, we allow both the field metric and the inflationary potential to be completely arbitrary.  We refer to any specific combination of a field metric and an inflationary potential as the \textit{inflationary Lagrangian}, or equivalently, the \textit{inflationary model}.

Before we present the background equations of motion, we introduce some notation.  Since we will be taking derivatives with respect to both the spacetime coordinates and the fields, we use Greek indices to represent quantities related to the spacetimes coordinates, $x^{\mu}$, and Latin indices to represent quantities related to the fields, $\phi_i$.  To denote the fields more compactly, we use boldface vector notation, i.e.,
\begin{align}
\label{phi vector}
\boldsymbol{\phi} \equiv (\phi_1,\phi_2), 
\end{align}
and we call $\boldsymbol{\phi}$ the field vector for short.  Note that despite calling $\boldsymbol{\phi}$ the field vector, the fields themselves do not transform as vectors, but rather they represent coordinates on the field manifold.  For true vectorial quantities lying in the tangent and co-tangent bundles of the field manifold, we also use boldface vector notation.  In addition, we use standard inner product notation.  The inner product of two vectors $\boldsymbol{A}$ and $\boldsymbol{B}$ is
\begin{align}
\label{vector prod defn}
\mathbf{A}^{\dag} \mathbf{B} \equiv \mathbf{A} \cdot \mathbf{B} \equiv G_{ij} A^i B^j,
\end{align}
and the norm of a vector $\boldsymbol{A}$ is
\begin{align}
\label{vector norm defn}
|\mathbf{A}| \equiv \sqrt{\mathbf{A}^{\dag} \mathbf{A}},
\end{align}   
where we use the symbol $^{\dag}$ on a naturally contravariant or covariant vector to denote its dual, e.g., $\dot{\boldsymbol{\phi}}^{\dag} \equiv (G_{ij}\dot{\phi}^j)$ and $\boldsymbol{\nabla}^{\dag} \equiv (G^{ij} \nabla_j)$.   We use this set of vector notation both for compactness and so that the background equations presented here can be applied to an arbitrary number of fields.  

Now we summarize the key background equations for multi-field inflation with a non-trivial field metric \cite{SasakiAndStewart-1995,NakamuraAndStewart-1996,NibbelinkAndVanTent-2000,NibbelinkAndVanTent-2001}.   The background density and pressure are found by varying the action in equation (\ref{action}) with respect to the spacetime metric, which gives 
\begin{align}
\label{density and pressure}
\rho = \frac{1}{2} |\dot{\boldsymbol{\phi}}|^2 + V,  \, \ \, \, \, \, \, \, \, \, \, \ \, \, \, \, \, \, \, \,  P = \frac{1}{2} |\dot{\boldsymbol{\phi}}|^2 - V.
\end{align}
The familiar Friedmann equation describing the evolution of the scale factor is derived from the $(0,0)$ component of Einstein's equations, which yields
\begin{align}
\label{Friedmann eqtn wrt t}
H^2 & = \frac{\rho}{3} = \frac{1}{3} \left(\frac{1}{2}|\dot{\boldsymbol{\phi}}|^2 + V\right),
\end{align}
where $H \equiv \frac{\dot{a}}{a}$ is the Hubble parameter and where the reduced Planck mass, $\bar{m} \equiv \frac{m_{Pl}}{\sqrt{8\pi}}$, has been set equal to one.  The equation of motion for the field vector is obtained by imposing zeroth-order covariant conservation of energy,
\begin{align}
\label{continuity equation}
\dot{\rho} + 3H (\rho + P) = 0.
\end{align}
Substituting equation (\ref{density and pressure}) into equation (\ref{continuity equation}) gives
\begin{align}
\label{Phi EoM wrt t}
\frac{D\dot{\boldsymbol{\phi}}}{dt} + 3H \dot{\boldsymbol{\phi}} + \boldsymbol{\nabla}^\dag V & = 0,
\end{align}
where $D$ acting on a contravariant vector $X^i$ means
\begin{align}
\label{D}
DX^i \equiv \nabla_j X^i \, d\phi^j  = dX^i + \Gamma^i_{jk} X^k d\phi^j, 
\end{align}
where $\Gamma^{i}_{jk}$ and $\nabla_j$ are the Levi-Civita connection and the covariant derivative, respectively, associated with the field metric.   $\frac{D\dot{\boldsymbol{\phi}}}{dt}$ represents the covariant rate of change of the field velocity vector with respect to the field manifold, but we call it the acceleration of the field vector for short.  In this paper, we use $D$ and the covariant derivative $\boldsymbol{\nabla}$ to make the equations of motion simpler and manifestly covariant with respect to the field metric.

From here forward, we depart from the standard approach of working in terms of the comoving time, $t$.  Instead, we work in terms of the dimensionless parameter $N$, which represents the logarithmic growth of the scale factor and is related to $t$ by 
\begin{align}
\label{N defn}
dN \equiv d \ln a = H \, dt.
\end{align}
Sasaki and Tanaka \cite{SasakiAndTanaka-1998} were the first to repeatedly use $N$ as the time variable in their equations of motion and to recognize its advantages. Our three primary reasons for using $N$ are 
\begin{enumerate}
\item Because $N$ represents the number of $e$-foldings of the scale factor, it is more directly linked to observables; 
\item It simplifies both the background and perturbed equations of motion; and  
\item It makes the equations of motion dimensionless (since $\boldsymbol{\phi}$ is expressed in units of $\overline{m}$) and hence makes it easier to compare the relative sizes of various terms and parameters in the theory.
\end{enumerate}
These three advantages make it easier to extract physical meaning from the equations of motion.    In working in terms of $N$, we use the short-hand notation 
\begin{align}
\label{d/dN short-hand}
' \equiv \frac{d}{dN}
\end{align}  
to represent differentiation with respect to $N$. 

We now re-cast the set of background equations for multi-field inflation using the time variable $N$.  Using equation (\ref{N defn}), the Friedmann equation (\ref{Friedmann eqtn wrt t}) is written in our notation as
\begin{align}
\label{Friedmann eqtn v1}
H^2 & = \frac{V}{\left(3 - \frac{1}{2}|\boldsymbol{\phi}'|^2\right)}.
\end{align}
To re-cast equation (\ref{Phi EoM wrt t}) in terms of $N$, we change variables and use equation (\ref{Friedmann eqtn v1}), obtaining    
\begin{align}
\label{Phi EoM wrt N v1}
\frac{D\boldsymbol{\phi}'}{dN} + \left(3 + (\ln H)'\right) \boldsymbol{\phi}' + \left(3 - \frac{1}{2} |\boldsymbol{\phi}'|^2\right) \boldsymbol{\nabla}^\dag \ln V & = 0.
\end{align}
To further simplify these two equations of motion, we introduce the canonical slow-roll parameter $\epsilon$, defined as
\begin{align}
\label{epsilon defn}
\epsilon \equiv - \frac{\dot{H}}{H^2} = - (\ln H)',
\end{align}
which represents how much the inflationary expansion deviates from perfect exponential growth.  Combining the logarithmic derivative of equation (\ref{Friedmann eqtn v1}) and equation (\ref{Phi EoM wrt N v1}) yields \cite{SasakiAndTanaka-1998}
\begin{align}
\label{epsilon in terms of field speed}
\epsilon = \frac{1}{2} |\boldsymbol{\phi}'|^2.
\end{align}
Therefore, the parameter $\epsilon$ can also be interpreted simply in terms of the dimensionless speed of the field vector:
\begin{align}
\label{v defn}
v \equiv |\boldsymbol{\phi}'| = \sqrt{2\epsilon}.
\end{align}
Substituting this result into equation (\ref{Friedmann eqtn v1}), the Friedmann equation reduces to \cite{NibbelinkAndVanTent-2000}
\begin{align}
\label{Friedmann eqtn v2}
H^2 & = \frac{V}{(3 - \epsilon)}.
\end{align}
Similarly, equation (\ref{Phi EoM wrt N v1}) simplifies to
\begin{align}
\label{Phi EoM wrt N v2}
\frac{\boldsymbol{\eta}}{(3 - \epsilon)} + \boldsymbol{\phi}' + \boldsymbol{\nabla}^\dag \ln V & = 0,
\end{align}
where $\boldsymbol{\eta}$ represents the covariant acceleration of the field vector and is defined as
\begin{align}
\label{eta defn}
\boldsymbol{\eta} \equiv \frac{D \boldsymbol{\phi}'}{dN}.
\end{align}
We use the symbol $\boldsymbol{\eta}$ in analogy to Nibbelink and Van Tent's \cite{NibbelinkAndVanTent-2000} multi-field slow-roll vector $\boldsymbol{\eta}^{(2)} \equiv \frac{D\dot{\boldsymbol{\phi}}/dt}{H |\dot{\boldsymbol{\phi}}|}$, which was inspired by the standard single-field slow-roll parameter $\eta \equiv \frac{\frac{d^2V}{d\phi^2}}{V} \approx - \left( \frac{\ddot{\phi}}{H\dot{\phi}} - \epsilon\right)$.  

Note that in equation (\ref{Phi EoM wrt N v2}), the field metric appears both in the first term through the Levi-Civita connection and in the third term to raise the index of the covariant gradient operator.  Therefore, the two background fields in $\boldsymbol{\phi}$ can be coupled through the field metric if it is non-trivial, as well as through the gradient of $\ln V$.   

Equation (\ref{Phi EoM wrt N v2}) (along with equation (\ref{epsilon in terms of field speed})) governs the evolution of the fields.   To solve this equation, the final ingredient we need is a particular choice of initial conditions for $\boldsymbol{\phi}$ and $\boldsymbol{\phi}'$.  In two-field inflation, each potential choice of initial conditions corresponds to a different position and roll direction in the two-dimensional field space.  That is, we can view the inflationary Lagrangian as specifying all possible trajectories and the initial conditions as picking one particular trajectory to follow.   

The result of this extra field degree of freedom is that the role of initial conditions in two-field inflation is more complicated than in the single-field case.   First, the landscape of inflationary dynamics that can arise from two fields is potentially much richer and more complex.   Second, there is an uncountable number of initial conditions and hence trajectories corresponding to each inflationary Lagrangian.  As a result, it is possible for two very similar sets of initial conditions to give rise to very different inflationary dynamics.  Fortunately, in some inflationary models, attractor solutions may make the inflationary dynamics essentially independent of the initial conditions.  However, in other models, we must be wary that the inflationary dynamics---and hence observables like the power spectra---may be very sensitive to the initial conditions.   Looking ahead, this means that it is often not sufficient to test the viability of a two-field inflationary model using only one or a handful of initial conditions.  We mention these complications now because it is important to consider them when finding the background solution(s) to the equations of motion.   Later, in Section \ref{Applications} of this paper, we discuss the ramifications of these complications, and we illustrate for the first time how to incorporate sufficient consideration of initial conditions into constraining two-field models of inflation.

Finally, the above equations and considerations apply to finding the background solution for any two-field or indeed any multi-field model of inflation.

\subsubsection{Background Field Kinematics}
\label{kinematics}

In this section, we present a framework for understanding the background field kinematics.  Such a framework will be useful for understanding and classifying the kinematics of a wide variety of two-field inflationary models.  It also helps provide a direct link between the inflationary Lagrangian and the inflationary dynamics.  Moreover, there are other important benefits of such a framework.
\begin{enumerate}
\item It allows us to appreciate the \textit{separate} impacts of different kinematical quantities on key observables.  In the past, these quantities were lumped together and assumed to be small under the standard two-field slow-roll approximation, hence their separate effects had not been fully explored.
\item It allows us to infer how the field perturbations evolve and to predict general features in the power spectra, as we show later in this paper.
\item It helps us to work backwards to determine what features an inflationary model needs to have in order to be consistent with observational data.
\end{enumerate}

In multi-field inflation, we find that the three primary kinematical vectors of interest are
\begin{enumerate}
\item The field vector, $\boldsymbol{\phi}$; 
\item The field velocity, $\boldsymbol{\phi}'$; and 
\item The covariant field acceleration, $\boldsymbol{\eta} \equiv \frac{D\boldsymbol{\phi}'}{dN}$, which was defined through equations (\ref{D}) and (\ref{eta defn}).  
\end{enumerate}
This set of vectors has the intuitive appeal of being directly analogous to the position, velocity, and acceleration vectors in Newtonian mechanics.  The main differences worth emphasizing are (1) that here the fields are to be viewed as the ``position'' coordinates on the field manifold that is induced by the field metric, and (2) that the kinematical vectors are manifestly covariant with respect to the field manifold, which may have non-trivial geometry.

For two-field models of inflation, the above set of three vectors constitutes a set of 6 scalar quantities, with 3 scalars for each of the two given fields.  Along with these three kinematical vectors, we can associate a field basis, where the $\mathbf{e}_{1}$ basis vector points in the direction of the first field, $\phi_1$, while the $\mathbf{e}_{2}$ basis vector points in the direction of the second field, $\phi_2$.  

Although we could consider the field kinematics in the given field basis, there is a second basis in which it is more useful to consider the background kinematics.  This second basis is motivated by the fact that the field perturbations can be decomposed into perturbations parallel and perpendicular to the field trajectory, and that the former constitute bona fide density perturbations, while the latter do not.  In this basis, the $\mathbf{e}_{\parallel}$ basis vector points in the direction of the field velocity, while the $\mathbf{e}_{\perp}$ basis vector points orthogonal to the field trajectory, in the direction of $(\mathbf{I} - \mathbf{e}_{\parallel} \mathbf{e}_{\parallel}^\dag) \boldsymbol{\eta}$, where $\mathbf{I}$ is the $2 \times 2$ identity matrix.  Because this basis is induced by the inflaton vector kinematics, we call it the \textit{kinematical basis}.  This basis has been used before in two-field inflation, initially by \cite{GordonEtAl-2000}, and was extended to multi-field inflation by \cite{NibbelinkAndVanTent-2000,NibbelinkAndVanTent-2001}.  To denote the components of a general vector $\mathbf{A}$ in this basis, we use the notation
\begin{align}
\label{vec cmpts in kine basis}
A_{\parallel} \equiv \mathbf{e}_{\parallel} \cdot \mathbf{A}, \, \, \, \, \, \, \, \, \, \, \, \, \, A_{\perp} \equiv \mathbf{e}_{\perp} \cdot \mathbf{A},
\end{align}
and to denote a particular component of a general matrix $\mathbf{M}$, we use the notation
\begin{align}
\label{matrix cmpts in kine basis}
M_{\parallel \perp} \equiv \mathbf{e}_{\parallel}^\dag \mathbf{M} \mathbf{e}_{\perp}, \, \, \, \, \, \, \, \, \, \, \, \, \, \, \,  etc. 
\end{align}

Let us consider our original kinematical vectors, but in the kinematical basis.  First, the field vector, $\boldsymbol{\phi}$, decomposes into components $\phi_{\parallel}$ and $\phi_{\perp}$.  In this paper, we will not need to use this decomposition of the field vector, so we do not consider it further.\footnote{The decomposition of the background field vector into components $\phi_{\parallel}$ and $\phi_{\perp}$ can indeed be useful.  For example, Nibbelink and Van Tent \cite{NibbelinkAndVanTent-2001} calculated the adiabatic part of gravitational potential spectrum for multi-field quadratic potentials  in the conventional slow-roll limit and found that the particular solution depends on  $\phi_{\perp}$.}  Second, in this basis, the field velocity points along the $\mathbf{e}_{\parallel}$ basis vector.  Its components in this basis are $\boldsymbol{\phi}' = (v,0)$, where we defined $v$ earlier as the field speed.  Lastly, there is the field acceleration, which decomposes into a component parallel to the field trajectory, $\eta_{\parallel}$, and a component perpendicular to the field trajectory, $\eta_{\perp}$.  

We now collect the results of this decomposition into a set of three scalar quantities to represent the main field kinematics in any two-field model of inflation.  The first quantity in this trio is the field speed, $\textit{v}$, which typically appears in the equations of motion via the quantity $\epsilon = \frac{1}{2} v^2$; we use $\textit{v}$ and $\epsilon$ interchangeably as our first kinematical scalar quantity.  For the second quantity, we choose the term $\frac{\eta_{\parallel}}{v} = (\ln v)'$ because $\eta_{\parallel}$ often appears in the equations of motion in the combination $\frac{\eta_{\parallel}}{v}$.  The quantity $\frac{\eta_{\parallel}}{v}$ has a physical meaning: it measures the logarithmic rate of change of the field speed, $v$, and for this reason, we call it the \textit{speed up rate}.  The third quantity we include in our trio is $\frac{\eta_{\perp}}{v}$, as $\eta_{\perp}$ often appears in the equations of motion in the combination $\frac{\eta_{\perp}}{v}$.  The quantity $\frac{\eta_{\perp}}{v}$ measures how fast the field velocity is changing direction with respect to the field manifold.  We can see this by considering the rate of change of the basis vectors: 
\begin{align}
\label{De_p/dN}
\frac{D\mathbf{e}_{\parallel}}{dN} = \frac{\eta_{\perp}}{v} \, \mathbf{e}_{\perp}, \, \, \, \, \, \, \,  \, \, \, \, \, \, \, \, \, \, \, \frac{D\mathbf{e}_{\perp}}{dN} = - \frac{\eta_{\perp}}{v} \, \mathbf{e}_{\parallel}.
\end{align}
Since $\mathbf{e}_{\parallel}$ also represents the direction of the field trajectory, this means that the quantity
\begin{align}
\label{turn rate}
\left|\frac{D\mathbf{e}_{\parallel}}{dN}\right| = \frac{\eta_{\perp}}{v}
\end{align}
tells us how quickly the field trajectory is changing direction along the field manifold.  We therefore call $\frac{\eta_{\perp}}{v}$ the \textit{turn rate}.  Note that unlike the speed up rate, which can be either positive or negative depending on whether the field speed is increasing or decreasing, the turn rate is always positive.   One can therefore think of our three kinematical scalars as simply representing the field speed ($v$) and the rate of change of the magnitude ($\frac{\eta_{\parallel}}{v}$) and direction ($\frac{\eta_{\perp}}{v}$) of the field velocity.  Our three kinematical quantities in the kinematical basis can be concisely written as $v$ and the vector $\frac{\boldsymbol{\eta}}{v}$, and they are summarized in Table 1.  

\begin{table}[t]
\centering
\renewcommand{\arraystretch}{1.5}
\begin{tabular}{|cccccc|}
\hline
\multicolumn{6}{|c|}{\normalsize \rule{0cm}{0.5cm} Table 1.  Covariant Field Vector Kinematics} \tabularnewline[0.5ex] 
\hline \multicolumn{6}{|c|}{$\, $}\tabularnewline [-3.7ex] \hline 
\multicolumn{6}{|c|}{\normalsize Main Kinematical Vectors} \tabularnewline \hline 
\multicolumn{2}{|x{1.07in}|}{Field Vector}&\multicolumn{2}{x{1.07in}}{Field Velocity}&\multicolumn{2}{|x{1.07in}|}{Field Acceleration} \tabularnewline \hline
\multicolumn{2}{|c|}{\rule{0cm}{0.53cm}\normalsize $\boldsymbol{\phi}$}& \multicolumn{2}{c}{\normalsize $\boldsymbol{\phi}'$}&\multicolumn{2}{|c|}{$\normalsize \boldsymbol{\eta} \equiv$ \large $\frac{D\boldsymbol{\phi}'}{dN}$} \tabularnewline [1.1ex] \hline
\multicolumn{6}{|c|}{$\, $}\tabularnewline [-3.7ex] \hline 
\multicolumn{6}{|c|}{\normalsize Main Kinematical Scalars} \tabularnewline \hline 
\multicolumn{2}{|x{1.07in}|}{Field Speed}&\multicolumn{2}{x{1.07in}}{Speed Up Rate}&\multicolumn{2}{|x{1.07in}|}{Turn Rate} \tabularnewline \hline
\multicolumn{2}{|c|}{\large \rule{0cm}{0.65cm}$v$}& \multicolumn{2}{c}{\Large$\frac{\eta_{\parallel}}{v}$ \normalsize $= (\ln v)'$}&\multicolumn{2}{|c|}{\Large $\frac{\eta_{\perp}}{v}$ \normalsize $=$ \large $\left|\frac{D\mathbf{e}_{\parallel}}{dN}\right|$}\tabularnewline [1.7ex]
\hline
\end{tabular}
\end{table}

Of our three kinematical scalars, the first two---the field speed and the speed up rate---also characterize single-field models of inflation.  But the third kinematical scalar, the turn rate, is necessarily zero in single-field inflation since the field trajectory has no ability to turn in a one-dimensional field space.   Therefore, it is the turn rate that distinguishes between single-field and multi-field models of inflation.  Indeed, other authors have recognized that the  `turning' of the field trajectory is the true marker of multi-field behavior (e.g., \cite{LythAndRiotto-1998,GordonEtAl-2000,NibbelinkAndVanTent-2000,NibbelinkAndVanTent-2001}).  Here, we are building on this work by formalizing the idea of a quantity that precisely describes how quickly the background field trajectory is changing direction, and we do so in a completely covariant manner that can be applied to any arbitrary inflationary Lagrangian, including those with non-canonical kinetic terms.  

Taking this idea one step further, we introduce a new way of understanding the background kinematics.   We propose viewing the first two kinematical scalars---the field speed and the speed up rate---as characterizing the `single-field type' behavior of any inflationary model.  We propose using the third kinematical scalar, the turn rate, to characterize the degree of multi-field behavior in any inflationary model.  This provides an intuitive way to parse and analyze the kinematical behavior of any model of inflation into parts shared in common with single-field models of inflation and parts unique to multi-field inflation.   Moreover, with this understanding, we introduce a new idea: the ratio of the speed up rate to the turn rate, $\frac{\eta_{\perp}}{v} / \frac{\eta_{\parallel}}{v}$, can be used to indicate the relative `proportion' of multi-field to single-field behavior in any inflationary scenario.  That is, when $\frac{\eta_{\perp}}{v} \ll \frac{\eta_{\parallel}}{v}$, the background trajectory has very little curvature and hence resembles a single-field trajectory, whereas when $\frac{\eta_{\perp}}{v} \gtrsim \frac{\eta_{\parallel}}{v}$, the trajectory has substantial curvature with respect to the field manifold, indicating significant multi-field behavior.   We will later show that this same ratio,  $\frac{\eta_{\perp}}{v} / \frac{\eta_{\parallel}}{v}$, also indicates how much the evolution of adiabatic density modes is affected by mode sourcing.   Hence, the ratio of the turn rate to speed up rate can be used to signify the relative importance of multi-field effects for both the unperturbed and perturbed solutions. 

The full power of this kinematical framework will become apparent later, when we show how it can provide important insight into the evolution of field perturbations and into the power spectra, and how it can help us connect features of the inflationary Lagrangian to the spectral observables.

\subsubsection{Slow-roll and Slow-turn Approximations}
\label{SRA and STA}

In this section, we consider approximations to the background equations of motion.  Using the above kinematical framework, we generalize the single-field slow-roll approximation to two-field inflation and introduce a new approximation: the slow-turn approximation.  At the end of this section, we use these approximations to derive approximate expressions for the background equations.

Our natural starting point is the single-field slow-roll approximation, which consists of two simplifying assumptions: (1) the potential dominates the energy density ($V \gg \dot{\phi}^2$), and (2) the field acceleration is small enough to be neglected ($\ddot{\phi} \ll 3H \dot{\phi}$) in the equation of motion for the field vector (\ref{Phi EoM wrt t}).  When both conditions are met, the field is slowly changing or ``slow-rolling'' with respect to the Hubble time.  And therefore, by equations (\ref{Friedmann eqtn wrt t}) and (\ref{Phi EoM wrt t}), the potential must change slowly, too.  Together, these two conditions ensure that the potential serves as an effective cosmological constant, driving nearly exponential growth of the scale factor.  Nearly exponential expansion in turn guarantees a nearly scale-invariant spectrum of adiabatic density fluctuations, in good agreement with observational data.  

To generalize the slow-roll approximation to two-field inflation, we follow a common approach in generalizing the first slow-roll condition, but not in generalizing the second condition.  We generalize the first slow-roll condition as the inflationary expansion is nearly exponential, which can be mathematically expressed as
\begin{align}
\label{SRA Cond 1}
\epsilon \equiv -(\ln H)' = \frac{1}{2} v^2 \ll 1.
\end{align}
Equation (\ref{SRA Cond 1}) shows that the field speed, our first kinematical scalar, must be small for the inflationary expansion to be nearly exponential.  Just like its single-field counterpart, when the first slow-roll condition in equation (\ref{SRA Cond 1}) is satisfied, the potential does indeed dominate the energy density, which is clear from equation (\ref{Friedmann eqtn v2}). 

In generalizing the second slow-roll condition, however, we depart from the common approach of individually constraining each of the matrix components of the Hessian of $V$ (e.g., \cite{MukhanovAndSteinhardt-1997,BartoloEtAl-2001a,WandsEtAl-2002,DiMarcoAndFinelli-2005,ByrnesAndWands-2006}), i.e.,
\begin{align}
\label{Std SRA Cond 2}
\left|\frac{\nabla_i \nabla_j V}{V}\right| \ll 1.
\end{align}
We also do not follow the other common approach (e.g., \cite{SasakiAndTanaka-1998,NibbelinkAndVanTent-2000,NibbelinkAndVanTent-2001}), which is to generalize the second slow-roll condition as 
\begin{align}
\label{Std SRA Cond 2 wrt t}
\left|\frac{D\boldsymbol{\dot{\phi}}}{dt}\right| \ll 3H |\boldsymbol{\dot{\phi}}|.
\end{align} 
In our formalism, equation (\ref{Std SRA Cond 2 wrt t}) would be equivalent to the condition
\begin{align}
\label{Std SRA Cond 2 wrt N}
\left|\frac{\boldsymbol{\eta}}{3 - \epsilon}\right| \ll v.
\end{align}
The reason why we depart from these two standard approaches as represented by equations (\ref{Std SRA Cond 2}) and (\ref{Std SRA Cond 2 wrt N}) is because they are more stringent than the minimum condition needed to ensure that the potential is slowly changing.  Also, they lump together and simultaneously constrain two different aspects of the background kinematics---the speed up rate and the turn rate.   These two rates have very different impacts on the inflationary dynamics.  Moreover, the two rates may not necessarily be small at the same time, and different approximations can be made depending on whether one or both of the two rates are small.  So departing from convention, we treat the turn rate as distinct from the speed up rate to embrace the facts that not only the individual sizes of the two rates matters but also their relative sizes to each other matters.   

Instead, we redefine the second slow-roll condition more narrowly as the deviation from exponential expansion is slowly changing.  Using equations (\ref{epsilon defn}) and (\ref{epsilon in terms of field speed}), it can be shown that this is equivalent to requiring the speed up rate, our second kinematical scalar, to be small, 
\begin{align}
\label{SRA Cond 2}
\left|\frac{\eta_{\parallel}}{v}\right| \ll 1.
\end{align} 
This less restrictive version of the second slow-roll condition in equation (\ref{SRA Cond 2}) along with equation (\ref{SRA Cond 1}) is sufficient to guarantee that the potential is slowly changing, which can be seen from equation (\ref{Friedmann eqtn v2}).  

As for the turn rate, our third kinematical scalar, we instead endow it with its own separate condition and set of approximations.  If the turn rate is sufficiently small such that
\begin{align}
\label{STA Cond}
\frac{\eta_{\perp}}{v}  \ll 1,
\end{align}
then the field trajectory changes direction slowly, and we say that the field vector is \textit{slowly turning} or is exhibiting \textit{slow-turn} behavior.  

These distinctions between slow-roll and slow-turn behavior are more important than they might initially appear.  As we argued earlier, the field speed and speed up rate represent single-field type behavior, while the turn rate represents the degree of multi-field behavior.  Therefore, the importance of our alternative framework is that we have separated the limits on single-field-type behavior (our slow-roll conditions) from the limits on multi-field behavior (our slow-turn condition).   We later illustrate the full benefits of this disaggregation by showing that these two limiting cases of behavior have different implications for the evolution of perturbations and the power spectra.

Table 2 summarizes our conditions for slow-roll and slow-turn behavior.  Since two-field models often exhibit both slow-roll and slow-turn behavior at the same time, we call the combined slow-roll and slow-turn limits the SRST limit for short.  

\begin{table}[t]
\centering
\renewcommand{\arraystretch}{1.5}
\begin{tabular}{|x{1.635in}|x{1.635in}|}
\hline
\multicolumn{2}{|c|}{\normalsize \rule{0cm}{0.5cm} Table 2.  Kinematical Limits} \tabularnewline[0.5ex] 
\hline \multicolumn{2}{|c|}{$\, $}\tabularnewline[-3.7ex] \hline  \tabularnewline[-3.4ex]  \normalsize Slow-Roll Conditions & \normalsize Slow-Turn Condition \tabularnewline[0.5ex] \hline
\rule{0cm}{0.55cm} (1) \, \large$\epsilon$ \normalsize $= \frac{1}{2} v^2 \ll 1,$ & \multirow{2}{*}{\rule{0cm}{0.55cm} \,  \Large$\frac{\eta_{\perp}}{v}$\normalsize $ \, \ll 1 \, \, \, \, \, $} \tabularnewline [1ex]
(2) \, \Large$\left|\frac{\eta_{\parallel}}{v}\right|$\normalsize $\, \ll 1 \, \, \, \, \, \, \, \, \, $ &  \tabularnewline [1.6ex]
\hline
\end{tabular}
\end{table}

Now armed with our slow-roll and slow-turn conditions, we return to the background equations to see how they simplify in the slow-roll and slow-turn limits.   We denote the lowest-order approximation to a function $f$ by the notation $f^{(1)}$ and the next-to-lowest-order approximation by $f^{(2)}$, where $f^{(2)}$ includes \textit{both} the lowest and next-to-lowest-order terms.  The Friedmann equation (\ref{Friedmann eqtn v2}) depends only on $\epsilon$ and not on the speed up or turn rates.  To lowest-order in the slow-roll limit, it reduces to
\begin{align}
\label{Friedmann eqtn in SRA}
(H^{(1)})^2 = \frac{1}{3} V.
\end{align}
The background field equation (\ref{Phi EoM wrt N v2}) can be re-arranged as
\begin{align}
\label{Phi EoM wrt N v3}
\boldsymbol{\phi}' + \boldsymbol{\nabla}^\dag \ln V = - \frac{\boldsymbol{\eta}}{(3 - \epsilon)}. 
\end{align}
Given the slow-roll and slow-turn conditions in equations (\ref{SRA Cond 2}) and (\ref{STA Cond}), the right-hand side of equation (\ref{Phi EoM wrt N v3}) therefore represents deviations from the SRST limit: in the slow-roll limit, $\eta_{\parallel}$ can be neglected, while in the slow-turn limit, $\eta_{\perp}$ can be neglected.  In the full SRST limit, equation (\ref{Phi EoM wrt N v3}) reduces to
\begin{align}
\label{Phi EoM wrt N in SRST}
\boldsymbol{\phi}'^{(1)} = - \boldsymbol{\nabla}^\dag \ln V. 
\end{align}

\begin{figure*}[t]
\normalsize
\centering
(a) $\,$ First-Order SRST Approximation
\vskip 2 pt
\includegraphics[height=63.7mm]{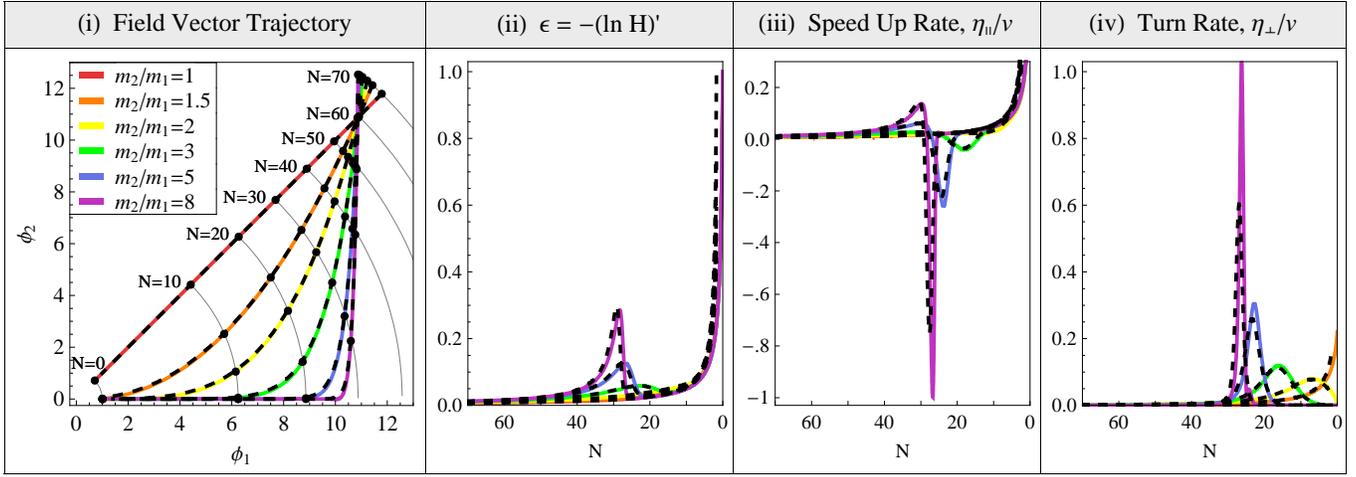}
\vskip 10 pt
(b) $\,$ Second-Order SRST Approximation
\vskip 2 pt
\includegraphics[height=63.7mm]{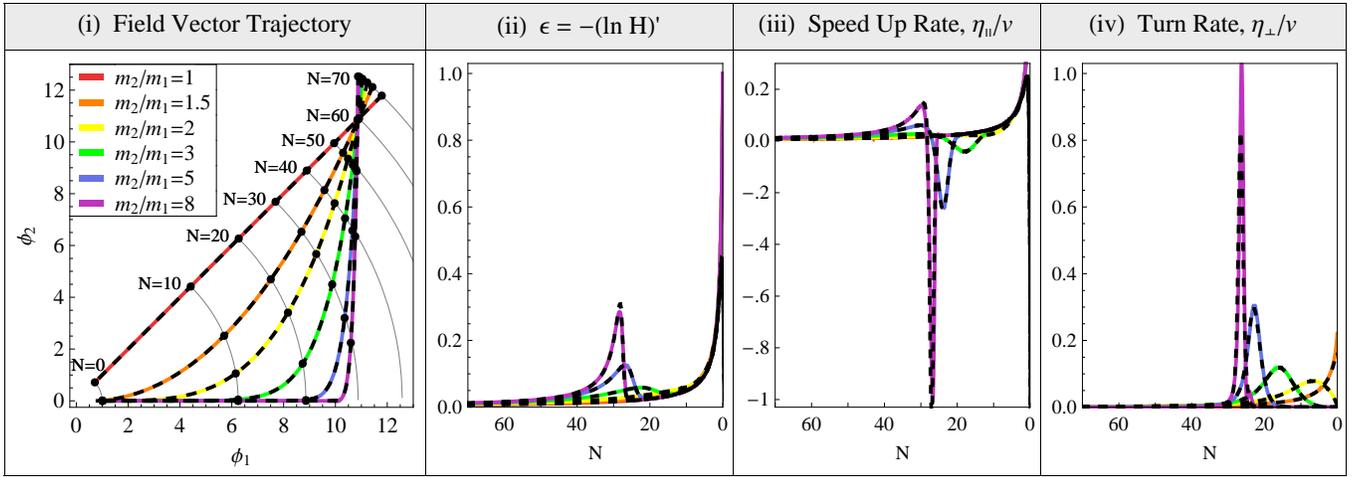}
\label{SRST Approxs}
\caption{The exact (colored lines) and approximate (black dashed lines) solutions are depicted for both the (a) first-order and (b) second-order SRST approximations.  Shown are the field trajectory and the three kinematical scalars (with $\epsilon$ used in place of $v$) for six different values of the mass ratio $\frac{m_2}{m_1}$ for the double quadratic potential $V = \frac{1}{2} m_1^2 \phi_1^2 + \frac{1}{2} m_2^2 \phi_2^2$ with canonical kinetic terms.   The same initial conditions were assumed $60$ $e$-folds before the end of inflation, and the $x$-axis for plots (ii) - (iv) represents the number of $e$-folds before inflation ends.  Only the trajectory corresponding to $\frac{m_2}{m_1}=8$ violates the slow-roll and slow-turn conditions and only for less than 2 $e$-folds.  Overall, the SRST approximation is a good approximation as long as the gradient of $\ln V$ is not too large and is not changing rapidly in magnitude or direction.}
\end{figure*}
 
When the field vector is in the SRST limit, we can find approximations for each of our three key kinematical quantities directly in terms of the potential.  Starting with equation (\ref{Phi EoM wrt N in SRST}), we find that to first-order in the SRST limit, the field speed is
\begin{align}
\label{v in SRST}
v^{(1)} = |\boldsymbol{\nabla} \ln V|,
\end{align} 
and hence
\begin{align}
\label{epsilon in SRST}
\epsilon^{(1)} = \frac{1}{2} |\boldsymbol{\nabla} \ln V|^2.
\end{align}
Differentiating equation (\ref{Phi EoM wrt N in SRST}) gives
\begin{align}
\label{eta in SRST}
\boldsymbol{\eta}^{(1)} = - \mathbf{M} \, \boldsymbol{\phi}'^{(1)} = \mathbf{M} \, \boldsymbol{\nabla}^\dag \ln V , 
\end{align}
where we define the \textit{mass matrix}, $\mathbf{M}$, as
\begin{align}
\label{M defn}
\mathbf{M} \equiv \boldsymbol{\nabla}^\dag \boldsymbol{\nabla}\ln V.
\end{align}
(As an aside, we define the mass matrix differently from Nibbelink and Van Tent \cite{NibbelinkAndVanTent-2000}, who defined a mass matrix for general multi-field inflation as $\mathbf{M}^2 \equiv \boldsymbol{\nabla} \boldsymbol{\nabla}^\dag V$, because when using $N$ as the time variable, it is more natural to define the mass matrix as a dimensionless quantity.)  From equation (\ref{eta in SRST}), to first-order in the SRST limit, the speed up rate is 
\begin{align}
\label{speed up rate in SRST}
\left(\frac{\eta_{\parallel}}{v}\right)^{(1)} = - M^{(1)}_{\parallel \parallel}, 
\end{align}
where 
\begin{align}
\label{M cmpts in SRST}
M^{(1)}_{\parallel \parallel} \equiv (\mathbf{e}_{\parallel}^{(1)})^\dag \mathbf{M} \mathbf{e}_{\parallel}^{(1)}
\end{align}
and 
\begin{align}
\label{e para in SRST}
\mathbf{e}_{\parallel}^{(1)} = - \frac{\boldsymbol{\nabla}^\dag \ln V}{|\boldsymbol{\nabla} \ln V|}.
\end{align}  
Similarly, the turn rate can be approximated by
\begin{align}
\label{turn rate in SRST}
\left(\frac{\eta_{\perp}}{v}\right)^{(1)} = - M^{(1)}_{\parallel \perp}, 
\end{align}
where $M^{(1)}_{\parallel \perp}$ is defined in analogy to $M^{(1)}_{\parallel \parallel}$ and where  $\mathbf{e}_{\perp}^{(1)}$ is orthogonal to $\mathbf{e}_{\parallel}^{(1)}$, in the direction that makes the turn rate positive.   These coefficients of our mass matrix are related to quantities originally defined and used by \cite{GordonEtAl-2000} and \cite{WandsEtAl-2002} for two-field inflation with canonical kinetic terms and by \cite{NibbelinkAndVanTent-2000} for multi-field inflation with non-canonical kinetic terms.  These authors first thought to project the Hessian of $V$ onto the kinematical basis vectors, as these quantities appear naturally in the equations of motion for the field perturbations when working in the kinematical basis.  Nibbelink and Van Tent \cite{NibbelinkAndVanTent-2000} also effectively related their mass matrix back to their kinematical parameters to first-order in the SRST limit, similarly to what we have done above.

Let us step back and consider the full importance of equations (\ref{v in SRST}), (\ref{speed up rate in SRST}), and (\ref{turn rate in SRST}).  These equations show that the gradient and the covariant Hessian of $\ln V$ provide insight into the background field kinematics: the norm of the gradient approximates the field speed, while the $(\parallel,\parallel)$ and $(\parallel,\perp)$ components of the mass matrix approximate the speed up rate and turn rate, respectively.  That is, from only the inflationary Lagrangian and the field coordinates, we can estimate the background field kinematics without solving the equations of motion.  We will later show that this very important bridge between the background kinematics and the inflationary Lagrangian allows one to connect certain features of the Lagrangian to certain features in the power spectra.

When more accuracy is desired, it is useful to have next-to-lowest-order approximations for the various kinematical quantities.  The second-order expressions for the field velocity and acceleration can be obtained by applying the operator $\left[\mathbf{I} - \frac{1}{(3-\epsilon)} \frac{D}{dN}\right] \approx \left[\mathbf{I} + \frac{1}{3} \boldsymbol{\nabla}^\dag \ln V \boldsymbol{\nabla} \right]$ to the corresponding first-order expressions.  This gives the compact expressions
\begin{align}
\label{velocity and eta in SRST O2}
\boldsymbol{\phi}'^{(2)} & = - \left[\mathbf{I} + \frac{1}{3} \mathbf{M}\right] \boldsymbol{\nabla}^\dag \ln V, \nonumber \\
\boldsymbol{\eta}^{(2)} & = - \left[\mathbf{M} + \frac{1}{3} \mathbf{M}^2 + \frac{1}{3} \boldsymbol{\nabla}^\dag \ln V \, \boldsymbol{\nabla} \mathbf{M} \right] \boldsymbol{\phi}'^{(2)}, \\ 
& = \left[\mathbf{M} + \frac{2}{3} \mathbf{M}^2 + \frac{1}{3} \boldsymbol{\nabla}^\dag \ln V \, \boldsymbol{\nabla} \mathbf{M} \right] \boldsymbol{\nabla}^\dag \ln V. \nonumber
\end{align}
Therefore, the second-order expressions for our kinematical quantities are
\begin{widetext}
\begin{align}
\label{kine quantities in SRST O2}
v^{(2)} = & |\boldsymbol{\nabla} \ln V|\left[1 + \frac{1}{3} M_{\parallel \parallel} \right], \nonumber \\
\epsilon^{(2)} \approx & \frac{1}{2} |\boldsymbol{\nabla} \ln V|^2 \left[1 + \frac{2}{3} M_{\parallel \parallel} \right], \\
\left(\frac{\eta_{\parallel}}{v}\right)^{(2)} = & - M_{\parallel \parallel} - \frac{1}{3}(M_{\parallel \parallel})^2 - \frac{1}{3}(M_{\parallel \perp})^2 - \frac{1}{3} (\boldsymbol{\nabla}^\dag\ln V \boldsymbol{\nabla} \mathbf{M})_{\parallel \parallel}, \nonumber \\ 
\left(\frac{\eta_{\perp}}{v}\right)^{(2)} = & - M_{\parallel \perp} - \frac{1}{3} M_{\parallel \parallel} M_{\parallel \perp} - \frac{1}{3} M_{\parallel \perp} M_{\perp \perp} - \frac{1}{3} (\boldsymbol{\nabla}^\dag\ln V \boldsymbol{\nabla} \mathbf{M})_{\perp \parallel},  \nonumber 
\end{align}
\end{widetext}
where it is implied that the matrix components are with respect to the second-order expressions for the kinematical basis vectors.  Lastly, using equation (\ref{kine quantities in SRST O2}), the Friedmann equation to next lowest-order in slow-roll is
\begin{align}
\label{Friedmann eqtn in SRA O2}
(H^{(2)})^2 = \frac{V}{3} \left(1 + \frac{1}{6} |\boldsymbol{\nabla} \ln V|^2 \right).
\end{align}

Finally, we emphasize that although we have introduced approximations for the slow-roll and slow-turn limits, we will not restrict ourselves to the small subset of inflationary models that exhibit strictly SRST behavior up until close to the end of inflation.  Indeed, some common types of two-field models can temporarily violate the slow-roll and/or slow-turn conditions for a handful of $e$-folds, only to satisfy the conditions again thereafter.  Fortunately, it turns out that we can use the slow-roll and slow-turn approximations at first- or second-order even when the field vector velocity is moderately large or is changing moderately fast in magnitude or direction, as shown in Figure 1.  Therefore, we can apply either the first- or second-order slow-roll and slow-turn approximations to a wide range of two-field inflationary scenarios.

\subsection{Perturbed Equations}
\label{perturbed equations}

We now turn our attention to the field perturbations.  Using our formalism and working in terms of gauge-invariant quantities defined in Section \ref{metric perturbations}, we simplify the standard equation for the evolution of the field perturbations in Section \ref{field perturbations}.  We show that the evolution of modes is determined by the mass matrix, $\mathbf{M}$, plus typically small corrections.  Thereafter, in Sections \ref{adiabatic modes} and \ref{entropy modes}, we decompose the field perturbations into adiabatic and entropy modes, and then build on the work of \cite{GordonEtAl-2000,DiMarcoEtAl-2002} to derive covariant equations for the evolution of both mode types in the case of a completely arbitrary field metric.  We show how the evolution of these two modes and the relative degree of mode sourcing (the \textit{multi-field effects}) can be inferred from the background field kinematics and the curvature of the field manifold.  In addition, for the super-horizon limit, we present a simple exact equation of motion for the adiabatic modes and two semi-analytic approximations for the amplitude of entropy modes.  We also explain why previous analytic estimates which assume that the effective entropy mass and kinematical quantities are approximately constant in the super-horizon limit \cite{BartoloEtAl-2001a,DiMarcoAndFinelli-2005,LalakEtAl-2007} generally produce large errors in estimating the power spectra and hence why a semi-analytic approach is needed.  Finally, in Section \ref{curvature modes}, we use these results to derive expressions for the super-horizon evolution of the related curvature and isocurvature modes.

\subsubsection{Metric Perturbations}
\label{metric perturbations}

Since the primordial density perturbations are small, we work to linear order in both the spacetime metric and the scalar field perturbations.  At linear order, the scalar field perturbations decouple from the vector and tensor metric perturbations, so we only need to consider the coupling between the scalar field and metric perturbations \cite{Bardeen-1980,Stewart-1990}.  Considering only scalar perturbations to the metric, the most general metric for a perturbed (3+1)-dimensional Friedmann-Robertson-Walker spacetime can be written as \cite{Bardeen-1980,KodamaAndSasaki-1984,MukhanovEtAl-1992} 
\begin{align}
\label{perturbed metric}
ds^2 = &-(1 + 2A)dt^2 + 2a(t) \partial_iB dx^i dt \nonumber \\
& + a^2(t)\left[(1 - 2\psi)\delta_{ij} + 2\partial_i \partial_j E\right]dx^i dx^j,
\end{align}
where the functions $A$, $B$, $\psi$, and $E$ completely parametrize the set of all possible scalar metric perturbations (and where here only the scripts $i$ and $j$ refer to the spatial spacetime coordinates, not the fields). 

To avoid the complications of working in a particular gauge, we work instead in terms of so-called gauge-invariant quantities.  Such an approach also has the benefit of ensuring that we work only in terms of physical quantities.  Considering the four scalar metric perturbations, two linear combinations of them represent gauge modes, and a third is linearly related to the others since scalar field theories produce no anisotropic stress to linear order \cite{MukhanovEtAl-1992}.  That leaves us needing to choose a single gauge-invariant quantity to represent the scalar metric perturbations.  The gauge-invariant quantity we use is the Bardeen variable $\Psi$ \cite{Bardeen-1980,MukhanovEtAl-1992}, defined as
\begin{align}
\label{Bardeen variable}
\Psi \equiv \psi + (aH) \left[(aH)E' - B\right], 
\end{align}
which equals the metric perturbation in the longitudinal gauge.  

To represent the field perturbations in gauge-invariant form, we work in terms of the multi-field version of the Mukhanov-Sasaki variable \cite{Sasaki-1986,Mukhanov-1988},
\begin{align}
\label{Mukhanov-Sasaki variable}
\boldsymbol{\delta \phi}_f \equiv \boldsymbol{\delta \phi} + \psi \, \boldsymbol{\phi}', 
\end{align}
which equals the field vector perturbation in the flat gauge.  This choice has a very important benefit: it decouples the field perturbations from the metric perturbation (but not vice versa), eliminating the need to solve a coupled set of field and metric perturbation equations.   Because of this and the fact that we will work in terms of the power spectra of the scalar field perturbations rather than of the metric perturbations, we will not have to consider $\Psi$ any further.  Therefore, in the remainder of this section, we consider only the field perturbations.

\subsubsection{The Field Perturbation Equation}
\label{field perturbations}

For multi-field inflation, the evolution equation for the field perturbations can be found by perturbing the equation of motion for the background fields (see, e.g., \cite{NakamuraAndStewart-1996}).  The standard result in Fourier space \cite{NakamuraAndStewart-1996} is expanded as
\begin{widetext}
\begin{align}
\label{Delta Phi EoM wrt t}
\frac{D^2 \boldsymbol{\delta \phi}_f}{dt^2} + 3H \frac{D{\boldsymbol{\delta \phi}}_f}{dt} \, +  \left(\frac{k}{a}\right)^2 \boldsymbol{\delta \phi}_f  = - \left[\boldsymbol{\nabla}^\dag \boldsymbol{\nabla} V - \left(3 - \frac{\dot{H}}{H^2}\right)\dot{\boldsymbol{\phi}}\dot{\boldsymbol{\phi}}^\dag - \frac{1}{H} \frac{D\dot{\boldsymbol{\phi}}}{dt} \dot{\boldsymbol{\phi}}^\dag - \frac{1}{H} \dot{\boldsymbol{\phi}} \frac{D\dot{\boldsymbol{\phi}}^\dag}{dt} - \mathbf{R}(\boldsymbol{\dot{\phi}},\boldsymbol{\dot{\phi}}) \right] \boldsymbol{\delta \phi}_f,
\end{align}
\end{widetext}  
where $k$ is the comoving wavenumber and the matrix $\mathbf{R}(\boldsymbol{\dot{\phi}},\boldsymbol{\dot{\phi}})$ is defined as \cite{NibbelinkAndVanTent-2000}
\begin{align}
\label{R defn}
R^a_{\, \, d}(\boldsymbol{\dot{\phi}},\boldsymbol{\dot{\phi}}) \equiv R^a_{\, \, bcd} \dot{\phi}^b \dot{\phi}^c,
\end{align}
where $R^a_{\, \, bcd}$ is the Riemann curvature tensor associated with the field metric.  To simplify equation (\ref{Delta Phi EoM wrt t}), we change variables using equation (\ref{N defn}), use equation (\ref{epsilon defn}), and substitute equation (\ref{Phi EoM wrt N v2}) for each instance of $\boldsymbol{\phi}'$.  Also, we simplify the matrix $\mathbf{R}(\boldsymbol{\dot{\phi}},\boldsymbol{\dot{\phi}})$ using the Bianchi identities, which for a general two-dimensional metric yield
\begin{align}
\label{2D Bianchi Result}
R_{abcd} = \frac{1}{2} R (G_{ac}G_{bd} - G_{ad}G_{bc}),
\end{align}
where $R$ is the Ricci scalar, which equals twice the Gaussian curvature, or equivalently, the product of the two principle curvatures of the field manifold.  Substituting equation (\ref{2D Bianchi Result}) into equation (\ref{R defn}) and using equation (\ref{epsilon in terms of field speed}) gives
\begin{align}
\label{Ricci Scalar}
\mathbf{R}(\boldsymbol{\dot{\phi}},\boldsymbol{\dot{\phi}}) & = \frac{1}{2} H^2 R \left(\boldsymbol{\phi}' \boldsymbol{\phi'}^\dag - 2\epsilon \mathbf{I} \right) = - \epsilon H^2 R \, \mathbf{e}_{\perp} \mathbf{e}_{\perp}^\dag,
\end{align}
where in the last step, we used the completeness relation for our kinematical basis vectors.  Finally, after substituting equation (\ref{Ricci Scalar}) into equation (\ref{Delta Phi EoM wrt t}) and using equation (\ref{Friedmann eqtn v2}), we find
\begin{align}
\label{Delta Phi EoM wrt N}
\frac{1}{(3 - \epsilon)} \frac{D^2\boldsymbol{\delta \phi}_f}{dN^2} + & \frac{D\boldsymbol{\delta \phi}_f}{dN} +  \left(\frac{k^2}{a^2 V}\right) \boldsymbol{\delta \phi}_f \nonumber \\ & = - \left[\mathbf{\tilde{M}} + \frac{\boldsymbol{\eta}\boldsymbol{\eta}^\dag}{(3-\epsilon)^2}\right] \boldsymbol{\delta \phi}_f,
\end{align}
where the \textit{effective mass matrix},\footnote{For comparison, Nibbelink and Van Tent defined an effective mass matrix as  $\tilde{\mathbf{M}}^2 \equiv \boldsymbol{\nabla} \boldsymbol{\nabla}^\dag V - \mathbf{R}(\boldsymbol{\dot{\phi}},\boldsymbol{\dot{\phi}})$ \cite{NibbelinkAndVanTent-2001}.} $\mathbf{\tilde{M}}$, is defined as
\begin{align}
\label{M eff defn}
\mathbf{\tilde{M}} \equiv \mathbf{M} + \frac{1}{(3-\epsilon)} \epsilon R \mathbf{e}_{\perp} \mathbf{e}_{\perp}^\dag.
\end{align}
The beauty of this more compact version of the field perturbation equation is that it reveals the mode evolution is determined primarily by the mass matrix, plus corrections arising from the curvature of the field manifold\footnote{We make the typical assumption that the curvature of the field manifold is not too large.  More specifically, we assume $|R| \lesssim 1$ and therefore that $\epsilon |R| \ll 1$ in the slow-roll limit.} and usually negligible corrections from the field acceleration.  

We can show that these corrections are typically small by working in the kinematical basis. First, note that contracting $\mathbf{e}_{\parallel}$ with the terms in brackets on the right-hand side of equation (\ref{Delta Phi EoM wrt N}) and using the lowest-order SRST approximation in equation (\ref{eta in SRST}) for the uncontracted vector $\boldsymbol{\eta}$ yields
\begin{align}
\label{(M + eta eta) parallel cmpt in SRST}
\left[\mathbf{M} + \frac{\boldsymbol{\eta}\boldsymbol{\eta}^\dag}{(3-\epsilon)^2} \right] \mathbf{e}_{\parallel} & \approx \left[1 - \frac{2\epsilon^{(1)}}{9} \left(\frac{\eta_\parallel}{v}\right) \right] \mathbf{M} \mathbf{e}_{\parallel}^{(1)} \approx \mathbf{M} \mathbf{e}_{\parallel}^{(1)},
\end{align}
where in the final step, we ignored the second term in brackets because it is suppressed by two SRST terms: $\epsilon$ times the speed up rate.  

Second, consider the ($\perp,\perp$) components of the same terms in brackets on the right-hand side of equation (\ref{Delta Phi EoM wrt N}).   The term involving the field acceleration will be much smaller than $M_{\perp \perp}$ as long as the background field vector is not turning rapidly.  Rewriting this term as 
\begin{align}
\label{eta eta perp cmpt}
\left(\frac{\eta_{\perp}}{3-\epsilon}\right)^2 = M_{\perp \perp} \left[\frac{1}{(3-\epsilon)^2} \left(\frac{2\epsilon}{M_{\perp \perp}}\right) \left(\frac{\eta_{\perp}}{v}\right)^2\right],
\end{align}
we can see that if $\epsilon$ is no more than an order of magnitude greater than $M_{\perp \perp}$,\footnote{In \cite{NibbelinkAndVanTent-2001}, it was effectively assumed that $M_{\perp \perp}$ is the same order as $\epsilon$, but we find that in scenarios where both fields are important, more typically $M_{\perp \perp}$ tends to be within an order of magnitude of $\epsilon$.} then this term can be neglected as long as the field vector is not turning very rapidly.  But this result also holds even in those cases where $M_{\perp \perp} \ll \epsilon \ll 1$, as in such two-field scenarios, this means that the second field has a negligible effect on the background field dynamics and hence the turn rate is minuscule, i.e., $\frac{\eta_{\perp}}{v} \lll 1$.  A more direct argument follows from the fact that in the conventional slow-roll approximation, $M_{\perp \perp} \gg (\boldsymbol{\nabla} \ln V \, \mathbf{e}_{\perp})^2 = \left(\frac{\eta_{\perp}}{3-\epsilon}\right)^2$, where the last equality follows from projecting equation (\ref{Phi EoM wrt N v2}) onto $\mathbf{e}_{\perp}$.    

Combining these observations, we can conclude that the term  $\frac{\boldsymbol{\eta} \boldsymbol{\eta}^\dag}{(3-\epsilon)^2}$ is effectively suppressed by two SRST parameters relative to $\mathbf{\tilde{M}}$.  Therefore, in the SRST limit, the perturbed equation of motion can be approximated by
\begin{align}
\label{Delta Phi EoM wrt N w M only}
\frac{1}{(3-\epsilon)} \frac{D^2\boldsymbol{\delta \phi}_f}{dN^2} + \frac{D\boldsymbol{\delta \phi}_f}{dN} +  \left(\frac{k^2}{a^2 V}\right) \boldsymbol{\delta \phi}_f \approx - \mathbf{\tilde{M}} \, \boldsymbol{\delta \phi}_f,
\end{align} 
where either the first-order SRST approximation
\begin{align}
\label{M eff in SRST}
\mathbf{\tilde{M}}^{(1)} \equiv \mathbf{M} + \frac{1}{3} \epsilon^{(1)} R \,\mathbf{e}_{\perp}^{(1)} (\mathbf{e}_{\perp}^{(1)})^\dag,
\end{align}
or the second-order SRST approximation
\begin{align}
\label{M eff in SRST O2}
\mathbf{\tilde{M}}^{(2)} \equiv \mathbf{M} + \frac{1}{3} \left[\epsilon^{(2)} + \frac{1}{3}(\epsilon^{(1)})^2\right] R \,  \mathbf{e}_{\perp}^{(2)} (\mathbf{e}_{\perp}^{(2)})^\dag
\end{align}
can be used in place of $\mathbf{\tilde{M}}$, as desired (and where it is understood that the factor $\frac{1}{(3-\epsilon)}$ on the left-hand side of equation (\ref{Delta Phi EoM wrt N w M only}) is to be expanded to the same order for consistency).  If the kinetic terms are canonical, then the effective mass matrix simply reduces to the mass matrix, $\mathbf{M}$, to either order in the SRST limit.

We conclude this section by considering the perturbed equation of motion for modes with wavelengths significantly larger than the causal horizon.  In this limit, the mode wavelengths satisfy $\left(\frac{k}{aH}\right)^2 \ll 1$, and the equation of motion (\ref{Delta Phi EoM wrt N w M only}) reduces even further.  First,  the sub-horizon term, $\left(\frac{k^2}{a^2 V}\right) \boldsymbol{\delta \phi}_f$, can be neglected  since $\frac{k^2}{a^2V} \sim \left(\frac{k}{aH}\right)^2 \ll 1$.  Second, for modes significantly outside the horizon, the acceleration of the field perturbation vector can be neglected whenever the background field vector is in the SRST limit \cite{TaruyaAndNambu-1997,NibbelinkAndVanTent-2001}.  The reason why is that according to the separate universe formalism (see \cite{TaruyaAndNambu-1997,RigopoulosAndShellard-2003}, in particular), the combination of the background fields and the field perturbations is indistinguishable from the background outside the horizon.  Thus, it can be shown that for super-horizon modes, if the background field vector is in the SRST limit, so is the perturbed field vector \cite{TaruyaAndNambu-1997,NibbelinkAndVanTent-2001}.  Therefore, the super-horizon evolution of the field perturbations can be approximated to first-order by
\begin{align}
\label{Delta Phi EoM wrt N in SRST super-horz}
\frac{D\boldsymbol{\delta \phi}_f}{dN} \approx - \mathbf{\tilde{M}}^{(1)} \, \boldsymbol{\delta \phi}_f.
\end{align}
where the first-order approximation for the effective mass matrix is given by equation (\ref{M eff in SRST}).

Interestingly, equation (\ref{Delta Phi EoM wrt N in SRST super-horz}) is similar in form to equation (\ref{eta in SRST}).  Indeed, to first-order in the SRST limit, the super-horizon evolution of both $\boldsymbol{\delta \phi}_f$ and $\boldsymbol{\phi}'$ are determined by the effective mass matrix, $\mathbf{\tilde{M}}$.  This can be seen by re-writing equation (\ref{eta in SRST}) as
\begin{align}
\label{eta in SRST in terms of M eff}
\boldsymbol{\eta}^{(1)} = - \mathbf{\tilde{M}}^{(1)} \boldsymbol{\phi}'^{(1)}, 
\end{align}
where we have used that $\mathbf{\tilde{M}} \boldsymbol{\phi}' = \mathbf{M} \boldsymbol{\phi}'$. This means that in the super-horizon SRST limit, we simply need to understand the matrix coefficients of  $\mathbf{\tilde{M}}$ in order to understand the behavior of both the unperturbed and perturbed fields.  It is this very important commonality that allows us to predict the super-horizon behavior of the field perturbations simply by knowing the field kinematics and the field manifold curvature.   And, as we will later see, it is the three unique coefficients of $\mathbf{\tilde{M}}$ (or equivalently, the speed up rate, the turn rate, and the effective entropy mass), along with $H$ and $\epsilon$, that fully represent all the main features of a two-field inflationary model.

\subsubsection{Adiabatic Modes}
\label{adiabatic modes}

We now examine the evolution of the field perturbations in greater detail, but in the kinematical basis.  In rotating to the kinematical basis, the modes naturally separate into adiabatic modes and entropy modes.  Adiabatic modes are field perturbations along the trajectory, and we use the notation $\delta \phi_{\parallel} \equiv \mathbf{e}_{\parallel} \cdot \boldsymbol{\delta \phi}_f$ to represent them, dropping the subscript $f$ for simplicity.  They correspond to perturbations in the total energy density, or equivalently, in the curvature of constant time hypersurfaces (curvature perturbations).  The second type of modes in this basis, the entropy modes, are orthogonal to the field trajectory, and we denote them as $\delta \phi_{\perp} \equiv \mathbf{e}_{\perp} \cdot \boldsymbol{\delta \phi}_f$.  In contrast to adiabatic modes, entropy modes do not represent bona fide density perturbations, but rather correspond to relative fluctuations in the two different fields that leave the density or curvature unperturbed (isocurvature perturbations).  We will discuss the relationships between adiabatic and entropy modes and the curvature (density) and isocurvature perturbations in more detail in Section \ref{curvature modes}.

Evolution equations for both adiabatic and entropy modes have been derived for two-field inflationary models under the assumption of canonical kinetic terms \cite{GordonEtAl-2000} and for the non-canonical field metric $\mathbf{G} = \mathrm{diag}(1, e^{b(\phi_1)})$ \cite{DiMarcoEtAl-2002}, which can be used to describe scalar-tensor theories in the Einstein frame.\footnote{More recently, evolution equations for non-canonical kinetic terms that cannot be expressed as $g^{\mu \nu} G_{\ij} \partial_{\mu} \phi^i \partial_{\nu} \phi^j$ have been found \cite{LangloisAndRenaux-Petel-2008}, which allows for scenarios in which the entropy modes propagate at the effective speed of sound.  But we adhere to models where the non-canonical kinetic terms can be expressed as   $g^{\mu \nu} G_{\ij} \partial_{\mu} \phi^i \partial_{\nu} \phi^j$ and hence where both the adiabatic and entropy modes propagate at the speed of light.}   Here, we build upon this work by 
\begin{enumerate}
\item Deriving evolution equations for a completely arbitrary field metric and doing so in covariant form; 
\item Discussing how the evolution of adiabatic and entropy modes can be inferred from the background kinematics and the curvature of the field manifold;
\item Deriving an exact expression for the evolution of adiabatic modes and two approximate semi-analytic solutions for the entropy modes in the super-horizon limit; and 
\item Explaining why the common assumption that a parameter called the effective entropy mass is approximately constant is generally not accurate for estimating the amplitude of entropy modes and hence for estimating the power spectra. 
\end{enumerate}

We start by deriving a covariant evolution equation for the adiabatic modes.   In contrast to other approaches, we derive the evolution equation exclusively in terms of our kinematical quantities.  This alternative approach is worthwhile because it allows us to directly see how the background kinematics affect the evolution of adiabatic modes.  To derive such an expression starting from equation (\ref{Delta Phi EoM wrt N}), we need to find an expression relating the coefficients of $\mathbf{\tilde{M}}$ in the kinematical basis to our kinematical quantities, defined in Section \ref{kinematics}.  We can find expressions for the matrix coefficients $\tilde{M}_{\parallel \parallel}$ and $\tilde{M}_{\parallel \perp}$ by differentiating the background field equation (\ref{Phi EoM wrt N v2}) and then projecting the resulting equation onto our two kinematical basis vectors.  Using the results of these projections, equation (\ref{Delta Phi EoM wrt N}), and the covariant derivatives of the kinematical basis vectors in equation (\ref{De_p/dN}), the evolution equation for the adiabatic perturbations can be written as
\begin{widetext}
\begin{align}
\label{Adiabatic EoM}
\frac{1}{(3 - \epsilon)} \delta \phi_{\parallel}'' + & \delta \phi_{\parallel}' + \left\{\left(\frac{k^2}{a^2V}\right) - \left(\frac{\eta_{\parallel}}{v}\right) \left[\frac{1}{(3-\epsilon)} \left(\ln \frac{\eta_{\parallel}}{v}\right)' + 1 + \frac{1}{(3-\epsilon)} \left(\frac{\eta_{\parallel}}{v}\right)\right] \right\} \delta \phi_{\parallel} \nonumber \\
& = 2 \left(\frac{\eta_{\perp}}{v}\right) \left\{\frac{1}{(3-\epsilon)} \delta \phi_{\perp}' +  \left[\frac{1}{(3-\epsilon)} \left(\ln \frac{\eta_{\perp}}{v}\right)' + 1 + \frac{1}{(3-\epsilon)} \left(\frac{\eta_{\parallel}}{v}\right)\right]\delta \phi_{\perp} \right\}.
\end{align}
\end{widetext}

Equation (\ref{Adiabatic EoM}) shows that the sub-horizon term, the speed up rate, and the turn rate primarily control the evolution of the adiabatic modes.   Notice that when the turn rate is non-zero, the entropy modes source the adiabatic modes.  But when the turn rate is zero, the adiabatic modes decouple from the entropy modes, and we recover the equation of motion for the single-field case.   This shows that the turn rate is not only the marker of multi-field behavior for the unperturbed fields, but also is the marker of multi-field behavior for the perturbed fields.  Here, we are defining multi-field behavior for the perturbed fields to mean sourcing of the adiabatic modes by the entropy modes (though we recognize that the multi-field case is additionally distinguished from the single-field case by the presence of entropy modes).

Now focusing on the super-horizon limit, $\left(\frac{k}{aH}\right)^2 \ll 1$, we can derive a particularly elegant and simple expression for the growing adiabatic modes.  Neglecting the subhorizon term in equation (\ref{Adiabatic EoM}) and regrouping the remaining terms, we can write the super-horizon equation of motion as 
\begin{align}
\label{Adiabatic EoM super-horz}
& \left[\frac{1}{(3-\epsilon)} \frac{d}{dN} + 1 + \frac{2}{(3-\epsilon)}\left(\frac{\eta_{\parallel}}{v}\right)\right] \left(\frac{\delta \phi_{\parallel}}{v}\right)' \\ 
& \, = \left[\frac{1}{(3-\epsilon)} \frac{d}{dN} + 1 + \frac{2}{(3-\epsilon)}\left(\frac{\eta_{\parallel}}{v}\right)\right] \left(2\frac{\eta_{\perp}}{v} \frac{\delta \phi_{\perp}}{v}\right).  \nonumber
\end{align}
Setting the right-hand side of equation (\ref{Adiabatic EoM super-horz}) to zero and solving the resulting homogeneous equation yields the homogeneous solutions for the growing and decaying modes.  The latter will be strongly suppressed by the quasi-exponential expansion, and hence generally does not contribute much to the adiabatic density power spectrum at the end of inflation. As it turns out, the evolution equation for the growing adiabatic mode---both the complementary and particular parts of the solution---can be picked off of equation (\ref{Adiabatic EoM super-horz}) by recognizing that the two terms in brackets on the left and right-hand sides of equation (\ref{Adiabatic EoM super-horz}) are identical.  From this recognition, we can conclude that the growing super-horizon adiabatic modes are described by 
\begin{align}
\label{Adiabatic EoM super-horz v1}
\left(\frac{\delta \phi_{\parallel}}{v}\right)' = 2 \frac{\eta_{\perp}}{v} \left(\frac{\delta \phi_{\perp}}{v}\right),
\end{align}
or equivalently, using $(\ln v)' = \frac{\eta_{\parallel}}{v}$, by
\begin{align}
\label{Adiabatic EoM super-horz v2}
\delta \phi_{\parallel}' & = \left(\frac{\eta_{\parallel}}{v}\right) \, \delta \phi_{\parallel} + 2 \left(\frac{\eta_{\perp}}{v}\right) \, \delta \phi_{\perp}.
\end{align}
We emphasize that equations (\ref{Adiabatic EoM super-horz v1}) and (\ref{Adiabatic EoM super-horz v2}) describing the growing adiabatic modes are exact in the super-horizon limit.   If additionally the fields are in the SRST limit, the speed up and turn rates in equation (\ref{Adiabatic EoM super-horz v2}) can be replaced by the approximations in equations (\ref{speed up rate in SRST}) and (\ref{turn rate in SRST}), respectively, yielding
 \begin{align}
\label{Adiabatic EoM in SRST super-horz}
\delta \phi_{\parallel}' & \approx - M^{(1)}_{\parallel \parallel} \, \delta \phi_{\parallel} - 2 M^{(1)}_{\parallel \perp} \, \delta \phi_{\perp}.
\end{align}
The same SRST expression can also be obtained by starting from equation (\ref{Delta Phi EoM wrt N in SRST super-horz}) and using equations (\ref{De_p/dN}) and (\ref{turn rate in SRST}).  

Equation (\ref{Adiabatic EoM super-horz v2}) shows that in the super-horizon limit, the evolution of the growing adiabatic modes is controlled simply by the speed up rate and the turn rate.  The first term on the right-hand side of the equation implies that the larger the speed up rate, the faster the `intrinsic' evolution of the modes.  The second term, which involves both the turn rate and the amplitude of the entropy modes, represents the sourcing of adiabatic modes by the entropy modes.  The faster the turn rate, the more the entropy modes source the adiabatic modes.   When both kinematical rates are small---that is, when the background fields are in the SRST limit---the adiabatic modes evolve slowly.  However, when either rate is large, the adiabatic modes evolve significantly.

But it is not just the absolute sizes of the speed up and turn rates that determine the evolution of adiabatic modes; the relative sizes of the two rates to each other also matters.  In fact, the ratio of these two kinematical scalars---in conjunction with the ratio of the two mode amplitudes---determines the relative contribution of the entropy mode sourcing to the total growth of adiabatic modes.  We can therefore categorize the inflationary dynamics in a region of spacetime into three kinds of physical behavior, depending on the strength of the entropy mode sourcing.   First, when $\frac{\eta_{\parallel}}{v} \delta \phi_{\parallel} \gg \frac{\eta_{\perp}}{v} \delta \phi_{\perp}$, the sourcing of adiabatic modes by entropy modes is negligible.   In this limit, the solution to equation (\ref{Adiabatic EoM super-horz v2}) is
\begin{align}
\label{Adiabatic EoM single-field}
\delta \phi_{\parallel} \propto v,
\end{align}
and hence the adiabatic modes evolve essentially independently.   Since equation (\ref{Adiabatic EoM single-field}) becomes exact in the single-field case, anytime the evolution of the growing adiabatic modes can be well-approximated by equation (\ref{Adiabatic EoM single-field}), we say that the inflationary dynamics for the adiabatic mode are \textit{effectively single-field}.    The second case is when $\frac{\eta_{\parallel}}{v} \delta \phi_{\parallel} \sim \frac{\eta_{\perp}}{v} \delta \phi_{\perp}$.  In this regime, sourcing must be taken into account, and we say that the scenario is inherently multi-field since the evolution of adiabatic cannot be approximated by the single-field equation of motion.  Finally, when $\frac{\eta_{\parallel}}{v} \delta \phi_{\parallel} \ll \frac{\eta_{\perp}}{v} \delta \phi_{\perp}$, the sourcing effects are the dominant driving force behind the growth of adiabatic modes.  In this limit, we say that the multi-field effects are strong.  Again, this shows how critical the ratio $\frac{\eta_{\perp}}{v} / \frac{\eta_{\parallel}}{v}$ is in determining the relative contribution of the multi-field effects---in other words, the importance of the mode sourcing.  We summarize these three cases in Table 3.

\setlength{\tabcolsep}{6pt}
\begin{table}[t]
\centering
\renewcommand{\arraystretch}{1.5}
\begin{tabular}{|x{1.05in}|x{1.0in}|x{1.0in}|}
\hline
\multicolumn{3}{|c|}{\normalsize\rule{0cm}{0.5cm}Table 3. Super-Horizon Growth of Adiabatic Modes} \tabularnewline[0.5ex] 
\hline \multicolumn{3}{|c|}{$\, $}\tabularnewline[-3.7ex] \hline  \tabularnewline[-3.4ex]  \normalsize Condition & \multicolumn{2}{c|}{\normalsize Physical Behavior} \tabularnewline[0.5ex] 
\hline
\vskip 0pt $\, $\Large$\frac{\eta_{\parallel}}{v}$\normalsize $\delta \phi_{\parallel} \gg \,$\Large$\frac{\eta_{\perp}}{v}$\normalsize $\delta \phi_{\perp}$ & \multicolumn{2}{p{2.0in}|}{\normalsize Entropy mode sourcing is negligible. The evolution is effectively single-field: $\delta \phi_{\parallel} \propto v$.
} \tabularnewline[0.6ex]
\hline
\vskip 0pt $\, $\Large$\frac{\eta_{\parallel}}{v}$\normalsize $\delta \phi_{\parallel} \sim \,$\Large$\frac{\eta_{\perp}}{v}$\normalsize $\delta \phi_{\perp}$ & \multicolumn{2}{p{2.0in}|}{\normalsize Entropy mode sourcing is appreciable.  Multi-field effects must be taken into account.
} \tabularnewline[0.6ex]
\hline
\vskip 0pt $\, $\Large$\frac{\eta_{\parallel}}{v}$\normalsize $\delta \phi_{\parallel} \ll \,$\Large$\frac{\eta_{\perp}}{v}$\normalsize $\delta \phi_{\perp}$ & \multicolumn{2}{p{2.0in}|}{\normalsize Entropy mode sourcing predominates. Multi-field effects are strong.
} \tabularnewline[0.6ex]
\hline
\end{tabular}
\end{table}

Thus, we have shown how the background kinematics control the evolution of adiabatic modes. This is a prime example of the usefulness of viewing the speed up and turn rates as distinct quantities that serve as markers of very different physical behavior.

\subsubsection{Entropy Modes}
\label{entropy modes}

In this section, we derive a covariant equation of motion for the entropy modes, as well as approximate semi-analytic expressions that are valid in the super-horizon slow-turn limit.  In tandem, we discuss how the evolution of entropy modes can be inferred from the background kinematics and the curvature of the field manifold.  

We start from equation (\ref{Delta Phi EoM wrt N}) and use both equation (\ref{De_p/dN}) and the projections of the derivative of equation (\ref{Phi EoM wrt N v2}) onto both $\mathbf{e}_{\parallel}$ and $\mathbf{e}_{\perp}$.  After some algebra, the equation of motion for the entropy modes becomes 
\begin{widetext}
\begin{align}
\label{entropy EoM}
\frac{1}{(3-\epsilon)} \delta \phi_{\perp}'' + \delta \phi_{\perp}' + 
\left[\left(\frac{k^2}{a^2V}\right) + M_{\perp \perp} + \frac{\epsilon R}{(3-\epsilon)} - \frac{3(1-\epsilon)}{(3-\epsilon)^2} \left(\frac{\eta_{\perp}}{v}\right)^2\right] \delta \phi_{\perp} = - \frac{2}{(3-\epsilon)} \left(\frac{\eta_{\perp}}{v}\right) \left[\delta \phi_{\parallel}' - \left(\frac{\eta_{\parallel}}{v}\right) \delta \phi_{\parallel}\right].
\end{align}
\end{widetext}
In contrast to the equation of motion for adiabatic modes (\ref{Adiabatic EoM}), the evolution of entropy modes is controlled by six quantities: the sub-horizon term, $M_{\perp \perp}$, $\epsilon$, $R$, the turn rate, and the speed up rate.  So the curvature of the field manifold, $R$, does affects the evolution of entropy modes, but not the evolution of adiabatic modes.  Interestingly, $\epsilon$ rather weakly affects the evolution of entropy modes but with one exception: it strongly modulates the effect of the curvature of the field manifold on the entropy modes.  In addition, notice that again the turn rate controls the sourcing of one mode type by the other.   

Now we consider equation (\ref{entropy EoM}) in the super-horizon limit, both with and without making assumptions about the background kinematics.  In the general super-horizon limit, the sub-horizon term vanishes, and we can substitute equation (\ref{Adiabatic EoM super-horz v2}) for the term in brackets on the right-hand side of equation (\ref{entropy EoM}) to obtain
\begin{align}
\label{entropy EoM super-horz}
\frac{\delta \phi_{\perp}''}{(3 - \epsilon)} + \delta \phi_{\perp}' = - \mu_{\perp} \delta \phi_{\perp},
\end{align}
where $\mu_{\perp}$ is the \textit{effective entropy mass} and is defined as
\begin{align}
\label{Entropy Mass}
\mu_{\perp} \equiv M_{\perp \perp} + \frac{\epsilon R}{(3 - \epsilon)} + \frac{9-\epsilon}{(3-\epsilon)^2} \left(\frac{\eta_{\perp}}{v}\right)^2.
\end{align} 
Equation (\ref{entropy EoM super-horz}) shows that the entropy modes evolve independently in the super-horizon limit, even in the presence of an arbitrary non-canonical field metric.  This is a very important result: because the entropy modes evolve independently in this limit, we can find their amplitude and also determine the evolution of adiabatic modes without solving a set of fully coupled equations.    

Equation (\ref{entropy EoM super-horz}) also reveals that the evolution of entropy modes is determined by a quantity called the effective entropy mass.  The concept of an effective entropy mass for a general two-field potential was first introduced in \cite{GordonEtAl-2000} and was also used in \cite{BartoloEtAl-2001a,TsujikawaEtAl-2002,DiMarcoAndFinelli-2005}, for example.  We are building upon this work by (1) redefining the effective entropy mass so that it is dimensionless and can be more directly compared against our kinematical and other parameters in the theory, and by (2) deriving a covariant expression for it in the case of an arbitrary two-field potential with a completely arbitrary field metric.  Moreover, in the remainder of this section, we will extend this work on the effective entropy mass by:
\begin{enumerate}
\item Exploring the absolute and relative sizes of the terms in the effective entropy mass in greater detail than has been done before, and clarifying which terms in $\mu_{\perp}$ can be neglected depending on the background kinematics and the curvature of the field manifold; 
\item Showing that assuming the effective entropy mass is constant generally produces large errors in estimating the amplitude of entropy modes and hence in estimating the power spectra; and  
\item Deriving semi-analytic approximations for the amplitude of entropy modes, which are based on approximations to the effective entropy mass.
\end{enumerate} 

First, we examine each of the three terms in the effective entropy mass.  The first term in equation (\ref{Entropy Mass}) is the $(\perp,\perp)$ coefficient of the mass matrix (the covariant Hessian of $\ln V$) and hence is an indication of the local curvature of the surface $f(\phi_1,\phi_2) = \ln V(\phi_1,\phi_2)$ along the entropic direction.   When the $(\perp,\perp)$ coefficient of the mass matrix is positive, as is most typical, this term will suppress the amplitude of entropy modes after horizon exit.  However, when $M_{\perp \perp}$ is negative, this term will fuel the growth of entropy modes.  

The second term in equation (\ref{Entropy Mass}) involves the Ricci scalar, $R$, of the field manifold.  When $R$ is positive---which occurs when the field manifold is locally elliptical---the field curvature helps to suppress the entropy modes.  However, when $R$ is negative---which occurs when the surface is locally hyperbolic---the field curvature fuels the growth of entropy modes. An example of a non-canonical field metric that always produces a negative field curvature is $\mathbf{G} = \mathrm{diag}(1, e^{b(\phi_1)})$, which describes scalar-tensor theories in the Einstein frame.\footnote{This does not mean that the net result of having this non-canonical field metric is an increase in the post-horizon amplitude of entropy modes, as the field metric also affects the value of the other two terms in the effective entropy mass.} 

The third term in the effective entropy mass is equal to the turn rate squared times a numerical factor between 1 and 2 that depends on the value of $\epsilon$.  Because of the sign in front of this term, it always damps the entropy modes.  In the limit of slow-turning, this term has a negligible effect.  In the limit of fast-turning ($\frac{\eta_{\perp}}{v} \gtrsim 1$), it causes a rapid suppression of the entropy modes.    Interestingly, a large turn rate is frequently accompanied by a boost in the magnitude of $M_{\perp \perp}$, due to the rapid rotation of basis vectors projecting out different combinations of the mass matrix's coefficients.

Now we consider the relative sizes of the three terms in the effective entropy mass.  The size of the Ricci scalar term, $\frac{\epsilon R}{(3-\epsilon)}$, relative to $M_{\perp \perp}$ depends on the field metric and the potential, so we cannot make any universal statements about their relative sizes.  However, if the kinetic terms are canonical, then the Ricci scalar term vanishes.  Also, we note that it is common for some of the most popular field metrics to produce field curvature terms that are at least an order of magnitude less than $M_{\perp \perp}$ when the fields are in the SRST limit, but this is not true for all field metrics.  Interestingly, the Ricci scalar is multiplied by a factor of $\epsilon$, so its effect on the entropy modes is strongly controlled by the field speed; when all else is equal, this curvature term becomes much more important near the end of inflation and whenever else $\epsilon$ is large.  Hence we expect the curvature term to typically be more important when the slow-roll conditions are violated.  As for the relative size of the third term in $\mu_{\perp}$, this term at leading order is $\left(\frac{\eta_{\perp}}{v}\right)^2$, so based on the arguments we made in Section \ref{field perturbations}, it is one order higher in the SRST expansion than $M_{\perp \perp}$.  This term can therefore be ignored whenever the field vector is slowly turning.  

Next, since the super-horizon equation of motion for the entropy modes (\ref{entropy EoM super-horz}) is homogeneous, we can consider finding solutions to it. As there are no exact analytical solutions to equation (\ref{entropy EoM super-horz}), we explore approximations.  In what follows, we assume that the curvature term is the same order of magnitude as $M_{\perp \perp}$, though depending on the model, the curvature term may be negligible and hence may be dropped from various order approximations; we leave it to the reader to determine when this is possible.  Now the most obvious approximation to invoke is the SRST approximation applied to both the unperturbed and perturbed fields. Under this approximation, the third term in the effective entropy mass can be neglected, as well as the acceleration of the amplitude of entropy modes,\footnote{By ignoring the acceleration of the amplitude entropy modes, we have effectively ignored the decaying modes, which are usually rapidly suppressed in this limit.} yielding
\begin{align}
\label{entropy EoM super-horz in SRST}
\delta \phi_{\perp}' \approx - \tilde{M}^{(1)}_{\perp \perp} \delta \phi_{\perp},
\end{align}
where recall that $\tilde{M}_{\perp \perp} = M_{\perp \perp} + \frac{\epsilon R}{(3-\epsilon)}$ and where we have used $\mathbf{e}_{\perp}^{(1)}$ to find the ($\perp,\perp$) component of $\mathbf{\tilde{M}}^{(1)}$.  But the above equation actually holds more generally, as long as the background field is slowly turning and $\epsilon R \ll 1$; slow-roll is not needed for this approximation to be valid.  So in the slow-turn limit and assuming $\epsilon R \ll 1$, the term which controls the evolution of entropy modes is just the first-order slow-turn approximation to the effective entropy mass, $\mu_{\perp}^{(1)} = \tilde{M}^{(1)}_{\perp \perp}$.   Hence, equation (\ref{entropy EoM super-horz in SRST}) gives us a simple metric, $\tilde{M}^{(1)}_{\perp \perp}$, that we can use to find and compare the evolution of entropy modes across widely different inflationary scenarios.

Now in using equation (\ref{entropy EoM super-horz in SRST}) to derive an analytical approximation for the entropy mode amplitude, the effective entropy mass has often been treated as constant \cite{BartoloEtAl-2001a,TsujikawaEtAl-2002,DiMarcoAndFinelli-2005,LalakEtAl-2007} , which yields the following approximation
\begin{align}
\delta \phi_{\perp}(N) = [\delta \phi_{\perp}]_* e^{- \mu_{\perp}^* (N-N_*)},
\end{align}
where we use $*$ to denote that a quantity is to be evaluated at horizon exit.  The rational behind this approximation is that as long as the background field vector is in the SRST limit, the effective entropy mass can be treated as roughly constant after horizon exit, because like the standard slow-roll parameters, the effective entropy mass is slowly changing.  But this assumption is tantamount to assuming that the standard slow-roll parameters continue to remain roughly constant over many $e$-folds, which is often problematic.  (See \cite{NibbelinkAndVanTent-2000} for a short discussion of this general issue.)   Indeed, with some exceptions, we find that the effective entropy mass changes significantly between horizon exit and the end of inflation, and often well before the SRST approximation breaks down.  Tsujikawa \textit{et al.} \cite{TsujikawaEtAl-2002} noticed this problem in exploring double inflation scenarios.  Lalak \textit{et al.} \cite{LalakEtAl-2007} also acknowledged similar limitations, and hence their numerical analysis used the full equations of motion in order to accurately follow the evolution of the power spectra.  As to the origin of the typically significant increase in the effective entropy mass during inflation, we find that much of this increase can often be attributed to the inevitable large drop in the potential energy density and also in many cases to the rotation of the kinematical basis vectors over many $e$-folds of inflation.  The former phenomenon is due to the fact that as $\epsilon \approx \frac{1}{2} |\boldsymbol{\nabla} \ln V|^2$ increases, the potential decreases more quickly, and hence $M_{\perp \perp}$ significantly increases in magnitude.  

These findings have important implications.  In most cases, the entropy modes will be damped more strongly than previously analytic estimates would predict \cite{BartoloEtAl-2001a,DiMarcoAndFinelli-2005}.  For some two-field scenarios, this leads to rather modest inaccuracies in estimating the curvature (density)power spectrum, while for others, it leads to unusably large errors.  The size of these inaccuracies often depends significantly on the initial conditions, so it is even less common that this assumption can be used to estimate the density power spectrum for all possible initial conditions for a given inflationary Lagrangian.  More problematically, it leads to even larger inaccuracies in estimating the isocurvature and cross spectra, which obviously depend sensitively on the amplitude of entropy modes at the end of inflation.  In particular, this assumption often over-estimates the isocurvature and cross spectra by one to several orders of magnitude. 

Therefore, we do not assume the effective entropy mass is constant.  Instead, we integrate over the effective entropy mass to estimate the super-horizon amplitude of entropy modes, obtaining the approximation
\begin{align}
\label{entropy EoM super-horz in SRST integrated}
\delta \phi_{\perp} \approx \left[\delta \phi_{\perp}\right]_* e^{- \int_{N_*}^{N} \tilde{M}^{(1)}_{\perp \perp}(\tilde{N}) \, d\tilde{N}}.
\end{align}
Equation (\ref{entropy EoM super-horz in SRST integrated}) provides a good approximation whenever the background field vector is slowly turning, regardless of the values of $\epsilon$ and of the speed up rate.  

When more accuracy is desired or the turn rate is moderately large, then the following second-order approximation can be used.  Assuming that $\left(\frac{\eta_{\perp}}{v}\right)^2 \approx (M^{(1)}_{\parallel \perp})^2$ is significant relative to but still significantly less than $M_{\perp \perp}$, the equation 
\begin{align}
\label{entropy EoM super-horz in SRST O2}
\delta \phi_{\perp}' \approx - \mu_{\perp}^{(2)} \delta \phi_{\perp},  
\end{align}
where $\mu_{\perp}^{(2)}$ is equal to
\begin{align}
\label{Entropy Mass in SRST O2}
\mu_{\perp}^{(2)} \equiv & \, \tilde{M}_{\perp \perp} + \frac{1}{3} (\tilde{M}_{\parallel \perp})^2 + \frac{1}{3}(\tilde{M}_{\perp \perp})^2 \nonumber \\ & + \frac{1}{3} (\boldsymbol{\nabla}^\dag \ln V \boldsymbol{\nabla} \tilde{M})_{\perp \perp},
\end{align}
is the second-order approximation to the equation of motion (\ref{entropy EoM super-horz}).  That is, differentiating this expression to find $\delta \phi_{\perp}''$ yields equation (\ref{entropy EoM super-horz}) when equation (\ref{entropy EoM super-horz}) itself is expanded to second-order in the SRST limit.   In equation (\ref{Entropy Mass in SRST O2}), the second-order SRST approximations for the background fields and the kinematical basis vectors are to be used, so that both the unperturbed and perturbed fields are calculated to the same order in the SRST expansion.  Solving equation (\ref{entropy EoM super-horz in SRST O2}) yields the same integral expression as in equation (\ref{entropy EoM super-horz in SRST integrated}) after making the replacement $\tilde{M}^{(1)}_{\perp \perp} \rightarrow \mu_{\perp}^{(2)}$.  Based on the definition of $\mu_{\perp}^{(2)}$ in equation (\ref{Entropy Mass in SRST O2}), technically $\mu_{\perp}^{(2)}$ is not the second-order SRST approximation for the effective entropy mass; however, we still denote the quantity on the right-hand side of equation (\ref{Entropy Mass in SRST O2}) with the symbol $\mu_{\perp}^{(2)}$ to avoid introducing too many new symbols and terms.  Our results are equivalent to those in \cite{ByrnesAndWands-2006} for the case of canonical kinetic terms and to those in \cite{DiMarcoEtAl-2002} for the non-canonical field metric $\mathbf{G} = \mathrm{diag}(1,e^{b(\phi_1)})$.  

From the above discussion and approximations, we can therefore predict the behavior of the entropy modes simply from the background kinematics and the geometry of the field manifold.  When the turn rate is small, the amplitude of entropy modes will be determined by $M_{\perp \perp}$, and we expect the entropy modes to be gradually suppressed (or enhanced).  Conversely, when the turn rate is large, then the entropy modes will be rapidly suppressed.  If the Ricci scalar term is significant relative to $M_{\perp \perp}$, it will also either suppress or enhance the modes, depending on its sign.  In the slow-roll limit, the Ricci scalar term is small, but when $\epsilon$ is large, then usually the Ricci scalar term and the term $M_{\perp \perp}$ will be larger in magnitude, so the entropy modes will be more quickly suppressed (or enhanced).

\begin{figure*}[t]
\centering
\includegraphics[height=175.5mm]{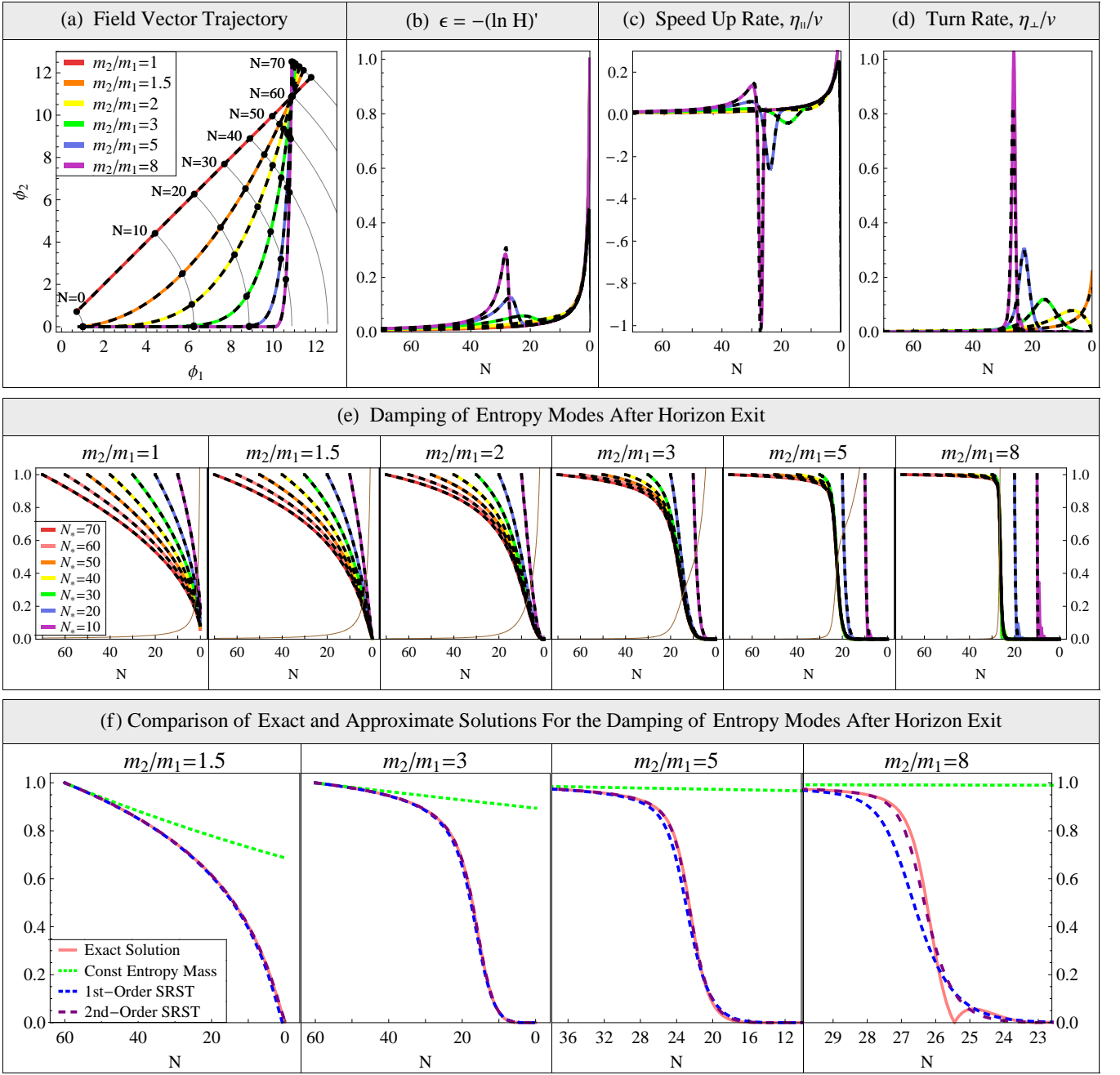}
\label{Entropy Modes Plots} 
\caption{The accuracy of three different approximations for the super-horizon evolution of entropy modes for six different values of the mass ratio $\frac{m_2}{m_1}$ for the double quadratic potential $V = \frac{1}{2} m_1^2 \phi_1^2 + \frac{1}{2} m_2^2 \phi_2^2$ with canonical kinetic terms.   The same initial conditions were assumed $60$ $e$-folds before the end of inflation, and the $x$-axis for plots (b) - (f) represents the number of $e$-folds before inflation ends.  In Figures 2(a)-(d), the exact solutions (thick colored lines) and the second-order SRST approximation (dashed black lines) are shown for (a) the field vector trajectory, (b) $\epsilon$, (c) the speed up rate, and (d) the turn rate.  In Figure 2(e), the exact solutions (thick colored lines) and the second-order SRST approximation (dashed black lines) are shown for the amplitude of entropy modes that exit the horizon $N_*=10,20,30,40,50,60,$ and $70$ $e$-folds before the end of inflation. The dimensionless effective entropy mass (thin brown line) is overlaid for comparison.  In Figure 2(f), the exact solution (thick colored lines) and three different approximations (dashed lines) for the post-horizon damping of the entropy mode that exits the horizon $N_*=60$ $e$-folds before the end of inflation is shown for four of the six different trajectories.  The three approximations are the assumption that the effective entropy mass is constant after horizon exit (dotted green line); the first-order slow-turn approximation for both the background and perturbed field vectors (dashed blue line); and the second-order SRST approximation for both the background and perturbed field vectors (longer dashed purple line).  Here, by the end of inflation, and often much sooner, the assumption that the effective entropy mass can be treated as constant greatly over-estimates the amplitude of entropy modes. 
}
\end{figure*} 

\clearpage

Lastly, we illustrate the accuracy of the constant entropy mass and our first-order and second-order semi-analytic approximations for the super-horizon amplitude of entropy modes. Figure 2 shows the accuracy of these approximations for six related inflationary scenarios with different speed up and turn rate profiles.  Figure 2(f) shows that assuming the effective entropy mass is constant after horizon exit significantly over-estimates the amplitude of entropy modes for all six trajectories by the end of inflation, and most often, much sooner.  The second approximation, the first-order slow-turn approximation, provides an excellent approximation for the damping of entropy modes when the field vector is slowly turning.  Its accuracy is insensitive to $\epsilon$ and to the speed up rate, as is clear from its equally good accuracy near the end of inflation for models with small turn rates.  However, when the field trajectory is moderately to rapidly turning, it does not estimate the amplitude of entropy modes as well.  We will later show that the curvature power spectrum is most sensitive to the amplitude of entropy modes when the background field vector is rapidly turning, so this second approximation is not sufficiently accurate to estimate the power spectra in the case of moderately fast turning.  Finally, the best approximation for the damping of entropy modes is given by the second-order SRST approximation, which produces more accurate estimates of the damping when the trajectory is turning moderately fast.

\subsubsection{Curvature and Isocurvature Perturbations}
\label{curvature modes}

Before we proceed to calculate the power spectra, we need to relate our field perturbations to the quantities whose power spectra we want to calculate.  The power spectrum of greatest interest is usually the spectrum of density perturbations.   As a proxy for the density power spectrum, often the power spectrum of a quantity called the comoving curvature perturbation is calculated instead, since the curvature power spectrum is easy to calculate and the two power spectra are identical up to numerical factors after inflation ends.  We use this common strategy and work in terms of the comoving curvature perturbations and the associated isocurvature perturbations.  In this section, we define curvature and isocurvature perturbations, relate them to the adiabatic and entropy modes, and find equations of motion for both perturbation types.  

The curvature perturbation is defined as the perturbation in the curvature of constant time hypersurfaces.  The curvature perturbation in the comoving gauge, $\mathcal{R}$, was introduced by \cite{Lukash-1980,Bardeen-1980,KodamaAndSasaki-1984,Lyth-1985}, and it represents a gauge-invariant quantity.  During inflation, it can be shown that the comoving perturbation equals \cite{SasakiAndTanaka-1998}
\begin{align}
\label{comoving R in terms of adiabatic perturbation}
\mathcal{R} = \frac{\delta \phi_{\parallel}}{v},
\end{align}  
where recall that $\delta \phi_{\parallel}$ represents the gauge-invariant quantity $\mathbf{e}_{\parallel} \cdot \boldsymbol{\delta \phi}_f$.  By contrast, isocurvature perturbations represent relative fluctuations in the two different fields that leave the total curvature unperturbed and hence they are related to entropy perturbations.  As in \cite{WandsEtAl-2002}, we define the isocurvature perturbations as
\begin{align}
\label{isocurv defn}
\mathcal{S} \equiv \frac{\delta \phi_{\perp}}{v}, 
\end{align} 
so that the curvature and isocurvature power spectra have similar power at horizon exit.

To find the spectra of these two quantities at the end of inflation from their spectra at horizon exit, we need to find evolution equations for the curvature and isocurvative perturbations in the super-horizon limit.  We start by finding an evolution equation for the curvature perturbations in the super-horizon limit.  Recall that we already found an expression for the super-horizon evolution of $\frac{\delta \phi_{\parallel}}{v}$ in equation (\ref{Adiabatic EoM super-horz v1}).  Substituting equations (\ref{comoving R in terms of adiabatic perturbation}) and (\ref{isocurv defn}) into equation (\ref{Adiabatic EoM super-horz v1}), we find  
\begin{align}
\label{comoving R EoM}
\mathcal{R}' = 2 \left(\frac{\eta_{\perp}}{v}\right) \mathcal{S}.
\end{align}
Equation (\ref{comoving R EoM}) is exact in the super-horizon limit, as we did not invoke any slow-roll or slow-turn approximations in deriving equation (\ref{Adiabatic EoM super-horz v1}). According to equation (\ref{comoving R EoM}), the super-horizon evolution of the curvature perturbation depends only on the covariant turn rate of the background trajectory relative to the field manifold and on the amplitude of the isocurvature perturbations; it is insensitive to $\epsilon$ and the speed up rate.  This matches previous results that the curvature perturbation evolves when the background trajectory is curved (e.g., \cite{LythAndRiotto-1998,GordonEtAl-2000,NibbelinkAndVanTent-2000,NibbelinkAndVanTent-2001}).  When all else is equal, the greater the turn rate, the more the comoving curvature perturbation evolves.  Conversely, whenever the field trajectory is not changing direction, the comoving curvature perturbation is conserved in the super-horizon limit, just like in single-field inflation.  All of this again proves that the turning of the background field trajectory is the true marker of multi-field effects (e.g., \cite{LythAndRiotto-1998,GordonEtAl-2000,NibbelinkAndVanTent-2000,NibbelinkAndVanTent-2001}).  Integrating equation (\ref{comoving R EoM}) gives
\begin{align}
\label{comoving R EoM integrated}
\mathcal{R} =  \mathcal{R}_* + \int_{N_*}^N 2 \left(\frac{\eta_{\perp}}{v}\right) \mathcal{S} \, d\tilde{N}.
\end{align}
 
Next, we derive a super-horizon evolution equation for the isocurvature perturbations.  Using equation (\ref{isocurv defn}) and $(\ln v)' = \frac{\eta_{\parallel}}{v}$, we find 
\begin{align}
\label{isocurv EoM}
\mathcal{S}' = \frac{1}{v} (v\mathcal{S})' - \frac{\eta_{\parallel}}{v} \mathcal{S} = \frac{1}{v} \left[\delta \phi_{\perp}' - \frac{\eta_{\parallel}}{v}\delta \phi_{\perp}\right].
\end{align} 
Thus, an expression for $\mathcal{S}$ can be found using the equation of motion for the entropy modes (\ref{entropy EoM super-horz}) and the speed up rate.  If an exact solution for $\delta \phi_{\perp}$ is known analytically or can be found from equation (\ref{entropy EoM super-horz}), then an exact solution for the isocurvature modes can be found using equation (\ref{isocurv defn}) directly.  Otherwise, the approach of Wands \textit{et al.} \cite{WandsEtAl-2002} can be used, which involves parametrizing the super-horizon evolution of isocurvature modes as
\begin{align}
\label{isocurv EoM in SRST}
\mathcal{S}' & = \beta \mathcal{S}.
\end{align} 
The above expression also holds in general on large scales \cite{MalikEtAl-2002}. We can use the above expression by finding approximations for $\beta$ in the SRST limit.  From the first-order expressions for the evolution of entropy modes (\ref{entropy EoM super-horz in SRST}) and the speed up rate (\ref{speed up rate in SRST}), we find that to first-order
\begin{align}
\label{isocurv mass in SRST}
\beta^{(1)} = \tilde{M}^{(1)}_{\parallel \parallel} - \tilde{M}^{(1)}_{\perp \perp}.
\end{align}
Similarly, to second-order, $\beta$ is approximated by
\begin{align}
\label{isocurv mass in SRST O2}
\beta^{(2)} = & \, - \left(\frac{\eta_{\parallel}}{v}\right)^{(2)} - \mu_{\perp}^{(2)},  \nonumber \\ 
= & \, \tilde{M}_{\parallel \parallel} + \frac{1}{3} (\tilde{M}_{\parallel \parallel})^2 + \frac{1}{3} (\boldsymbol{\nabla}^\dag \ln V  \boldsymbol{\nabla} \mathbf{\tilde{M}})_{\parallel \parallel} \\
& \, - \tilde{M}_{\perp \perp} - \frac{1}{3} (\tilde{M}_{\perp \perp})^2 - \frac{1}{3} (\boldsymbol{\nabla}^\dag \ln V  \boldsymbol{\nabla} \mathbf{\tilde{M}})_{\perp \perp}, \nonumber 
\end{align}
which follows from equations (\ref{kine quantities in SRST O2}) and (\ref{entropy EoM super-horz in SRST O2}).  In equations (\ref{isocurv mass in SRST}) and (\ref{isocurv mass in SRST O2}), the same order SRST approximations for the background fields and basis vectors are to be used.  As an aside, note the degree of symmetry in the terms in the second and third lines of equation (\ref{isocurv mass in SRST O2}): the terms on the third line can be obtained from the terms on the second line by the substitution $\mathbf{e}_{\parallel} \rightarrow i \mathbf{e}_{\perp}$.  Finally, to find a solution for the amplitude of isocurvature modes, we integrate equation (\ref{isocurv EoM in SRST}) to get
\begin{align}
\label{isocurv EoM integrated}
\mathcal{S} = \mathcal{S}_* e^{\int_{N_*}^{N} \beta \, d\tilde{N}},
\end{align}
where one of the two approximations for $\beta$ is to be used, depending on the accuracy required.  

Equation (\ref{isocurv EoM integrated}) and the approximations for $\beta$ show that the amplitude of isocurvature perturbations depends on the integral of negative the sum of the effective entropy mass and the speed up rate.  Through the effective entropy mass, whenever the integral of $M_{\perp \perp}$ or $\epsilon R$ is large and positive (negative), the isocurvature modes will be suppressed (amplified).  Also, since the effective entropy mass depends on the turn rate at second-order in the SRST approximation, whenever the turn rate becomes substantial, the isocurvature modes will also be damped.  As for the speed up rate, it affects the isocurvature modes simply because the isocurvature modes are related to entropy modes by a factor of $\frac{1}{v}$.  So when the integral of the speed up rate is large and positive---that is, when the field speed has grown substantially---the isocurvature modes are much smaller.  Conversely, when the speed up rate is large and negative---for example, due to a sudden drop in the field speed---the isocurvature modes will increase.   Alternatively, we can interpret the meaning of equation (\ref{isocurv EoM integrated}) from a more geometrical perspective: the amplitude of isocurvature modes depends at lowest-order on the integral of the difference of an effective measure of the curvatures of the surface $f(\phi_1, \phi_2) = \ln V(\phi_1,\phi_2)$ along the adiabatic and entropic directions and on $\frac{1}{3} \epsilon^{(1)}$ times the curvature of the field manifold.  

Now we use these results for the isocurvature modes to find an equation of motion for the curvature modes.  Plugging equation (\ref{isocurv EoM integrated}) into equation (\ref{comoving R EoM integrated}), we finally arrive at an expression for the super-horizon amplitude of curvature modes:
\begin{align}
\label{comoving R EoM integrated v1}
\mathcal{R} = \mathcal{R}_* + \mathcal{S}_* \int_{N_*}^N 2 \left(\frac{\eta_{\perp}}{v}\right) e^{- \int_{N_*}^{\tilde{N}} \beta \,d\tilde{\tilde{N}}} d\tilde{N}.
\end{align}

To understand equations (\ref{comoving R EoM integrated}) and (\ref{comoving R EoM integrated v1}), it is straightforward to carry over most of our separate observations about how the kinematics and the curvature of the field manifold affect the evolution of curvature and isocurvature modes.  However, there is one more complicated scenario that merits further discussion: the competing tendencies of the turn rate and the amplitude of isocurvature modes when the turn rate is large.  When the turn rate is large, the isocurvature modes will strongly source the curvature modes, but eventually the large turn rate will strongly suppress the isocurvature modes, thereby quenching further sourcing.  The net effect is that the curvature modes will increase dramatically but then level off very soon after.  Interestingly, because the amplitude of isocurvature modes tends to decrease after horizon exit, these sourcing effects have the potential to be even more dramatic if the large turn rate happens soon after horizon exit or whenever else the amplitude of isocurvature modes is large.  

Now in order to solve equation (\ref{comoving R EoM integrated v1}) analytically, some authors \cite{BartoloEtAl-2001a,DiMarcoAndFinelli-2005,LalakEtAl-2007} have assumed that both the turn rate and $\beta$ can be taken to be approximately constant.  We have already discussed the problems with assuming that the effective entropy mass is constant.  The same problems arise in assuming that the speed up rate, turn rates, and other slow-roll parameters are approximately constant over several e-folds.  So like the constant entropy mass approximation, assuming that these parameters can be approximated as constant has limited utility, as it frequently introduces large errors into estimates of the power spectra at the end of inflation.  

However, we can derive two semi-analytic approximations to the full expression in equation (\ref{comoving R EoM integrated v1}).  To first-order in the SRST limit, equation (\ref{comoving R EoM integrated v1}) can be approximated completely in terms of the cofficients of the effective mass matrix as
\begin{align}
\label{comoving R EoM integrated v2}
\mathcal{R} \approx \mathcal{R}_*  + \mathcal{S}_* \int_{N_*}^N (-2 \tilde{M}^{(1)}_{\parallel \perp}) e^{- \int_{N_*}^{N} (\tilde{M}^{(1)}_{\perp \perp} - \tilde{M}^{(1)}_{\parallel \parallel}) \, d\tilde{\tilde{N}}} d\tilde{N}.
\end{align}
And for more accuracy, the second-order SRST approximations for the turn rate (equation (\ref{kine quantities in SRST O2})) and for $\beta$ (equation (\ref{isocurv mass in SRST O2})) can be used instead, as we will do in Sections
\ref{Power Spectra} and \ref{Applications}.

Since it will serve as helpful short-hand later, an especially useful parameterization of the general relationship between curvature and isocurvature modes was introduced by \cite{AmendolaEtAl-2001} and extended by \cite{WandsEtAl-2002} and is expressed as 
\begin{align}
\label{transfer matrix}
\left(\begin{array}{c} \mathcal{R} \\ \mathcal{S} \end{array} \right) = & 
\left(\begin{array}{cc} 1 & T_{\mathcal{RS}} \\ 0 & T_{\mathcal{SS}} \end{array} \right)
\left(\begin{array}{c} \mathcal{R}_* \\ \mathcal{S}_* \end{array} \right),
\end{align}
where the transfer functions are defined as
\begin{align}
\label{transfer functions}
T_{\mathcal{RS}}(t_*,t) & \equiv \int_{t_*}^t \alpha(\tilde{t}) T_{\mathcal{SS}}(t_*,\tilde{t}) H(\tilde{t}) d\tilde{t}, \nonumber \\
T_{\mathcal{SS}}(t_*,t) & \equiv e^{\int_{t_*}^t \beta(\tilde{t}) H(\tilde{t}) d\tilde{t}}.
\end{align}
Comparing equations (\ref{isocurv EoM integrated}) and (\ref{comoving R EoM integrated v1}) to the above two equations, we can see that Wands \textit{et. al.}'s \cite{WandsEtAl-2002} function $\alpha$ equals twice the covariant turn rate and the function $\beta$ is as we defined earlier in equation (\ref{isocurv EoM in SRST}), with first- and second-order SRST approximations given by equations (\ref{isocurv mass in SRST}) and (\ref{isocurv mass in SRST O2}), respectively.   In the SRST limit and assuming canonical kinetic terms, our results agree with those in \cite{WandsEtAl-2002} to first-order and to those in \cite{ByrnesAndWands-2006} to second-order.  In Sections \ref{Power Spectra} and \ref{Applications}, we use this convenient framework as short-hand in our equations for the power spectra at the end of inflation.  However, we note that the above parameterization  (\ref{transfer matrix})  of the relationship between curvature and isocurvature modes in the super-horizon limit continues to hold even after inflation ends.

\subsection{The Power Spectra}
\label{Power Spectra}

We now use our results to calculate and interpret the power spectra at the end of inflation.  We build on previous results for simpler kinetic terms in \cite{BartoloEtAl-2001a,DiMarcoAndFinelli-2005,ByrnesAndWands-2006,LalakEtAl-2007} by treating the general case of completely arbitrary kinetic terms and doing so in a covariant manner.  We start by treating the field perturbation equation over its entire domain of validity in Section \ref{field perturbation equation in three regimes}, and we find solutions for the modes during the sub-horizon, horizon-crossing, and super-horizon regimes.  In Section \ref{pwr spec at horz exit}, we use these results and perform a rotation of bases in order to calculate the power spectra at horizon exit.  This rotation of basis gives rise to correlations between the curvature and isocurvature modes and hence a correlated cross spectrum.  In Section \ref{pwr spec at end}, we use our results for the evolution of curvature and isocurvature modes to find compact expressions for the power spectra at the end of inflation.  We discuss how these results reflect the background kinematics and the field manifold, and we discuss when multi-field effects are significant and when a two-field inflationary model can be dimensionally reduced to a single-field model.  We also use these results to argue that all two-field inflationary models can be reduced to just a handful of characteristic functions, representing all the kinematics and dynamics of the model.  Finally, we conclude by presenting simple expressions and a consistency condition for the canonical power spectrum observables in Section \ref{pwr spec observables}.

\subsubsection{Solving the Field Perturbation Equation}
\label{field perturbation equation in three regimes}

In Section \ref{perturbed equations}, we presented equations of motion for the field perturbations both in the given basis and in the kinematical basis, and we focused on how these equations simplify in the super-horizon limit.  In this section, we quantize the fields; solve the equation of motion over its entire domain; and use a rotation of bases to match the solutions across the boundaries.

To solve equation (\ref{Delta Phi EoM wrt N}), we follow an approach similar to \cite{NakamuraAndStewart-1996}.  We start by re-casting equation (\ref{Delta Phi EoM wrt N}) using the vector $\mathbf{q}$, where
\begin{align}
\label{q} 
\mathbf{q} \equiv a  \boldsymbol{\delta \phi}_f,
\end{align}
and using the conformal time, $\tau$, which is defined as 
\begin{align}
\label{tau}
d\tau \equiv \frac{dt}{a} = \frac{1}{aH}\, dN.
\end{align}
After some algebra, we obtain
\begin{widetext}
\begin{align}
\label{q EoM}
\frac{D^2 \mathbf{q}}{d\tau^2} + k^2\mathbf{q} = (aH)^2\left\{(2 - \epsilon) \mathbf{I} - (3-\epsilon)\left[\mathbf{\tilde{M}} + \frac{\boldsymbol{\eta \eta^\dag}}{(3-\epsilon)^2}\right]\right\} \mathbf{q}.
\end{align}
\end{widetext}
While there is no exact solution to equation (\ref{q EoM}) that is valid for all values of $\tau$, we can find separate solutions in the three standard regimes of interest and match the separate solutions on the boundaries.  

The first regime of interest is the sub-horizon limit, when the modes are well inside the horizon, which is defined by $\frac{k}{aH} \gg 1$ (or equivalently, by $-k\tau \rightarrow \infty$).  In this limit, the right-hand side of equation (\ref{q EoM}) is negligibly small, and equation (\ref{q EoM}) reduces to the Klein-Gordon equation.  The vector $\mathbf{q}$ thus describes a pair of decoupled simple harmonic oscillators, so the sub-horizon result from single-field inflation can simply be applied to each field.  The sub-horizon solution to equation (\ref{q EoM}) can thus be written as
\begin{align}
\label{q subhorz soln}
\mathbf{q} = \frac{1}{\sqrt{2k}} \left[\mathbf{a}(\mathbf{k}) e^{-ik\tau} + \mathbf{a}^\dag(-\mathbf{k}) e^{ik\tau}\right],     
\end{align}
where we have gone ahead and quantized the fields, introducing $a_i$ and $a_i^\dag$ as the particle annihilation and creation operators, respectively, for field $i$.  The annihiliation and creation operators for the two fields satisfy the canonical commutation relations,
\begin{align}
\label{CCRs}
[a_i(\mathbf{k}),a_j^\dag (\mathbf{k}')] = \delta_{ij} \delta^3(\mathbf{k} - \mathbf{k}'),
\end{align}
and the relation
\begin{align}
\label{vac}
a_i(\mathbf{k}) |0 \rangle = 0,
\end{align}
where we have assumed the usual Bunch-Davies vacuum \cite{BunchAndDavies-1978}.

The second regime centers around the time when the modes exit the causal horizon, which is defined by $\frac{k}{aH} \sim 1$.  Around horizon-crossing, the behavior of $\mathbf{q}$ changes rapidly, and the oscillatory behavior of the two modes starts to diminish.  In this regime, both the sub-horizon term $k^2 \mathbf{q}$ and the terms on the right-hand side of equation (\ref{q EoM}) have to be taken into account.  To find a solution during horizon-crossing, we change variables to $z \equiv -k\tau$. Using equation (\ref{tau}), to first order in slow-roll,  
\begin{align}
\label{z}
z = - \int \frac{k}{aH} \, dN \approx (1 + \epsilon)\frac{k}{aH},
\end{align}
where we used integration by parts and then neglected higher-order terms.  After using this change of variables and equation (\ref{z}), and keeping terms to the two lowest orders in the SRST expansion, we obtain
\begin{align}
\label{q EoM wrt z in SRST}
\frac{D^2 \mathbf{q}}{dz^2} + \mathbf{q} = \frac{1}{z^2} \left[2 \mathbf{I} + 3 (\epsilon \mathbf{I} - \mathbf{\tilde{M}}^{(1)})\right] \mathbf{q}.
\end{align}

To solve equation (\ref{q EoM wrt z in SRST}), we assume that during horizon-crossing, the time variation of the term in brackets on the right-hand side of the equation can be neglected.  Using this assumption, we can decouple the two modes by rotating to the basis that diagonalizes the effective mass matrix at horizon exit.  Here, we follow an approach similar to those used in references \cite{BartoloEtAl-2001a}, \cite{ByrnesAndWands-2006}, and  \cite{BartoloEtAl-2001b}.  We write the rotation matrix that diagonalizes the mass matrix as
\begin{align}
\label{U matrix}
\mathbf{U} = \left(\begin{array}{cc} \cos \tilde{\theta} & - \sin \tilde{\theta} \\ \sin \tilde{\theta} & \cos \tilde{\theta} \end{array}\right).
\end{align} 
Using the rotation matrix, the effective mass matrix can be diagonalized as
\begin{align}
\label{M in decoupled basis}
\mathbf{U^\dag \tilde{M}^{(1)}U} = \left(\begin{array}{cc} m_+ & 0 \\ 0 & m_- \end{array}\right), 
\end{align}
where the eigenvalues of the effective mass matrix are
\begin{align}
\label{M eigenvalues}
m_{\pm} = \frac{1}{2} \left[\mathrm{Tr}(\mathbf{\tilde{M}}^{(1)}) \pm \sqrt{[\mathrm{Tr}(\mathbf{\tilde{M}}^{(1)})]^2 - 4 \, \mathrm{Det}(\mathbf{\tilde{M}}^{(1)})}\right]_* .
\end{align}
Using equation (\ref{M eigenvalues}), we find the rotation angle for the rotation matrix to be
\begin{align}
\tan 2 \tilde{\theta} = \frac{2\tilde{M}^{(1)}_{12}}{\tilde{M}^{(1)}_{11} - \tilde{M}^{(1)}_{22}}.
\end{align}

Now we use equations (\ref{q EoM wrt z in SRST}) and (\ref{M in decoupled basis}) to find an equation of motion for the modes in the rotated basis, $\mathbf{\tilde{q}} \equiv (\tilde{q}_+,\tilde{q}_-) = \mathbf{U^{\dag} q}$.  We find
\begin{align}
\label{q EoM wrt z diagonalized}
\frac{D^2 \tilde{q}_{\pm}}{dz^2} + \left[1 - \frac{1}{z^2} \left(\nu_{\pm}^2 - \frac{1}{4}\right)\right] \tilde{q}_{\pm} \approx 0,
\end{align}
where 
\begin{align}
\nu_{\pm} \equiv \frac{3}{2} + \epsilon - m_{\pm}
\end{align}
and where we made the conventional assumption that the time variation of $\mathbf{U}$ along the field trajectory can be neglected for the couple of $e$-folds on either side of horizon exit, which is a valid assumption in the SRST limit (see, e.g., \cite{NakamuraAndStewart-1996,BartoloEtAl-2001a,NibbelinkAndVanTent-2001}).   Now that we have decoupled the modes, the solution for the modes can be written in terms of Hankel functions as
\begin{align}
\label{Hankel funcs}
\tilde{q}_{\pm} = \sqrt{\frac{\pi}{4k}} \sqrt{z} & [e^{i\frac{\pi}{2}(2+\epsilon - m_{\pm})} H^{(1)}_{\nu_\pm} (z) \tilde{a}_{\pm}(\mathbf{k}) \nonumber \\ & + e^{-i\frac{\pi}{2}(2+\epsilon - m_{\pm})} H^{(2)}_{\nu_\pm} (z) \tilde{a}_{\pm}^\dag (-\mathbf{k})],
\end{align}
where $\mathbf{\tilde{a}} \equiv \mathbf{U^{\dag}a}$ and where the overall normalization was determined from matching to the sub-horizon solution in equation (\ref{q subhorz soln}).  

Once the modes have passed significantly outside the horizon, the oscillatory behavior of the fields dies away and the fields are free to grow, decay, and/or to couple to each other.  After horizon exit but not too late that the SRST parameters have significantly evolved, the growing mode solutions can be found from the leading order term in the asymptotic expansion of the horizon-crossing solution in equation (\ref{Hankel funcs}).  The leading order term in this expansion is
\begin{align}
\label{Hankel funcs limit}
\tilde{q}_{\pm} \rightarrow \frac{i}{\sqrt{2k}} \left[1 + C (\epsilon - m_{\pm})\right] z^{-1-(\epsilon - m_{\pm})} b_{\pm}(\mathbf{k}),
\end{align}
where 
\begin{align}
C = 2 - \ln 2 - \gamma \approx 0.7296,
\end{align} 
$\gamma \approx 0.5772$ is the Euler-Mascheroni constant, and 
\begin{align}
\label{super-horz b operator}
b_{\pm}(\mathbf{k}) = e^{i\pi (\epsilon - m_{\pm})/2}\tilde{a}_{\pm}(\mathbf{k}) - e^{-i\pi(\epsilon - m_{\pm})/2}\tilde{a}_{\pm}^\dag (-\mathbf{k}). 
\end{align} 
Note that equation (\ref{super-horz b operator}) implies that the perturbations become classical soon after passing outside the horizon, since now $\mathbf{\tilde{q}}$ and its conjugate momenta commute \cite{NakamuraAndStewart-1996}. 

Finally, since the SRST parameters will  inevitably evolve some time after the modes exit the horizon, we will need to perform a second rotation in field space to the kinematical basis in order to determine the late-time behavior of the modes.  After performing the rotation, we will be able to use our second-order SRST semi-analytic approximations for the curvature and isocurvature modes that we derived for the super-horizon limit.  In Section \ref{pwr spec at end}, we illustrate this last series of steps using the transfer matrix formalism as short-hand.

\subsubsection{Power Spectra At Horizon Exit}
\label{pwr spec at horz exit}

Armed with these results, we now calculate the curvature and isocurvature power spectra and their correlated cross spectrum at horizon exit. 

We define the power spectrum, $P_{\mathcal{X}}$, of a quantity $\mathcal{X}$ as 
\begin{align}
\label{pwr spectrum defn}
\mathcal{P}_{\mathcal{X}} \delta^3(\mathbf{k} - \tilde{\mathbf{k}}) \equiv \frac{k^3}{2\pi^2} \langle \mathcal{X} (\mathbf{k}) \mathcal{X}^\dag (\tilde{\mathbf{k}}) \rangle,
\end{align}
and the cross spectrum, $C_{\mathcal{XY}}$, of the quantities $\mathcal{X}$ and $\mathcal{Y}$ as 
\begin{align}
\label{pwr spectrum defn}
\mathcal{C}_{\mathcal{XY}} \delta^3(\mathbf{k} - \tilde{\mathbf{k}}) \equiv \frac{k^3}{2\pi^2} \langle \mathcal{X} (\mathbf{k}) \mathcal{Y}^\dag (\tilde{\mathbf{k}}) \rangle. 
\end{align}

To calculate the spectra of curvature and isocurvature perturbations at horizon exit, we need to relate the perturbations in the decoupled basis to the curvature and isocurvature perturbations.  To relate the modes in the two bases, we follow a procedure similar that used by \cite{BartoloEtAl-2001a}, \cite{ByrnesAndWands-2006}, and \cite{BartoloEtAl-2001b}.  The curvature and isocurvature modes are related to the modes in the original given basis by a rotation matrix with rotation angle $\theta$, where $\tan \theta \equiv \frac{\phi_2'}{\phi_1'}$.  In turn, the modes in the given basis are related to the decoupled modes by the rotation matrix $\mathbf{U}$ defined in equation (\ref{U matrix}). Therefore, the curvature and isocurvature modes can be related to the decoupled modes by the combined transformation
\begin{align}
\label{kine to decoupled mode transformation}
\left(\begin{array}{c} a\delta \phi_{\parallel} \\ a\delta \phi_{\perp} \end{array} \right) = & 
\left(\begin{array}{cc} \cos (\tilde{\theta} - \theta) & - \sin (\tilde{\theta} - \theta) \\ \sin (\tilde{\theta} - \theta) & \cos (\tilde{\theta} - \theta) \end{array} \right)
\left(\begin{array}{c} \tilde{q}_+ \\ \tilde{q}_- \end{array} \right),
\end{align}
where the angle of the combined rotations can be expressed as
\begin{align}
\label{combined rotn angle}
\tan 2(\tilde{\theta} - \theta) = \frac{2\tilde{M}^{(1)}_{\parallel \perp}}{\tilde{M}^{(1)}_{\parallel \parallel} - \tilde{M}^{(1)}_{\perp \perp}}. 
\end{align}
Notice that the same two terms that determine the evolution of the curvature perturbations to first-order in the SRST expansion in equation (\ref{comoving R EoM integrated v2})---$2\tilde{M}_{\parallel \perp}$ and the combination $\tilde{M}_{\parallel \parallel}-\tilde{M}_{\perp \perp}$---also determine the net rotation angle between the kinematical and the decoupled bases. 

Now we find the power spectra at horizon exit.  Using the combined rotation angle to relate the modes in the two bases, the spectra of adiabatic and entropy modes are related to the spectra of decoupled modes by
\begin{widetext}
\begin{align}
\label{adia + entr spectra}
\langle \delta \phi_{\parallel} \delta \phi_{\parallel}^\dag \rangle & = \frac{1}{2a^2} [(1 + \cos 2 (\tilde{\theta} - \theta)) \langle \tilde{q}_+^2 \rangle + (1 - \cos 2 (\tilde{\theta} - \theta)) \langle \tilde{q}_-^2 \rangle ],  \nonumber \\
\langle \delta \phi_{\parallel} \delta \phi_{\perp}^\dag \rangle & = \frac{1}{2a^2} \sin 2 (\tilde{\theta} - \theta)) \left(\langle \tilde{q}_+^2 \rangle - \langle \tilde{q}_-^2 \rangle \right),  \\
\langle \delta \phi_{\perp} \delta \phi_{\perp}^\dag \rangle & = \frac{1}{2a^2} [(1 - \cos 2 (\tilde{\theta} - \theta)) \langle \tilde{q}_+^2 \rangle + (1 + \cos 2 (\tilde{\theta} - \theta)) \langle \tilde{q}_-^2 \rangle ]. \nonumber
\end{align}
\end{widetext}
After substituting equation (\ref{Hankel funcs limit}) and the expectation values 
\begin{align}
\label{expectation values of operators}
\langle b_{i}^\dag(\mathbf{k}) b_{j}(\tilde{\mathbf{k}}) \rangle = G_{ij} \delta^3 (\mathbf{k} - \tilde{\mathbf{k}})
\end{align}
into equation (\ref{adia + entr spectra}), the power spectra are given by
\begin{align}
\label{spectra at horz}
\mathcal{P}_{\mathcal{R}_*} & = \left(\frac{H_*}{2\pi}\right)^2 \frac{1}{2\epsilon_*} \left[1 + 2(C - 1) \epsilon - 2C\tilde{M}_{\parallel \parallel}\right]_*, \nonumber \\
\mathcal{C}_{\mathcal{R}\mathcal{S}_*} & = \left(\frac{H_*}{2\pi}\right)^2 \frac{1}{2\epsilon_*}[- 2C\tilde{M}_{\parallel \perp} ]_*, \\
\mathcal{P}_{\mathcal{S}_*} & = \left(\frac{H_*}{2\pi}\right)^2 \frac{1}{2\epsilon_*} \left[1 + 2(C - 1) \epsilon - 2C\tilde{M}_{\perp \perp}\right]_*.  \nonumber 
\end{align}
where it is implied that the terms are to be calculated to second-order in the SRST expansion; we have dropped the superscripts and chosen not to expand terms in order to de-clutter the expressions.  

Equation (\ref{spectra at horz}) shows that at horizon exit, the curvature and isocurvature spectra have similar power, while the correlated cross spectrum is down by a factor of approximately $2C$ times the turn rate ($\frac{\eta_{\perp}}{v} \approx -\tilde{M}^{(1)}_{\parallel \perp}$).  Thus, to lowest-order, the turn rate alone determines the size of the cross-spectrum relative to the curvature and isocurvature spectra at horizon exit.  This makes sense as we found earlier that the turn rate determines the strength of sourcing of one mode by the other.  If the turn rate vanishes when the modes exit the horizon, then those particular curvature and isocurvature modes will not be correlated at horizon exit, as in this case, the decoupled basis coincides with the kinematical basis.

Finally, we introduce the tensor power spectrum, $\mathcal{P}_{T}$, which represents the spectrum of gravitational waves.   We introduce the tensor power spectrum in order to calculate the tensor-to-scalar ratio, an important spectral observable.  The tensor spectrum is independent of the number of scalar fields, and to next-to-lowest-order in the slow-roll expansion, it is calculated to be \cite{StewartAndLyth-1993}
\begin{align}
\label{tensor pwr spec}
\mathcal{P}_{T_*} = 8\left(\frac{H_*}{2\pi}\right)^2 \left[1 + 2(C - 1) \epsilon\right]_*.
\end{align}
It is conserved for modes outside the horizon.

\subsubsection{Power Spectra At The End of Inflation}
\label{pwr spec at end}

We now calculate the three scalar spectra at the end of inflation.  

We can calculate the spectra at the end of inflation from the corresponding spectra at horizon exit.  To do so, we simply need to account for the further growth or decay of the modes after they exit the horizon.  To account for the mode evolution, we can use the semi-analytic expressions we found in Section (\ref{curvature modes}).  But rather than write out these expressions in full, we will use the transfer matrix formalism in equations (\ref{transfer matrix}) and (\ref{transfer functions}) as short-hand.  Using the transfer matrix formalism \cite{AmendolaEtAl-2001,WandsEtAl-2002} , the modes at the end of inflation are related to the modes at horizon exit by
\begin{align}
\label{transfer matrix v2}
\left(\begin{array}{c} \mathcal{R} \\ \mathcal{S} \end{array} \right) = & 
\left(\begin{array}{cc} 1 & T_{\mathcal{RS}} \\ 0 & T_{\mathcal{SS}} \end{array} \right)
\left(\begin{array}{c} \mathcal{R}_* \\ \mathcal{S}_* \end{array} \right).
\end{align}
where it is implied that the transfer functions are evaluated at the end of inflation.  Hence, the spectra at the end of inflation are related to the spectra at horizon exit by \cite{ByrnesAndWands-2006}
\begin{align}
\label{spectra in terms of transfer funcs}
\mathcal{P}_{\mathcal{R}} & = \mathcal{P}_{\mathcal{R}_*} + 2T_{\mathcal{RS}}\mathcal{C}_{\mathcal{RS}_*} + T_{\mathcal{RS}}^2 \mathcal{P}_{\mathcal{S}_*}, \nonumber \\
\mathcal{C}_{\mathcal{RS}} & = T_{\mathcal{SS}}\mathcal{C}_{\mathcal{RS}_*} + T_{\mathcal{RS}} T_{\mathcal{SS}} \mathcal{P}_{\mathcal{S}_*}, \\
\mathcal{P}_{\mathcal{S}} & = T_{\mathcal{SS}}^2 \mathcal{P}_{\mathcal{S}_*}. \nonumber
\end{align}

Substituing in equation (\ref{spectra at horz}) and keeping only the lowest order terms in the SRST expansion, the power spectra at the end of inflation are
\begin{align}
\label{pwr spec at end SRST}
\mathcal{P}_{\mathcal{R}} & = \left(\frac{H_*}{2\pi}\right)^2 \frac{1}{2 \epsilon_*}  (1+ T_{\mathcal{RS}}^2) , \nonumber \\
\mathcal{C}_{\mathcal{RS}} & = \left(\frac{H_*}{2\pi}\right)^2 \frac{1}{2 \epsilon_*} T_{\mathcal{RS}} T_{\mathcal{SS}}, \\
\mathcal{P}_{\mathcal{S}} & = \left(\frac{H_*}{2\pi}\right)^2 \frac{1}{2 \epsilon_*} T_{\mathcal{SS}}^2. \nonumber
\end{align}
In equation (\ref{pwr spec at end SRST}), all terms are calculated to lowest-order in the SRST expansion, including the transfer functions.  The single-field power spectra can be recovered from equation (\ref{pwr spec at end SRST}) by setting $T_{\mathcal{SS}}$ to zero, which also forces $T_{\mathcal{RS}} = 0$.

Next, we find the power spectra to second order in the SRST approximation.  But first, we introduce two new quantities that will allow us to write the expressions for the spectra and spectral observables more compactly.  We start by introducing a quantity called the correlation angle, which we define as
\begin{align}
\label{correlation angle}
\sin \Delta_N \equiv \frac{T_{\mathcal{RS}}}{\sqrt{1 + T_{\mathcal{RS}}^2}}, 
\end{align}
where we add the subscript $N$ to make it clear that $\Delta$ represents the correlation angle, not the spectra themselves.\footnote{Our correlation angle is inspired by the correlation angle introduced by Wands \textit{et. al} \cite{WandsEtAl-2002}, which was defined as $\cos \Delta \equiv \frac{C_{\mathcal{RS}}}{\sqrt{P_{\mathcal{R}} P_{\mathcal{S}}}} \approx \frac{T_{\mathcal{RS}}}{\sqrt{1 + T_{\mathcal{RS}}^2}}$.  But we have chosen to modify the definition of this quantity for two reasons.  First, Bartolo \textit{et. al.} \cite{BartoloEtAl-2001a} had already assigned the symbol $r_{\mathcal{C}}$ to the quantity $\frac{C_{\mathcal{RS}}}{\sqrt{P_{\mathcal{R}} P_{\mathcal{S}}}}$, to reflect the fact that it is similar in form to the tensor-to-scalar ratio, $r_T$.  Second, we needed a quantity to represent the angle between $\hat{e}_{\parallel}$ and $\boldsymbol{\nabla} N$,  so making this modification seemed to be the best compromise.}   We chose the subscript $N$ because the correlation angle represents the angle between $\mathbf{e}_{\parallel}$ and the gradient of the number of $e$-folds, $\boldsymbol{\nabla}  N$---that is, 
\begin{align}
\mathbf{e}_{\parallel} \cdot \boldsymbol{\nabla} N \propto \cos \Delta_N.
\end{align}
We can see this by considering the following facts.  For any multi-field model, the curvature spectrum equals \cite{SasakiAndStewart-1995}
\begin{align}
\label{multi field curv spec}
P_{\mathcal{R}} = \left(\frac{H_*}{2\pi}\right)^2 |\boldsymbol{\nabla} N|^2,
\end{align}
to lowest order.  Since
\begin{align}
\boldsymbol{\phi}' \cdot \boldsymbol{\nabla N} = 1,
\end{align}
then comparing equation (\ref{pwr spec at end SRST}) to equation (\ref{multi field curv spec}), it follows that for two-field inflation
\begin{align}
\boldsymbol{\nabla}^\dag N = \frac{1}{\sqrt{2\epsilon}} \left[\mathbf{e}_{\parallel} + T_{\mathcal{RS}} \mathbf{e}_{\perp}\right].
\end{align}
Therefore, the correlation angle is indeed the angle between $\boldsymbol{\nabla} N$ and $\mathbf{e}_{\parallel}$.  Also, it is useful to define a unit vector, $\mathbf{e}_N$, that points in the direction of $\boldsymbol{\nabla}^\dag N$, where based on the above expressions, $\mathbf{e}_N$ is related to the kinematical basis vectors by
\begin{align}
\label{e_N}
\mathbf{e}_{N} = \cos \Delta_N \, \mathbf{e}_{\parallel} + \sin \Delta_N \, \mathbf{e}_{\perp}.
\end{align}

Now combining equations (\ref{spectra at horz}) and (\ref{spectra in terms of transfer funcs}), and using our two new quantities (\ref{correlation angle}), and (\ref{e_N}), the second-order power spectra can be written as
\begin{widetext}
\begin{align}
\label{spectra at end}
\mathcal{P}_{\mathcal{R}} & = \left(\frac{H_*}{2\pi}\right)^2 \frac{1}{2 \epsilon_*} (1+ T_{\mathcal{RS}}^2) [1 + 2(C - 1) \epsilon - 2C\mathbf{e}_N^\dag \mathbf{\tilde{M}} \mathbf{e}_N]_*, \nonumber \\
\mathcal{C}_{\mathcal{RS}} & = \left(\frac{H_*}{2\pi}\right)^2 \frac{1}{2 \epsilon_*} T_{\mathcal{RS}} T_{\mathcal{SS}} [1 + 2(C - 1) \epsilon - 2C \mathbf{e}_N^\dag \mathbf{\tilde{M}} \mathbf{e}_{\perp} \sin^{-1} \Delta_N]_*,  \\
\mathcal{P}_{\mathcal{S}} & = \left(\frac{H_*}{2\pi}\right)^2 \frac{1}{2 \epsilon_*} T_{\mathcal{SS}}^2 [1 + 2(C - 1) \epsilon - 2C \mathbf{e}_{\perp}^\dag \mathbf{\tilde{M}} \mathbf{e}_{\perp}]_*.  \nonumber
\end{align}
\end{widetext}
For consistency, all terms are calculated to second-order in the SRST approximation, with the understanding that here we are using the second-order SRST approximations for the turn rate (equation (\ref{kine quantities in SRST O2})) and for $\beta$ (equation (\ref{isocurv mass in SRST O2})) to calculate the transfer functions.  We can also write the power spectra a bit more compactly using equation (\ref{tensor pwr spec}):
\begin{align}
\label{spectra in neat form}
\mathcal{P}_{\mathcal{R}} & = \frac{\mathcal{P}_T}{16 \epsilon_*} (1+T_{\mathcal{RS}}^2)[1 - 2C\mathbf{e}_N^\dag \mathbf{\tilde{M}} \mathbf{e}_N]_*, \nonumber \\
\mathcal{C}_{\mathcal{RS}} & = \frac{\mathcal{P}_T}{16 \epsilon_*} T_{\mathcal{RS}} T_{\mathcal{SS}} [1 - 2C \mathbf{e}_N^\dag \mathbf{\tilde{M}} \mathbf{e}_{\perp} \sin^{-1} \Delta_N]_*, \nonumber \\
\mathcal{P}_{\mathcal{S}} & = \frac{\mathcal{P}_T}{16 \epsilon_*} T_{\mathcal{SS}}^2 [1 - 2C \mathbf{e}_{\perp}^\dag \mathbf{\tilde{M}} \mathbf{e}_{\perp}]_* .
\end{align}
Note the high degree of symmetry among the second-order terms in equation (\ref{spectra in neat form}): each instance of the curvature perturbation in the spectra is accompanied by the unit vector $\mathbf{e}_N$, while each instance of the isocurvature perturbation is accompanied by the unit vector $\mathbf{e}_{\perp}$. 

Now we analyze the power spectra at the end of inflation.  At lowest-order, the power spectra are determined by just four quantities: $H$, $\epsilon$, $T_{\mathcal{SS}}$, and $T_{\mathcal{RS}}$.  Equation (\ref{pwr spec at end SRST}) shows that all three spectra are modulated by the factor $\left(\frac{H_*}{2\pi}\right)^2 \frac{1}{2\epsilon_*}$.  The differences among the spectra lie in the transfer functions, with the ratio $\frac{T_{\mathcal{SS}}}{T_{\mathcal{RS}}}$ effectively controlling the relative sizes of the three spectra.   In fact, this ratio directly controls the relative sizes of the isocurvature and cross spectra: if  $T_{\mathcal{RS}} \ge T_{\mathcal{SS}}$, then $\mathcal{C}_{\mathcal{RS}} \ge \mathcal{P}_{\mathcal{S}}$, otherwise $\mathcal{C}_{\mathcal{RS}} < \mathcal{P}_{\mathcal{S}}$.  Also, since $T_{\mathcal{SS}} < 1$ is virtually always true, we expect the curvature spectrum to be the largest of the three spectra.  Indeed, the smaller $T_{\mathcal{SS}}$ is, the smaller the cross spectrum ($\mathcal{C}_{\mathcal{RS}} \propto T_{\mathcal{SS}}$) and the isocurvature spectrum ($\mathcal{P}_{\mathcal{S}} \propto T_{\mathcal{SS}}^2$).

In turn, the two transfer functions depend on the kinematical profiles of the speed up rate, the turn rate, and the effective entropy mass, as we explained in detail in Section \ref{curvature modes}.   Typically, $T_{\mathcal{SS}} \ll 1$ by the end of inflation, since the speed up rate usually must be large and positive to end inflation, or equivalently, since the amplitude of isocurvature modes depends inversely on $v = \sqrt{2\epsilon}$.  For inflationary scenarios with large turn rates, typically  $T_{\mathcal{RS}} \gtrsim$ a few and $T_{\mathcal{SS}} \lll1$---for reasons we discussed earlier---resulting in $\mathcal{P}_{\mathcal{R}} \ggg  \mathcal{C}_{\mathcal{RS}} \ggg \mathcal{P}_{\mathcal{S}}$.   For these scenarios, we expect that the cross and isocurvature spectra will be hard to measure, if not undetectable, and that $\mathcal{P}_{\mathcal{R}} \ggg \mathcal{P}_{T}$.  On the opposite extreme, if the turn rate is very small and satisfies $\frac{\eta_{\perp}}{v} \ll \frac{\eta_{\parallel}}{v}$ during inflation, then we expect $T_{\mathcal{RS}}$ to be significantly less than one and $T_{\mathcal{SS}}$ to be larger than in the previous case but still at least an order of magnitude smaller than one.  In this limit, the curvature spectrum can be approximated by the corresponding single-field result, and the isocurvature and cross spectra will be smaller, with their relative sizes depending on the ratio of $\frac{T_{\mathcal{SS}}}{T_{\mathcal{RS}}}$.  However, unlike in the case of a large turn rate, the isocurvature and cross spectra will be still be appreciable and hence potentially measureable, as long as the isocurvature modes are not destroyed during the reheating process following inflation.  And if the turn rate vanishes exactly for all scales of interest, then we recover the exact single-field expression for the curvature spectrum, and the cross spectrum is exactly zero.  In this case, the isocurvature spectrum is still usually much smaller than the curvature spectrum. 

Taking these dependences together, this means we can trace the behavior of the power spectra back to five kinematical functions: $H$ and $\epsilon$, which set the overall scale of the spectra; and $\frac{\eta_{\parallel}}{v}$, $\frac{\eta_{\perp}}{v}$, and $\mu_{\perp}$, which determine the behavior of the transfer functions.  Or, being even more succinct, since $\epsilon = - (\ln H)'$ and $\frac{\eta_{\parallel}}{v} = \frac{(\ln H)''}{2 (\ln H)'}$, we can pare these functions down to a set of just three functions with their scale-dependences: $H$, $\frac{\eta_{\perp}}{v}$, and $\mu_{\perp}$, where the second two functions are unique to two-field inflation.   It is these three quantities that we can reconstruct from observational data, and hence that serve as the critical link between phenomenology and observational constraints.  To lowest order in the SRST limit, these quantities represent the value of $V$, $|\mathbf{\nabla} \ln V|$, and the coefficients of $\mathbf{\tilde{M}}$. In order words, we can think of these quantities as representing vital information about the value, the gradient, and the Hessian of $\ln V$, along with corrections from any non-trivlal geometry of the field manifold.   It is precisely these relationships among the kinematical functions, the Lagrangian, and the spectra that allow us to connect features in the spectra directly back to features in the inflationary Lagrangian.  We summarize these kinematical functions in Table 4, where we also indicate how they can be reconstructed from observational data.  

Finally, we discuss when a two-field model effectively looks like a single-field model---that is, when $\mathcal{C}_{\mathcal{RS}}$ and $\mathcal{P}_{\mathcal{S}}$ are vanishingly small and when $P_{\mathcal{R}}$ is identical to the single-field result.  Of course, after inflation ends the modes could be processed further, particularly during the 

\begin{widetext}
\begin{table}[t]
\renewcommand{\arraystretch}{1.5}
\begin{tabular}{|x{2.12in}|x{1.8in}|x{2.6in}|}
\hline
\multicolumn{3}{|c|}{\normalsize \rule{0cm}{0.55cm} Table 4.  The Quantities That Capture the Key Features of Any Two-Field Model of Inflation} \tabularnewline[1.4ex] 
\hline \multicolumn{3}{|c|}{$\, $}\tabularnewline[-3.7ex] \hline \normalsize \rule{0cm}{0.55cm} \textit{Quantity} & \normalsize \textit{Relation to the Lagrangian} & \normalsize \textit{Relation to the Spectral Observables} \tabularnewline[1.4ex] \hline
\rule{0cm}{0.55cm} \normalsize Hubble Rate, $H$ & \normalsize $\sqrt{\frac{V}{3}}$ & \normalsize $\pi \sqrt{\frac{\mathcal{P}_{T}}{2}}$  \tabularnewline[1.4ex] \hline
\rule{0cm}{0.55cm} \normalsize Field Speed, $v = \sqrt{2\epsilon} $ & \normalsize $|\boldsymbol{\nabla} \ln V|$ & \normalsize $\sqrt{-n_T}$  \tabularnewline[1.4ex] \hline
\rule{0cm}{0.55cm} \normalsize Speed Up Rate, \large $\frac{\eta_{\parallel}}{v}$ & \normalsize $- \tilde{M}_{\parallel \parallel}$ & \normalsize $\frac{n_T - n_{\mathcal{R}}}{2} + \frac{r_C^2}{1-r_C^2}  \left(\frac{2n_{\mathcal{C}}-n_{\mathcal{R}}-n_{\mathcal{S}}}{2} \right)$, $\,$ or $\, \, \frac{\alpha_T}{2n_T}$     \tabularnewline[1.4ex] \hline
\rule{0cm}{0.55cm} \normalsize Turn Rate, \large $\frac{\eta_{\perp}}{v}$ & \normalsize $- \tilde{M}_{\parallel \perp}$ & \normalsize $\frac{r_C}{\sqrt{1-r_C^2}} \left(\frac{n_{\mathcal{S}} - n_{\mathcal{C}}}{2}\right)$ \tabularnewline[1.4ex] \hline
\rule{0cm}{0.55cm} \normalsize Effective Entropy Mass, $\mu_{\perp}$ &  \normalsize $\tilde{M}_{\perp \perp}$ &  \normalsize $\frac{n_{\mathcal{S}} - n_T}{2}$ \tabularnewline [1.4ex]
\hline
\end{tabular}
\end{table}
\noindent \small TABLE 4. The above table lists the five quantities that represent the main features of any two-field model of inflation: $H$, $\epsilon$, $\frac{\eta_{\parallel}}{v}$, $\frac{\eta_{\perp}}{v}$, and $\mu_{\perp}$.  Alternatively, since $\epsilon = - (\ln H)'$ and $\frac{\eta_{\parallel}}{v} = \frac{(\ln H)''}{2 (\ln H)'}$, we can reduce them further to a set of three functions and their scale-dependences: $H$, $\frac{\eta_{\perp}}{v}$, and $\mu_{\perp}$, where the latter two are unique to two-field inflation.  All other quantities are derived from these fundamental quantities.  The second column shows that to lowest order in the SRST limit, these quantities represent the values of $V$, $|\mathbf{\nabla} \ln V|$, and $\mathbf{\tilde{M}}$, where $\mathbf{\tilde{M}} = \boldsymbol{\nabla}^\dag \boldsymbol{\nabla} \ln V + \frac{\epsilon R}{(3-\epsilon)}$ and $R$ is the Ricci scalar of the field manifold. The third column shows how these quantities relate to the spectral observables, to lowest order in the SRST limit.
\end{widetext}

\noindent reheating process.  However, because such post-inflationary processing is model-dependent, we do not consider it here, and leave the reader to append any post-inflationary processing to our calculations here.  If there is no post-inflationary processing of modes, a two-field model will look like a single-field model only if $T_{\mathcal{RS}} \approx 0$ and $T_{\mathcal{SS}} \lll 1$.  The former holds when $\frac{\eta_{\perp}}{v} \lll \frac{\eta_{\parallel}}{v}$.  The latter can be achieved only if the integral of the sum of $M_{\perp \perp}$ and of $\epsilon$ times the Ricci scalar of the field manifold is large and positive, due to tight constraints on the turn rate.   So given that the turn rate can be approximated by $- \tilde{M}_{\parallel \perp}$, we require $- \tilde{M}_{\parallel \perp} \lll 1$ over all measured scales and $\tilde{M}_{\perp \perp} \gg 1$ to hold for at least a couple of $e$-folds before the end of inflation.

\subsubsection{Spectral Observables and Consistency Condition} 
\label{pwr spec observables}

From the expressions for the power spectra, we find the spectral indices, the running of the spectral indices, the tensor-to-scalar ratio, and the cross-correlation ratio, and then derive a consistency condition among these quantities.  In this section, it is implied that all quantities are evaluated at horizon exit, so we drop the subscript $*$.  As before, our equations here do not include any model-dependent post-inflationary processing of the modes.
 
First, we find the spectral indices and the running of the spectral indices.  We define the spectral index of a spectrum, $\mathcal{P}_{\mathcal{X}}$, as
\begin{align}
\label{spectral index defn}
n_{\mathcal{X}} \equiv \frac{d\ln \mathcal{P}_{\mathcal{X}}}{d\ln k}.
\end{align}
For comparison, our definition of the curvature spectral index is related to the more commonly used scalar spectral index, $n_s$, by 
\begin{align}
\label{ns nR relationship}
n_{s} = 1 + n_{\mathcal{R}} = 1 + \frac{d\ln \mathcal{P}_{\mathcal{R}}}{d\ln k}.
\end{align}
We have chosen, however, to depart from convention and to use equation (\ref{spectral index defn}) to define all four spectral indices so that they can be more readily compared.  

There are a few expressions that come in handy in finding the spectral indices.  The derivative with respect to $k$ is related to the derivative with respect to $N$ by $\frac{d}{d\ln k} \approx (1 + \epsilon) \frac{d}{dN}$.   To find the running of the transfer functions, the following expressions are helpful:
\begin{align}
\label{transfer func derivs}
T_{\mathcal{SS}}' & = - \beta T_{\mathcal{SS}}, \nonumber \\
T_{\mathcal{RS}}' & = - 2 \frac{\eta_{\perp}}{v} - \beta T_{\mathcal{RS}}. 
\end{align}
We can use the above equations to derive another result:
\begin{widetext}
\begin{align}
\label{scale dep of T_RS/epsilon}
\left[\ln \left(\frac{1+T_{\mathcal{RS}}^2}{2\epsilon}\right)\right]' = 2  \left(\ln |\boldsymbol{\nabla} N|\right)' = - 2 \frac{\eta_{\parallel}}{v}\cos^2 \Delta_N - 4 \frac{\eta_{\perp}}{v} \sin \Delta_N \cos \Delta_N + 2 \mu_{\perp} \sin^2 \Delta_N \approx 2 \mathbf{e}_N^\dag \mathbf{\mathcal{M}} \mathbf{e}_N,
\end{align}
\end{widetext}
where $\mathbf{\mathcal{M}}$ is defined as
\begin{align}
\label{M mod matrix}
\mathbf{\mathcal{M}} & \equiv \left(\begin{array}{cc} - \left(\frac{\eta_{\parallel}}{v}\right)^{(2)} & - \left(\frac{\eta_{\perp}}{v}\right)^{(2)} \\ - \left(\frac{\eta_{\perp}}{v}\right)^{(2)} & \mu_{\perp}^{(2)} \end{array} \right),
\end{align}
Since to first-order in the SRST limit, $\mathbf{\mathcal{M}} = \mathbf{\tilde{M}}^{(1)}$, $\mathbf{\mathcal{M}}$ can be viewed as the second-order extension of the effective mass matrix.  Equation (\ref{scale dep of T_RS/epsilon}) shows that the scale dependence of $|\boldsymbol{\nabla} N|$ is determined by the ($\mathbf{e}_N,\mathbf{e}_N$) component of $\mathbf{\mathcal{M}}$.    

Using the above relations, we find the power spectra to first-order in the SRST limit:
\begin{align}
\label{ns}
n_{T} & = -2\epsilon, \nonumber \\ 
n_{\mathcal{R}} & = n_T + 2  \mathbf{e}_N^\dag \mathbf{\tilde{M}} \mathbf{e}_N, \\
n_{\mathcal{C}} & = n_T + 2 \mathbf{e}_N^\dag \mathbf{\tilde{M}} \mathbf{e}_\perp \sin^{-1} \Delta_N, \nonumber \\
n_{\mathcal{S}} & = n_T + 2 \mathbf{e}_\perp^\dag \mathbf{\tilde{M}} \mathbf{e}_\perp. \nonumber
\end{align}
Equation (\ref{ns}) shows that the deviations from scale invariance are determined by $\epsilon$ and the various coefficients of $\mathbf{\tilde{M}}$. In fact, the last three spectral indices in equation (\ref{ns}) are virtually identical up to the unit vectors used to project out particular components of the effective mass matrix.      The spectral index $n_T$ depends on $\epsilon$, while $n_\mathcal{S}$ depends on $\epsilon$ and $\tilde{M}_{\perp \perp}$ ($\approx \mu_\perp$).  But the dependence of other two spectral indices on the various coefficients of $\mathbf{\tilde{M}}$ is more complicated, as $\mathbf{e}_N$ depends strongly on $T_{\mathcal{RS}}$, a double integral expression involving the speed up rate, turn rate, and effective entropy mass.   For example, if multi-field effects are small ($T_{\mathcal{RS}} \ll 1$), then $\mathbf{e}_N$ will point mostly in the direction of $\mathbf{e}_{\parallel}$ and hence $n_{\mathcal{R}}$ will depend mostly on the matrix coefficient $\tilde{M}_{\parallel \parallel}$ ($\approx - \frac{\eta_{\parallel}}{v}$), just like in single-field inflation.  However, if multi-field effects are sufficiently large that $T_{\mathcal{RS}} > 1$---that is, if the sourcing of curvature modes by isocurvature modes accounts for more than half of the amplitude of curvatures modes at the end of inflation---then $n_{\mathcal{R}}$ will depend more on the coefficients $\tilde{M}_{\parallel \perp}$ ($\approx - \frac{\eta_{\perp}}{v}$) and $\tilde{M}_{\perp \perp}$ ($\approx \mu_\perp$) than on $\tilde{M}_{\parallel \parallel}$.   Similarly, the exact dependence of $n_{\mathcal{C}}$ on the coefficients $\tilde{M}_{\parallel \perp}$ and $\tilde{M}_{\perp \perp}$ depends strongly on $T_{\mathcal{RS}}$.

Now we calculate the spectral indices to second-order in the SRST limit.  In doing so, we arrive at the following intermediate step
\begin{align}
\label{ns vIntermed}
n_{T} & = \left[1 + \epsilon + (1 - C) \frac{d}{dN} \right](-2\epsilon), \nonumber \\ 
n_{\mathcal{R}} & = n_T + 2\left[1 + \epsilon - C \frac{d}{dN} \right]\mathbf{e}_N^\dag \mathbf{\mathcal{M}} \mathbf{e}_N, \\
n_{\mathcal{C}} & = n_T + 2\left[1 + \epsilon - C \frac{d}{dN} \right]\mathbf{e}_N^\dag \mathbf{\mathcal{M}} \mathbf{e}_\perp \sin^{-1} \Delta_N, \nonumber \\
n_{\mathcal{S}} & = n_T + 2\left[1 + \epsilon - C \frac{d}{dN}\right]\mathbf{e}_\perp^\dag \mathbf{\mathcal{M}} \mathbf{e}_\perp.  \nonumber
\end{align}
To act the differentiation operator on the matrix product, we use the first-order expressions $\frac{D}{dN} \approx - \boldsymbol{\nabla}^\dag \ln V \boldsymbol{\nabla}$ and the helpful result
\begin{align}
\label{De_N}
\frac{D\mathbf{e}_N}{dN}  \approx (\mathbf{e}_N^\dag \mathbf{\mathcal{M}} \mathbf{e}_N^{\perp}) \mathbf{e}_N^{\perp}, 
\end{align}
where
\begin{align}
\mathbf{e}_N^{\perp} = -\sin \Delta_N \mathbf{e}_{\parallel} + \cos \Delta_N \mathbf{e}_{\perp}.
\end{align}
Using these expressions, the second-order spectral indices can be written compactly as
\begin{widetext}
\begin{align}
\label{ns v2}
n_{T} & = \left[1 + \epsilon + (C-1) \mathcal{M}_{\parallel \parallel} \right] (-2\epsilon), \nonumber \\ 
n_{\mathcal{R}} & = n_T + 2 \mathbf{e}_N^\dag \left[(1 + \epsilon)\mathbf{\mathcal{M}} - 2 C (\mathbf{\mathcal{M}} \mathbf{e}_N^{\perp}) (\mathbf{\mathcal{M}} \mathbf{e}_N^{\perp})^\dag + C \boldsymbol{\nabla} \ln V \boldsymbol{\nabla} \mathbf{\mathcal{M}}\right] \mathbf{e}_N, \\
n_{\mathcal{C}} & = n_T + 2 \mathbf{e}_N^\dag \left[(1 + \epsilon)\mathbf{\mathcal{M}} + 2 C (\mathbf{\mathcal{M}} \mathbf{e}_N^\perp) (\mathbf{\mathcal{M}} \mathbf{e}_{\parallel})^\dag \sin^{-1} \Delta_N + C \boldsymbol{\nabla} \ln V \boldsymbol{\nabla} \mathbf{\mathcal{M}}\right] \mathbf{e}_\perp \sin^{-1} \Delta_N, \nonumber \\
n_{\mathcal{S}} & = n_T + 2 \mathbf{e}_\perp^\dag \left[(1 + \epsilon)\mathbf{\mathcal{M}} - 2 C (\mathbf{\mathcal{M}} \mathbf{e}_{\parallel}) (\mathbf{\mathcal{M}} \mathbf{e}_{\parallel})^\dag + C \boldsymbol{\nabla} \ln V \boldsymbol{\nabla} \mathbf{\mathcal{M}}\right] \mathbf{e}_\perp, \nonumber
\end{align}
\end{widetext}
where all terms are to be calculated to second-order and terms of higher order are dropped.   

The runnings of the spectral indices are defined as
\begin{align}
\label{running of spectral index defn}
\alpha_{\mathcal{X}} \equiv \frac{d n_{\mathcal{X}}}{d\ln k}.
\end{align}
We find them to first-order in the SRST limit by differentiating equation (\ref{ns}) and using equations (\ref{De_p/dN}) and (\ref{De_N}).  The results can be written compactly as
\begin{widetext}
\begin{align}
\label{alphas}
\alpha_{T} & = 4 \epsilon \tilde{M}_{\parallel \parallel}, \nonumber \\ 
\alpha_{\mathcal{R}} & = \alpha_T + 2 \mathbf{e}_N^\dag \left[2 (\mathbf{\tilde{M}} \mathbf{e}_N^{\perp}) (\mathbf{\tilde{M}} \mathbf{e}_N^{\perp})^\dag - \boldsymbol{\nabla} \ln V \boldsymbol{\nabla} \mathbf{\tilde{M}}\right] \mathbf{e}_N, \\
\alpha_{\mathcal{C}} & = \alpha_T + 2 \mathbf{e}_N^\dag \left[-2 (\mathbf{\tilde{M}} \mathbf{e}_N^\perp) (\mathbf{\tilde{M}} \mathbf{e}_{\parallel})^\dag \sin^{-1} \Delta_N - \boldsymbol{\nabla} \ln V \boldsymbol{\nabla} \mathbf{\tilde{M}}\right] \mathbf{e}_\perp \sin^{-1} \Delta_N, 
 \nonumber\\
\alpha_{\mathcal{S}} &  = \alpha_T + 2 \mathbf{e}_\perp^\dag \left[2 (\mathbf{\tilde{M}} \mathbf{e}_{\parallel}) (\mathbf{\tilde{M}} \mathbf{e}_{\parallel})^\dag - \boldsymbol{\nabla} \ln V \boldsymbol{\nabla} \mathbf{\tilde{M}}\right] \mathbf{e}_\perp. \nonumber
\end{align}
\end{widetext}

Next, we find the spectral observables that are based on ratios of the spectra: the tensor-to-scalar ratio and the cross-correlation ratio.  The tensor-to-scalar ratio, $r_T$, is defined as the ratio of the tensor power spectrum to the scalar (curvature) power spectrum.  From equations (\ref{tensor pwr spec}) and (\ref{spectra in neat form}), we find that to first-order in the SRST limit,
\begin{align}
\label{r_T}
r_T = & 16 \epsilon \cos^2 \Delta_N,
\end{align}
while to second-order,
\begin{align}
\label{r_T second-order}
r_T = & 16 \epsilon \cos^2 \Delta_N (1 + 2 C \mathbf{e}_{N}^\dag \mathbf{\tilde{M}} \mathbf{e}_N).
\end{align}
Modulo the factor of $\cos^2 \Delta_N$, equation (\ref{r_T}) is identical to the single-field result; that is, the single-field result provides only an upper bound on the two-field tensor-to-scalar ratio.  This result agrees with the result in \cite{ByrnesAndWands-2006} for the case of canonical kinetic terms.

In analogy to the tensor-to-scalar ratio, Bartolo \textit{et. al.} \cite{BartoloEtAl-2001a} introduced the quantity
\begin{align}
r_C \equiv \frac{\mathcal{C}_{\mathcal{RS}}}{\sqrt{\mathcal{P}_{\mathcal{R}}\mathcal{P}_{\mathcal{S}}}}
\end{align}
to represent the degree of cross-correlation in a two-field model.  We call this quantity the \textit{cross-correlation ratio}.  To lowest order, $r_C$ gives the correlation angle,
\begin{align}
\label{r_C first-order}
r_C = \sin \Delta_N, 
\end{align}
while to second-order, 
\begin{align}
r_C = & \sin \Delta_N \times \\
& \left[1  + C \cos^2 \Delta_N \left(\tilde{M}_{\parallel \parallel} - \tilde{M}_{\perp \perp} - 2 \tilde{M}_{\parallel \perp} \cot \Delta_N\right)\right]. \nonumber
\end{align}

We have just found relations for the spectral indices, running of the spectral indices, the tensor-to-scalar ratio, and the cross-correlation ratio.  These spectral observables give us important information about the physics of inflation.  The amplitude of the tensor spectrum gives $H$, while combinations of the spectral observables $n_T$, $n_{\mathcal{R}}$, $n_{\mathcal{C}}$, $n_{\mathcal{S}}$, $r_T$, and $r_C$ give us expressions for our four quantities that describe the shape of the inflation potential: $\epsilon \approx \frac{1}{2} |\boldsymbol{\nabla} \ln V|^2$, $\tilde{M}_{\parallel \parallel}$, $\tilde{M}_{\parallel \perp}$, $\tilde{M}_{\perp \perp}$.   First, the tensor spectral index determines $\epsilon$ via equation (\ref{ns}).  Next, the lowest-order difference between the isocurvature and tensor spectral indices yields 
\begin{align}
\label{consist cond M perp perp}
\tilde{M}_{\perp \perp} = \frac{n_{\mathcal{S}} - n_T}{2},
\end{align}
while the difference between the isocurvature and the cross spectral indices yields
\begin{align}
\label{consist cond M para perp}
\tilde{M}_{\parallel \perp} = \frac{r_C}{\sqrt{1 - r_C^2}} \left(\frac{n_{\mathcal{C}} - n_{\mathcal{S}}}{2}\right).
\end{align}
The third effective mass matrix coefficient is given by
\begin{align}
\label{consist cond M para para}
\tilde{M}_{\parallel \parallel} = \frac{n_{\mathcal{R}} - n_T}{2} +  \frac{r_C^2}{1 - r_C^2} \left(\frac{n_{\mathcal{R}}+n_{\mathcal{S}}-2n_{\mathcal{C}}}{2}\right),
\end{align}
to lowest order.  Together, this means that if we can measure the above mentioned observables, we can put constraints on certain features of the inflationary Lagrangian and hence work backwards to reconstruct the physics of inflation.   Table 4 summarizes the key relationships among the kinematics, Lagrangian, and spectral observables. 

Finally, we can combine these six spectral observables into a consistency condition for general two-field inflation.  Bartolo \textit{et. al.} \cite{BartoloEtAl-2001a} and Wands \textit{et. al.} \cite{WandsEtAl-2002} were the first to find consistency relations for two-field inflation with canonical kinetic terms.   In terms of our parameters and allowing for non-canonical kinetic terms, we find the lowest order consistency condition to be
\begin{align}
\label{r_T consistency cond first-order}
r_T = & -8n_T  (1 - r_C^2),
\end{align} 
which agrees with their results for canonical kinetic terms.  To extend the above result to second-order in the SRST limit, we substitute equations (\ref{ns}), (\ref{r_C first-order}), (\ref{consist cond M perp perp}), (\ref{consist cond M para perp}), and (\ref{consist cond M para para}) into equation (\ref{r_T second-order}) for $r_T$ to obtain
\begin{align}
\label{r_T consistency cond}
r_T = & -8n_T  (1 - r_C^2) \\ \times & \left[1 - \frac{1}{2}n_{T} + n_{\mathcal{R}} + \frac{r_C^2}{1+r_C^2} (n_{\mathcal{R}} + n_{\mathcal{S}} - 2 n_{\mathcal{C}})\right]. \nonumber 
\end{align} 

\begin{widetext}
\begin{center}
\begin{table}[t]
\renewcommand{\arraystretch}{1.5}
\begin{tabular}{|x{2in}|p{4.7in}|}
\hline
\multicolumn{2}{|c|}{\normalsize \rule{0cm}{0.55cm} Table 5. Summary of Key Quantities and How They Affect the Power Spectra} \tabularnewline[1.4ex] 
\hline \multicolumn{2}{|c|}{$\, $}\tabularnewline[-3.7ex] \hline \normalsize \rule{0cm}{0.55cm} \textit{Quantity} & \normalsize \rule{4.5cm}{0cm} \textit{Importance} \tabularnewline[1.4ex] \hline
\multirow{2}{*}{\normalsize $\epsilon$} & \rule{0cm}{0.55cm}\normalsize $\bullet$ Equals $-(\ln H)'$. \\ & \normalsize $\bullet$ Is related to the field speed via $\epsilon = \frac{1}{2} v^2$. \tabularnewline[1.4ex] \hline
\rule{0cm}{0.55cm} \large $\frac{\eta_{\parallel}}{v}$ & \normalsize $\bullet$ Determines the intrinsic growth rate of adiabatic modes. \tabularnewline[1.4ex] \hline
\multirow{2}{*}{\rule{0cm}{0.55cm} \large $\frac{\eta_{\perp}}{v}$} & \rule{0cm}{0.55cm}\normalsize $\bullet$ Is the marker of multi-field behavior for both the unperturbed and perturbed fields. \\ & \normalsize $\bullet$ Determines the degree of sourcing of adiabatic/curvature modes by entropy/isocurvature modes. \tabularnewline[1.4ex] \hline
\multirow{2}{*}{\rule{0cm}{1.02cm} \Large $\frac{\frac{\eta_{\perp}}{v}}{ \frac{\eta_{\parallel}}{v}}$} &  \rule{0cm}{0.55cm}\normalsize $\bullet$ Controls the relative contribution of entropy mode sourcing to the growth of adiabatic modes. \\ & \normalsize  $\bullet$ Can be viewed as indicating the relative importance of multi-field effects at a given time.  \tabularnewline[1.4ex] \hline
\rule{0cm}{0.55cm} \normalsize $\mu_{\perp}$ &  \rule{0cm}{0.55cm}\normalsize $\bullet$ Determines the damping of entropy modes. \tabularnewline[1.4ex] \hline
\multirow{2}{*}{\rule{0cm}{0.55cm} \normalsize $T_{\mathcal{SS}}$} & \rule{0cm}{0.55cm}\normalsize $\bullet$ Derivative quantity that depends on an integral of $-\mu_{\perp} - \frac{\eta_{\parallel}}{v}$. \\ & \normalsize $\bullet$ Represents the total damping of isocurvature modes after horizon exit. \tabularnewline[1.4ex] \hline
\multirow{2}{*}{\rule{0cm}{0.55cm} \normalsize $T_{\mathcal{RS}}$} &  \rule{0cm}{0.55cm}\normalsize  $\bullet$ Derivative quantity that depends on an integral of $\frac{\eta_{\perp}}{v}$ and $T_{\mathcal{SS}}$. \\ & \normalsize $\bullet$ Represents the total sourcing of curvature modes by isocurvature modes after horizon edit.  \tabularnewline[1.4ex] \hline
\rule{0cm}{0.55cm} \large $\frac{H_*^2}{\epsilon_*}$ & \normalsize $\bullet$ Sets the overall scale of the three scalar spectra. \tabularnewline[1.4ex] \hline
\rule{0cm}{0.55cm} \large $\frac{T_{\mathcal{SS}}}{T_{\mathcal{RS}}}$ & \normalsize $\bullet$ Determines the relative sizes of the three scalar spectra. \tabularnewline[1.4ex] \hline
\multirow{3}{*}{\rule{0cm}{0.55cm} \normalsize $\sin \Delta_N$} &  \rule{0cm}{0.55cm}\normalsize $\bullet$ Equals $\frac{T_{\mathcal{RS}}}{\sqrt{1 + T_{\mathcal{RS}}^2}}$. \\ & \normalsize $\bullet$ Gives an indication of the amount of cross-correlation. \\ & \normalsize $\bullet$ Determines which linear combinations of the coefficients of the effective mass matrix appear in the power spectra and the spectral observables. \tabularnewline[1.4ex] 
\hline
\end{tabular}
\end{table}
\end{center}
\end{widetext}

\noindent The above consistency condition agrees with the second-order result obtained by \cite{ByrnesAndWands-2006} for the case of canonical kinetic terms.  This consistency relation is a potentially powerful tool for testing the validity of two-field models of inflation.

\section{Applications}
\label{Applications}

In this section, we illustrate how to use our theoretical framework to understand and calculate the power spectra, and we show how to test two-field models of inflation against observational data.  

We demonstrate this by exploring four different classes of inflationary models.  Together, these four classes of models cover a wide range of kinematical behaviors and include an example with a non-canonical field metric.   For each class of models, we vary both the initial conditions and a characteristic parameter of the inflationary Lagrangian to understand the range of power spectra that can be generated.  To disentangle the separate contributions to the power spectra, we present a series of plots showing the background trajectories, our kinematical quantities, and the transfer functions.  Using these plots, we discuss how the transfer functions and the power spectra can be inferred from the background kinematics and the field metric.  Thereafter, we allow both the initial conditions and the characteristic Lagrangian parameter to vary, and we plot the results for a set of six spectral observables.  Finally, we use these results to determine which inflationary scenarios---that is, which combinations of the initial conditions and the inflationary Lagrangian---are consistent with observational constraints.  To our knowledge, this constitutes the first thorough and rigorous approach to understanding and constraining two-field models of inflation.

\subsection{Methods}

In this section, we provide an overview of our methods and general approach.

We start the investigation of each class of models by first varying the initial conditions, while holding the Lagrangian parameter constant, and we determine the effects on the power spectra.   Because the initial conditions can affect the inflationary dynamics and hence the power spectra, varying the initial conditions is essential in order to test and constrain two-field inflationary models against observations.  With an uncountable number of initial conditions, this is ostensibly a formidable task.  Indeed, though there have been a few attempts to consider initial conditions in single-field inflation, particularly by \cite{AlbrechtAndSorbo-2004,Tegmark-2004}, there have not been many attempts to do so in two-field inflation, though one notable exception in which the initial conditions receive more consideration is \cite{TsujikawaEtAl-2002}.  Fortunately, most inflationary models have attractor solutions and/or sampling a collection of initial conditions is sufficient to extrapolate the results to the set of all possible initial conditions.  In our analysis, we parameterize the initial conditions via a single parameter, $\theta$, which is the polar angle for the field vector 60 $e$-folds before the end of inflation.  When varying only the initial conditions, the six initial conditions we test for each model are $\theta = 0, \frac{\pi}{10}, \frac{2\pi}{10}, \frac{3\pi}{10}, \frac{4\pi}{10},$ and $\frac{\pi}{2}$.    

Next, we vary a characteristic parameter of the inflationary Lagrangian, while holding the initial conditions constant, and we find the resultant spectra.  We choose six representative values of that Lagrangian parameter to test.   We do this in order to understand how either the strength of one of the interaction parameters in the potential or the size of a parameter in the field metric impacts the spectra.   Varying a Lagrangian parameter also has the benefit of extending our analysis to an entire class of similar models, rather than just testing a single Lagrangian from that class of models.  

To conduct the above analyses, we perform all our calculations to second-order in the complete SRST approximation using our semi-analytic formulae.  However, we have checked these results all against exact numerical solutions and find good agreement for the scenarios tested.   The one exception is that the second-order calculations do not reflect any oscillations of the $\phi_2$ field about the $\phi_1$ axis, which occur in a very small subset of scenarios.  The effect of these oscillations is to introduce some tiny wiggles into the power spectra, but the full treatment of these oscillations is beyond the scope of this paper.  Also beyond the scope of this paper is dealing with the small subset of scenarios in which inflation will end before one of the fields has a chance to roll down the potential---that is, models that consist of two separate inflationary phases separated by a non-inflationary phase. For these scenarios, we simply set the initial conditions relative to the end of the first phase of inflation.  We refer the interested reader to \cite{PolarskiAndStarobinsky-1992} to see how these scenarios can be handled.

To complement the above mentioned analyses of the power spectra, we also plot the the background trajectories, kinematics, and transfer functions.  We do this in order to disentangle the multiple factors that affect the power spectra. We now describe this series of plots here to avoid having to use plot captions and to repeat the same information several times. For our plots of the background trajectories and the kinematical parameters ($\epsilon$, $\frac{\eta_{\parallel}}{v}$, $\frac{\eta_{\perp}}{v}$), we show the results for six different sample scenarios for the time period 70 $e$-folds before the end of inflation until the end of inflation.   The gray contour lines indicate the number of $e$-folds left before inflation ends. To depict the two transfer functions---$T_{\mathcal{SS}}$ and $T_{\mathcal{RS}}$---we plot the functions from horizon exit until the end of inflation, where the $x$-axis represents the number of $e$-folds before inflation ends.  We do this for seven sample mode wavelengths: for the modes that exit the horizon 70, 60, 50, 40, 30, 20, and 10 $e$-folds before the end of inflation (colored lines).  On the same plots, we overlay the value of the transfer function at the end of inflation (solid black lines).  These lines indicate the value of the transfer function at the end of inflation for the modes that exited the horizon at the time shown on the $x$-axis.   On the plots of $T_{\mathcal{SS}}$, we also overlay the value of the effective entropy mass (dashed black lines) during inflation.  For our spectral plots, we plot $\mathcal{P}_{\mathcal{R}}$ at horizon exit and $\mathcal{P}_{\mathcal{R}}$, $\mathcal{C}_{\mathcal{RS}}$, and $\mathcal{P}_{\mathcal{S}}$ at the end of inflation.  We show the spectra for the modes that exited the horizon between 70 $e$-folds before the end of inflation and the end of inflation; that is, the $x$-axis effectively represents the mode wavelength.   To calculate the spectra, we normalize the amplitude of the curvature power spectrum at the end of inflation to the 7-year WMAP + BAO + $H_0$ result $\Delta_{\mathcal{R}}^2(k_0) \approx 2.44 \times 10^{-9}$, where $k_0 = 0.002$ Mpc$^{-1}$ \cite{KomatsuEtAl-2010}, which we take to be 60 $e$-folds before the end of inflation.  This normalization in turn fixes the overall energy scale of the potential separately for each combination of the initial conditions and Lagrangian parameter, and hence also fixes the amplitudes of the isocurvature and cross spectra.  

After having separately varied the initial conditions and the Lagrangian parameter, we end our exploration of each class of models by allowing both parameters to vary.  The mesh size we use when allowing both the initial conditions and Lagrangian parameter to vary is 101 by 101, so we test a total of $10,201$ inflationary scenarios for each type of model.    For each of these 10,201 scenarios, we find the six spectral observables $r_T$, $r_C$, the isocurvature fraction, $n_T$, $n_{\mathcal{R}}$, and $\alpha_{\mathcal{R}}$,  where we define the isocurvature fraction, $f_{iso}$, as
\begin{align}
\label{isocurvature fraction}
f_{iso} \equiv \frac{\mathcal{P}_{\mathcal{S}}}{\mathcal{P}_{\mathcal{R}}}.
\end{align}
We pick these particular observables because they are among the ones that are most commonly considered and/or that are most likely to be well-constrained in the near future.  In calculating the spectral observables, we find them numerically from the power spectra, rather than using our semi-analytic formulae. The numerical errors are usually negligible, except in some cases for the running of the curvature spectral index, $\mathcal{\alpha}_{\mathcal{R}}$.   Lastly, we plot each of these scenarios in the $r_T - n_s$ plane, comparing them against the 95\% confidence limits derived from observational data.  To complement this plot, we provide a second plot showing exactly which of these scenarios are ruled by these observational constraints.  The observational constraints we use are the WMAP + BAO + $H_0$ 95\% confidence limits in \cite{KomatsuEtAl-2010} on the combination of $n_s = 1 + n_{\mathcal{R}}$ and $r_T$, which we approximate by the ellipse
\begin{align}
\label{ellipse constraint}
0.214 n_s^2 & +  0.00357 r_T^2 - 0.0450 n_s r_T \nonumber \\ & - 0.411 n_s + 0.0437r_T + 0.198 \le 0.
\end{align}

Finally, we note that for our aforementioned analyses, we have computed the spectra and spectral observables at the end of inflation and have therefore ignored any model-dependent processing of the modes that may occur after inflation ends.  What exactly happens to the spectra thereafter is unknown: there are many uncertainties associated with the end of inflation and with the reheating process during its aftermath.  Since the focus of the present paper is rather orthogonal to these issues---on the impact of having more than one inflaton field {\it during} inflation---our calculations do not take these issues into consideration.  Of course, some sort of assumptions are required in order to test an inflationary model against observations, so we have assumed that
\begin{enumerate}
\item Inflation ends abruptly when the scale factor stops accelerating, which is equivalent to the condition $\epsilon = 1$,\footnote{Specifically, when applying this criterion, we compute $\epsilon$ using the 
exact background equations.},
\item The curvature spectra are conserved across the inflationary boundary,\footnote{For a given two-field model, inflation does not end at a unique value of $V$, or equivalently, $H$. If inflation does not end at a unique energy density, the power spectra are not necessarily conserved across the inflationary boundary (e.g., \cite{DeruelleAndMukhanov-1995}).  However, in many cases of two-field inflation (including many of our examples below), attractor solutions exist and/or the evolution of the field vector at the end of inflation is essentially single-field, which diminishes any corrections to the power spectra. So we make the usual assumption that any corrections to the power spectra arising from the transition from inflation to the radiation-dominated era are negligibly small \cite{NakamuraAndStewart-1996}.}  and  
\item The isocurvature modes are not completely destroyed by the reheating process, and hence our calculations of the isocurvature spectra and fractions constitute upper limits.  
\end{enumerate}
Any additional assumptions about the nature of the end of inflation and reheating can simply be appended onto our calculations here, as desired.

\subsection{Multiplicative Double Polynomial Potential With Canonical Kinetic Terms}
\label{mult potl}

The first class of inflationary models we investigate is the multiplicative double polynomial potential with canonical kinetic terms.  We definite multiplicative double polynomial potential to mean all potentials of the form
\begin{eqnarray}
\label{Mult Potl}
V = M^4 |\phi_1|^n |\phi_2|^p.
\end{eqnarray}
This inflationary potential falls under the general category of multiplicative potential models, $V = \prod_i V_i(\phi_i)$, where $V_i(\phi_i)$ means that $V_i$ is a function only of the $i$th field.  Multiplicative models with canonical kinetic terms have the feature that their effective mass matrices are diagonal and hence the fields and field perturbations evolve essentially independently.   Here, we take the adjustable Lagrangian parameter to be the power, $p$, to which the $\phi_2$ field is raised.  We vary $p$, which we take to be a continuous variable greater than or equal to 1, while holding $n$ fixed at $n=2$.  Later, using the results in this section, it will be become evident that we can extrapolate from the $n=2$ case to all values of $n$ and hence consider the entire class of multiplicative double polynomial potentials.

\subsubsection{Varying the Initial Conditions}

For the first part of our analysis, we vary the initial conditions while holding the power $p$ fixed at $p=4$ and investigate the background trajectories and kinematics, the transfer functions, and the power spectra.   For the $\theta = 0^o$ and $\theta=90^o$ initial conditions, the potential is nonsensical if we set the corresponding field to zero, so instead we simply ignore the zero-ed field, and use the corresponding single-field potential for the non-zero field.
  
We find that varying the initial conditions results in trajectories with very little curvature that are very widely 

\begin{figure*}[t]
\begin{minipage}[t]{1.0\linewidth}
\centering
\Large
Figure 3.  Multiplicative Double Polynomial Potentials, $V(\phi_1,\phi_2) = M^4 \phi_1^2 |\phi_2|^p$, With Canonical Kinetic Terms
\large
\vskip 12 pt
Varying the Initial Condition, $\theta$, With the Lagrangian Parameter Fixed At $p = 4$
\normalsize
\vskip 12 pt
\includegraphics[height=213.5mm]{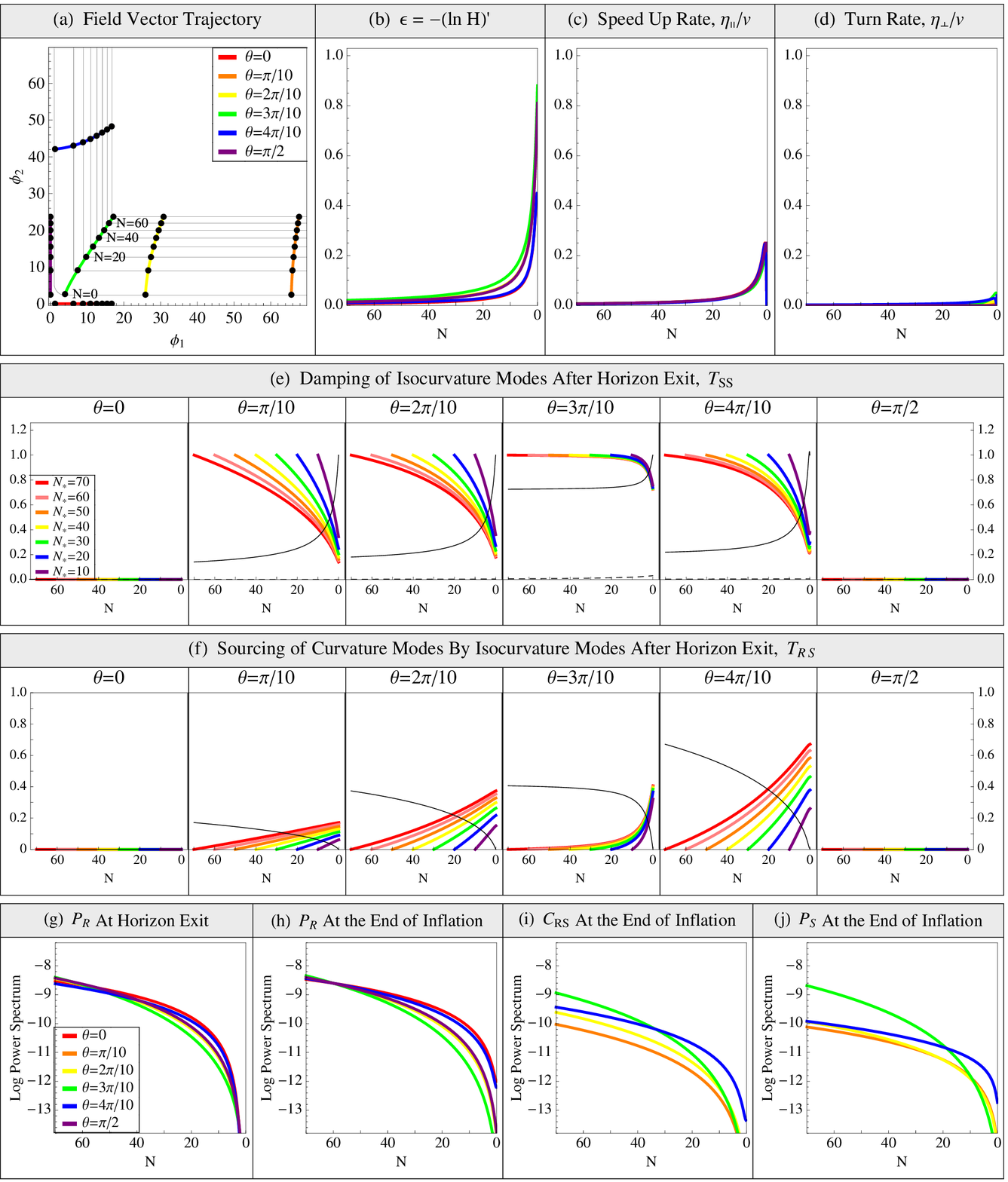}
\label{Mult VaryCoupling}
\end{minipage}
\end{figure*}

\clearpage

\noindent separated in field space, as shown in Figure 3(a).  With the exception of the field trajectory corresponding to $\theta \approx 55^o$, the other five trajectories depicted show that one field evolves significantly during the course of inflation, ending around the reduced Planck scale, while the other field only slightly evolves during the course of inflation.  Therefore, for most initial conditions, one of the two fields will dominate the background dynamics and primarily determine the kinematical parameters.  

Plots of the three kinematical parameters show typical slow-roll behavior up until the end of inflation.  As shown in Figure 3(b), $\epsilon$ is largest for those trajectories where both fields dominate, and it is also a little larger for trajectories corresponding to $\phi_2$ domination, reflecting the higher power to which the $\phi_2$ field is raised.  There is little variation in the speed up rate (Figure 3(c)), and the turn rate (Figure 3(d)) is very small for these models, typically about an order of magnitude less than the speed up rate.  Therefore, the scenarios produced by these models satisfy $\frac{\eta_{\parallel}}{v} \gg \frac{\eta_{\perp}}{v}$, and we expect the multi-field effects to be small.  These kinematics can also be gleaned by analyzing the lowest-order effective mass matrix.  This analysis (not shown) reveals that $\tilde{M}_{\parallel \parallel}^{(1)}$, which approximates the speed up rate, will always be the largest magnitude matrix coefficient and that the other matrix coefficients will virtually always be within an order of magnitude of each other, except near the very end of inflation or in the small subset of models in which the fields co-dominate.  

Now we consider the transfer functions.  Since $\tilde{M}_{\perp \perp} < 0$ for these models, they provide an example of scenarios in which the entropy modes actually grow in the super-horizon SRST limit.  However, since the speed up rate is positive and larger in magnitude than the effective entropy mass, the isocurvature modes will nevertheless be increasingly damped by the transfer function $T_{\mathcal{SS}}$ after they exit the horizon (Figure 3(e)).   The isocurvature modes are damped fairly steadily in the super-horizon limit, but the damping increases towards the end of inflation when both the effective entropy mass and speed up rate increase.  Also, the isocurvature modes are more damped in those scenarios in which the $\phi_2$ field dominates, and less damped in scenarios where the two fields co-dominate due to the smaller difference between the effective entropy mass and speed up rate in these scenarios.   Note that by the end of inflation, the isocurvature modes are still appreciable in all cases, as expected when the turn rate is so small and the effective entropy mass is negative.  (Except, of course, the isocurvature modes do not exist when $\theta = 0^o$ or $\theta = 90^o$.)

The second transfer function, $T_{\mathcal{RS}}$, is shown in Figure 3(f).  For the trajectories where one field dominates, $T_{\mathcal{RS}}$ increases steadily until the end of inflation, and it is larger as the initial angle increases.  For the trajectory where the fields co-dominate ($\theta \approx 55^0$), $T_{\mathcal{RS}}$ starts off significantly smaller, but increases much more rapidly at the very end of inflation. Interestingly, the more the sub-dominant field contributes to the inflationary dynamics, the more scale-invariant $T_{\mathcal{RS}}$ is.  In all cases, however, $T_{\mathcal{RS}}$ remains less than about 1, reflecting the fact that the turn rate is so small for these models and hence the total sourcing of curvature modes by isocurvature modes is modest.

Finally, we plot the set of power spectra for these scenarios.  The curvature power spectra at horizon exit (Figure 3(g)) and at the end of inflation (Figure 3(h)) show some variations due to the initial conditions, being more steeply sloped when the $\phi_2$ field is a dominant contributor to the background dynamics.  But they are close to  scale-invariant and lack any substantial scale-dependent features. The correlated cross spectrum (Figure 3(i)) and isocurvature spectrum (Figure 3(j)) are both smaller by about 1-2 orders of magnitude and are also close to scale-invariant, which is as expected given that the isocurvature modes are still appreciable at the end of inflation.  The two exceptions are the obvious absence of isocurvature and cross spectra for the single-field scenarios corresponding to $\theta = 0$ and $\theta = 90^o$, and that the isocurvature and cross spectra are virtually identical to the curvature spectra for the trajectory corresponding to $\theta \approx 55^o$, reflecting the fact that the two fields are co-dominant in this scenario.   These results match what we would predict in light of our analysis of the background kinematics and the transfer functions.

\subsubsection{Varying the Lagrangian Parameter}

Now we vary the power, $p$, to which $\phi_2$ is raised, while holding the initial condition fixed at $\theta = 45^o$. 

Varying the Lagrangian parameter $p$ also results in field trajectories with very little curvature (Figure 4(a)).  For the particular initial condition $\theta = 45^o$, if $p < 2$, the $\phi_1$ field dominates the inflationary dynamics, whereas if $p > 2$, then the $\phi_2$ field dominates; around $p \approx 2$, both fields contribute about equally. However, for this particular initial condition, the degree of domination of one field over the other is not as dramatic as some of the scenarios we investigated in the previous section.  As before, these scenarios show typical single-field slow-roll behavior up until the end of inflation, with $\epsilon$ (Figure 4(b)) increasing as $p$ increases but with little variation in the speed up rate (Figure 4(c)).  As before, the turn rate is very small for these models (Figure 4(d)) and smaller than the speed up rate.  The turn rate is smallest when both fields co-dominate, but is greatest when one field dominates but then the second field starts to pick up speed towards the end of inflation.  As in the previous section, an examination of the first-order effective mass matrix agrees with this analysis.
 
Now we investigate the two transfer functions.  The isocurvature modes are increasingly damped by the transfer function $T_{\mathcal{SS}}$ after they exit the horizon (Figure 4(e)), with slightly more damping as $p$ increases, with the exception of no damping in the scenario where $p=2$, where the fields exactly co-dominate.  For the transfer function $T_{\mathcal{RS}}$, its size mostly reflects the size of the corresponding turn rate, since $T_{\mathcal{SS}}$ depends more weakly on

\begin{figure*}[t]
\begin{minipage}[t]{1.0\linewidth}
\centering
\Large
Figure 4. Multiplicative Double Polynomial Potentials, $V(\phi_1,\phi_2) = M^4 \phi_1^2 |\phi_2|^p$, With Canonical Kinetic Terms
\large
\vskip 12 pt
Varying the Lagrangian Parameter, $p$, With the Initial Condition Fixed At $\theta = 45^o$
\vskip 12 pt
\normalsize
\includegraphics[height=213.5mm]{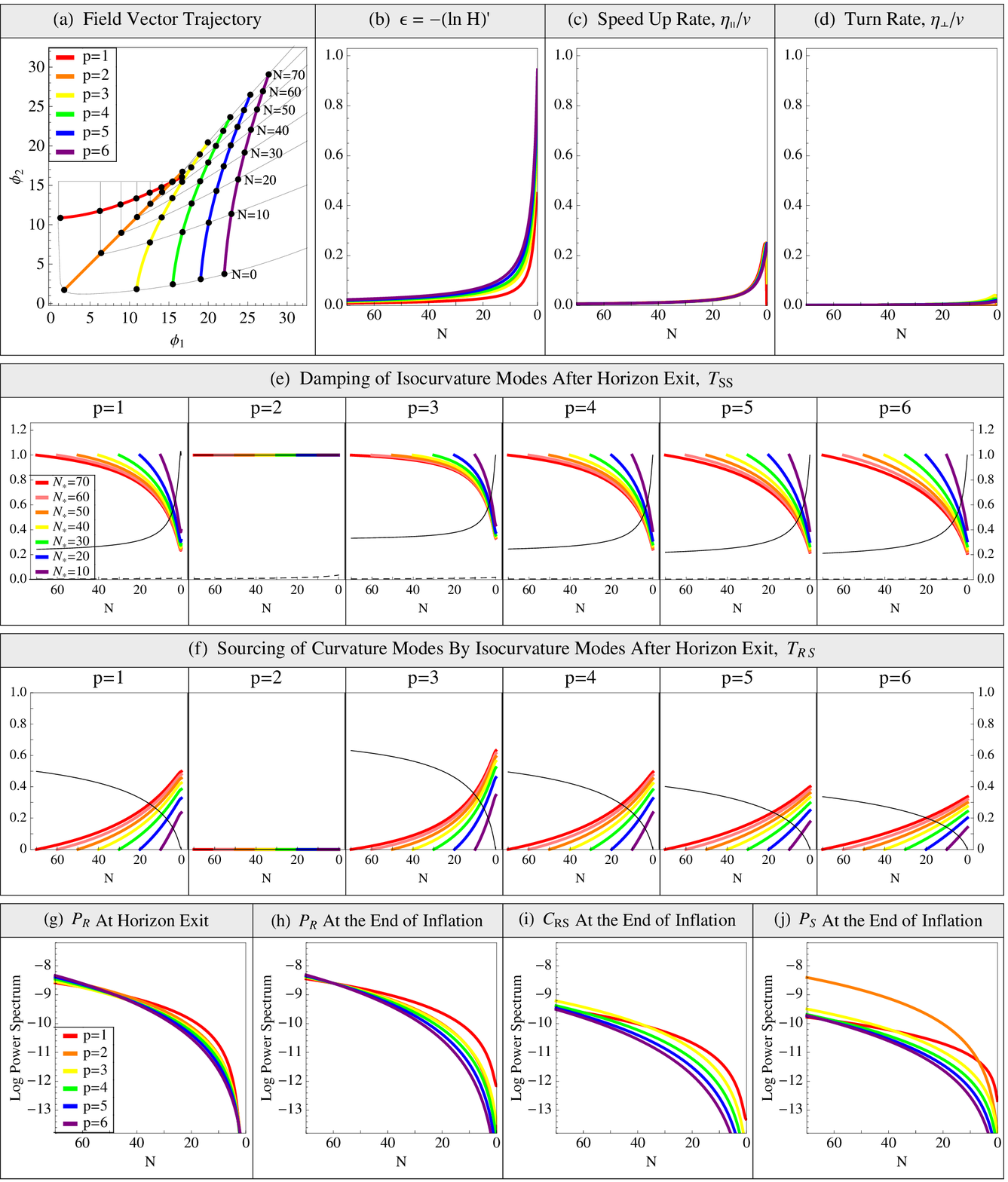}\label{Mult VaryInitConds}
\end{minipage}
\end{figure*}

\begin{figure*}[t]
\begin{minipage}[t]{1.0\linewidth}
\centering
\Large
Figure 5.  Multiplicative Double Polynomial Potentials, $V(\phi_1,\phi_2) = M^4 \phi_1^2 |\phi_2|^p$, With Canonical Kinetic Terms
\large
\vskip 12 pt
Varying Both the Initial Condition, $\theta$, and the Lagrangian Parameter, $p$
\normalsize
\vskip 12 pt
\includegraphics[height=210mm]{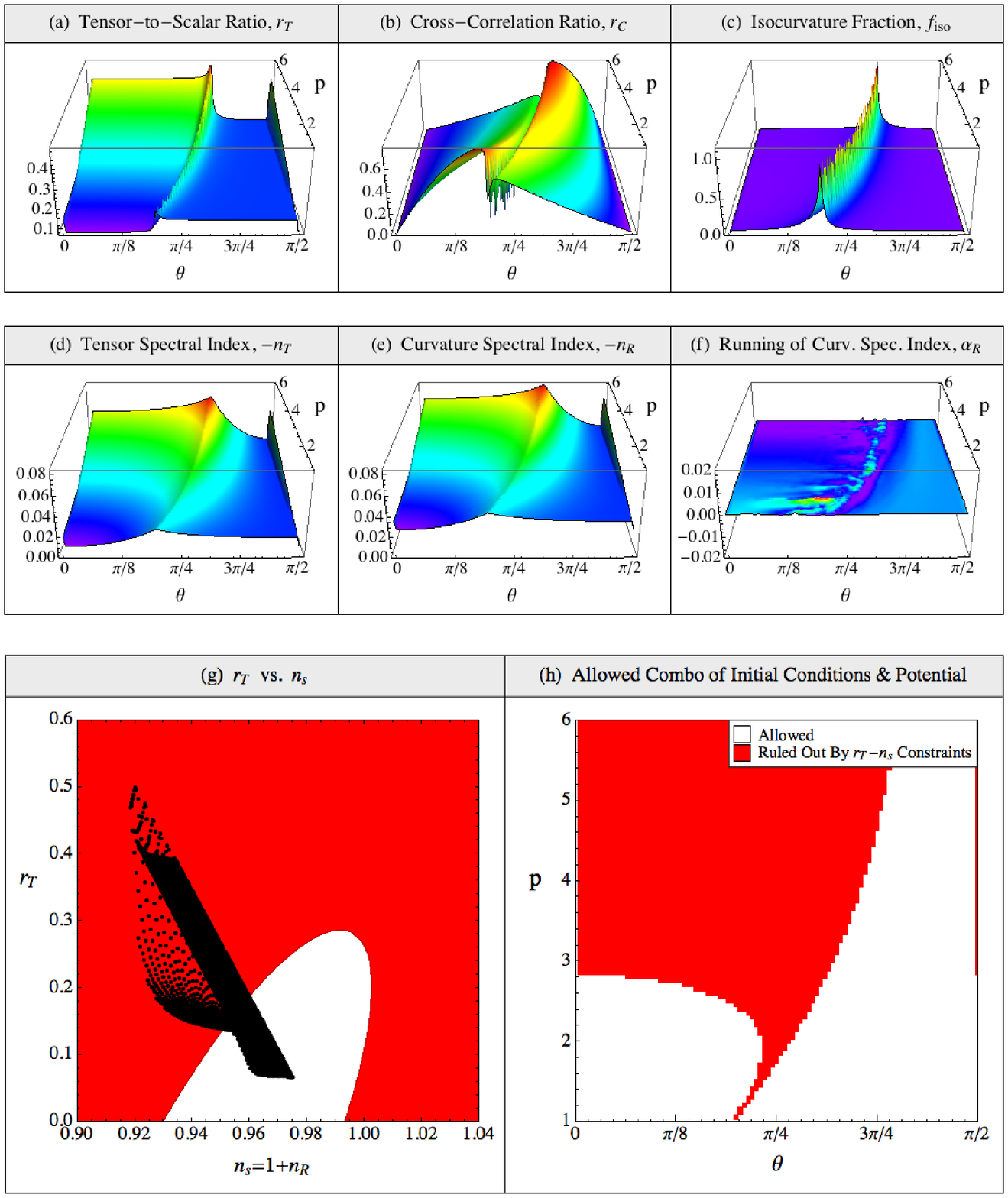}
\label{Mult VaryBoth}
\end{minipage}
\end{figure*}

\clearpage

\noindent $p$ than the turn rate does (Figure 4(f)).  So $T_{\mathcal{RS}}$ is smallest for those scenarios in which the fields co-dominate and is largest for those scenarios in which the sub-dominant field starts to pick up speed towards the end of inflation.  The rest of our observations from the previous section on how the behavior of $T_{\mathcal{SS}}$ and $T_{\mathcal{RS}}$ reflects the dominance of the fields and the field kinematics also applies here. 

The set of power spectra for these scenarios shows several similarities to those obtained when we varied the initial conditions.  The curvature spectra (Figures 4(g)-(h)) all are nearly scale-invariant with similar amplitudes, but they are more steeply sloped as the power $p$ increases, as expected.  As before, the cross spectrum (Figure 4(i)) and isocurvature spectrum (Figure 4(j))  are both smaller by about 1-2 orders of magnitude than the curvature spectra, reflecting the fact that the isocurvature modes are still significant at the end of inflation.  The one exception is for the trajectory corresponding to $p = 2$, which has a zero turn rate and zero damping of isocurvature modes, so the isocurvature spectrum is identical to the curvature spectrum and the cross spectrum vanishes.

\subsubsection{Spectral Observables}

Now we let the initial conditions and the power to which the $\phi_2$ field is raised both vary at the same time.  We determine how these variations affect our set of power spectrum observables, and we determine which scenarios are consistent with observational constraints.  

First, we consider the three quantities that depend on the relative amplitudes of the power spectra: $r_T$, $r_C$, and $f_{iso}$.  Figure 3(a) shows that except for the very small band of trajectories in which both fields about equally contribute to the inflationary dynamics, $r_T$ does not seem to depend on the initial conditions, but only on the power to which the dominant field is raised.  As $p$ increases, so does $r_T$.   By comparison, Figure 3(b) shows that the cross-correlation ratio---which also serves as a measure of how much the curvature modes are sourced by the isocurvature modes---exhibits more variation, increasing the more the sub-dominant field contributes to the inflationary dynamics.  The two ridges coincide with the trajectories with the largest turn rates, which correspond to scenarios in which the sub-dominant field starts to pick up more speed as the dominant field ends inflation.  To the left of the first ridge, $\phi_2$ dominates the inflationary dynamics, while to the right of the second ridge, $\phi_1$ dominates. The deep and narrow gorge-like feature between the two ridges corresponds to scenarios in which the fields are essentially equally co-dominant such that the turn rate is very small and hence the cross-correlation is tiny.  Only for this small subset of scenarios where the fields co-dominate is the isocurvature fraction large, as shown in Figure 3(c); otherwise, the isocurvature fraction is small.

Next, we consider the derivative quantities associated with the power spectra: the spectral indices $n_T$ and $n_{\mathcal{R}}$, and the running of the curvature spectral index, $\alpha_{\mathcal{R}}$.  The plots for both $-n_T \approx 2\epsilon$ (Figure 5(d)) and $-n_{\mathcal{R}}$ (Figure 5(e)) are very similar in profile, since the speed up rate and logarithmic running of $T_{\mathcal{RS}}$ produce very small deviations from scale invariance, which serves to slightly increase the value of $-n_{\mathcal{R}}$ relative to $-n_{T}$.   The values of these two spectral indices are greatest for those scenarios in which $\phi_2$ dominates or co-dominates, and they increase as $p$ increases.  Also, we note that the larger $p$ is, the larger $\theta$ must be for the fields to be co-dominant.   As $n_{\mathcal{R}}$ itself does not have much scale dependence, the running of the curvature spectral index is very small and negative for these models, as shown in Figure 5(f).  

Finally, we examine which inflationary scenarios produced by this type of model can be ruled out by the 95\% confidence limits on the combination of $n_s = 1 + n_{\mathcal{R}}$ and $r_T$.  According to these limits, the single-field potential $V \propto \phi^4$ is ruled out, but the single-field potential $V \propto \phi^2$ is still a viable candidate for describing inflation \cite{KomatsuEtAl-2010}.  Therefore, whenever the $\phi_1$ field, which is raised to the second power, dominates, the inflationary scenario is consistent with observational data.  This corresponds to higher values for $n_s$ and lower values for $r_T$, as shown in Figure 5(g), and this corresponds to the rightmost white region in Figure 5(h).  However, when the $\phi_2$ field dominates or co-dominates, whether the scenario is viable depends on the power to which the $\phi_2$ field is raised.  Scenarios in which $p \gtrsim 3.2$ and $\phi_2$ dominates or co-dominates are ruled out by observational constraints.  

Taken together, these results suggest that multiplicative double polynomial potentials with canonical kinetic terms produce nearly scale-invariant power spectra that are determined primarily by the field that dominates the 
inflationary dynamics, or by both fields if they are co-dominant, with corrections from the sub-dominant field.   As a result, the power to which the dominant field (or co-dominant fields) is raised determines whether the scenario is excluded by current observational data.  Also in these scenarios, the isocurvature and cross spectra is typically smaller, but possibly detectable, if unaffected by reheating and other post-inflationary processes.  Therefore, we can view the role of initial conditions in these models, in conjunction with the powers the two fields are raised to, as being to determine the dominant field (or whether the fields are co-dominant) and to set the size of the corrections from the sub-dominant field.

\subsection{Double Quadratic Potential With Canonical Kinetic Terms}

The second class of inflationary models we investigate is the double quadratic potential with canonical kinetic terms.  We define double quadratic potential to mean all potentials of the form
\begin{align}
\label{Quad Potl}
V = \frac{1}{2} m_1^2 \phi_1^2 + \frac{1}{2} m_2^2 \phi_2^2.
\end{align}
This inflationary model falls under the general category of additive potential models, $V = \sum_i V_i(\phi_i)$, where $V_i$ is a function only of the $i$th field.  Additive models with canonical kinetic terms have the feature that the given fields are non-interacting, so they interact only gravitationally.  Here, the adjustable Lagrangian parameter is the ratio of the masses, $\frac{m_2}{m_1}$, and we take $\frac{m_2}{m_1} \ge 1$.

\subsubsection{Varying the Initial Conditions}
\label{vary initial conds for double quadratic}

First, we vary the initial conditions while holding the mass ratio $\frac{m_2}{m_1}$ fixed.  We set $\frac{m_2}{m_1} = 5$ since this mass ratio produces trajectories with moderately fast turning behavior.

The trajectories produced by this model (Figure 6(a)) and associated kinematical quantities (Figures 6(b)-(d)) depend significantly on the initial conditions and roughly separate into three subsets of behavior.  The first subset of scenarios starts sufficiently close to the $\phi_1$ axis such that the $\phi_1$ field dominates the energy density during the last 60 $e$-folds of inflation.  For these trajectories, the field vector essentially rolls down the $\phi_1$ axis, producing typical slow-roll inflation with at most a small turn rate.  

The second and largest subset of trajectories consists of those that experience $\phi_2$ domination 60 $e$-folds before inflation ends, but have an initial angle $\theta \lesssim 80^o$.  For these trajectories, the more massive $\phi_2$ field initially dominates the energy density, driving the field vector towards the $\phi_1$ axis.  Eventually, once $\phi_2$ becomes sufficiently small, $\phi_1$ takes over as the dominant field, causing the trajectory to turn significantly in field space, before rolling down the $\phi_1$ axis and ending inflation.\footnote{We note that when using the exact equations of motion, the trajectories with $\phi \approx 70^o$ will oscillate about the $\phi_1$ axis around the time that the field vector turns sharply in field space.  As mentioned earlier, we will not be addressing these oscillations and any subsequent particle decays that arise, but rather simply point out that these oscillations arise as a consequence of the classical background dynamics for certain combinations of the initial conditions and sufficiently large values of the mass ratio.}  Since $\phi_2$ dominates the dynamics before the turn, while $\phi_1$ dominates the dynamics after the turn, these inflationary scenarios can be considered to have two distinct phases of inflation.  Exploring the kinematics of these scenarios in Figures 6(b)-(d), we find that as the field vector approaches the $\phi_1$ axis, $\epsilon$ increases substantially reflecting the fact that $\phi_2$ picks up speed.  But then as $\phi_2$ domination gives way to $\phi_1$ domination, $\epsilon$ drops significantly, essentially resetting itself for a phase of $\phi_1$ domination.  During the transition between the phases, the turn rate (Figure 6(d)) dramatically and transiently increases, and the trajectory turns significantly in field space; however, this peak in the turn rate occurs slightly after the large negative drop in the speed up rate.  In general, we find that the larger the initial angle $\theta$ is, the larger the speed up and turn rates are when they peak in magnitude, the narrower the peaks, and the closer these peaks occur to the end of inflation.  As this subset of scenarios represents the case in which the speed up rates and turn rates are comparable around similar times during inflation, we therefore expect the multi-field effects to be substantial.

The third subset of scenarios are also characterized by initial $\phi_2$ domination 60 $e$-folds before inflation ends but have initial angles $\theta \gtrsim 80^o$.  The difference is that for these scenarios, the $\phi_2$ field picks up enough speed for inflation to end while the $\phi_2$ field still dominates the energy density.  Eventually, a second phase of inflation driven by the $\phi_1$ field will occur sometime later.  As mentioned earlier, these more complicated scenarios are beyond the scope of this work, so for simplicity, we calculate the inflationary dynamics by treating these scenarios as if they consist of only one phase of inflation, with typical slow-roll behavior and at most a very small turn rate.  We refer the interested reader to \cite{PolarskiAndStarobinsky-1992} to see in detail how one might handle these more complicated scenarios.

Now we consider the two transfer functions $T_{\mathcal{SS}}$ (Figure 6(e)) and $T_{\mathcal{RS}}$ (Figure 6(f)).  For all three subsets of scenarios, whenever the fields are in the super-horizon SRST limit, $T_{\mathcal{SS}}$ decays gradually.  However, for the second subset of scenarios, during the transition between inflationary phases, $T_{\mathcal{SS}}$ varies dramatically. Initially, the drop in $\epsilon$ (or equivalently, the large negative speed up rate) strongly enhances the isocurvature modes, and then the subsequent large turn rate strongly suppresses the modes.   After the turn, $\tilde{M}_{\perp \perp}$ remains high and the speed up rate gradually increases again, so that $T_{SS} \lll 1$ for all modes that exit the horizon both before and after the turn.   This is exactly the behavior for $T_{\mathcal{SS}}$ that we expect in inflationary scenarios with large turn rates.  The second transfer function, $T_{\mathcal{RS}}$, satisfies $T_{\mathcal{RS}} \ll 1$ for the first and third subsets of scenarios, because of their small turn rates.  But $T_{\mathcal{RS}}$ becomes larger than one by the end of inflation and exhibits significant scale-dependent features for the second subset of scenarios.  For these scenarios, $T_{\mathcal{RS}}$ starts off small, but when the turn rate rises dramatically, so too does $T_{\mathcal{RS}}$.  The steepness of the increase in the turn rate is indeed reflected in the steepness of the rise in $T_{\mathcal{RS}}$, and after the sharp turn, $T_{\mathcal{RS}}$ levels off.  Also, the value of $T_{\mathcal{RS}}$ at the end of inflation is largest for those modes that exit the horizon around the time the turn rate is large.  For modes that exit the horizon after the sharp turn, the value of $T_{\mathcal{RS}}$ at the end of inflation is negligibly small.

Lastly, we examine the resultant power spectra in Figures 6(g)-(i).  For the second subset of scenarios, the curvature power spectrum is strongly scale-dependent at horizon exit, largely reflecting the transient rise and fall of $\epsilon$ during the transition between inflationary phases.  Interestingly, despite the strong scale dependence of the curvature spectrum at horizon exit, this scale-dependence is largely blunted by the sourcing of curvature modes by isocurvature modes, so the curvature spectrum has a much weaker scale dependence at the end of inflation.  The net result is that the two phases of inflation are distinctly evident in the final power spec- 

\begin{figure*}[t]
\begin{minipage}[t]{1.0\linewidth}
\centering
\Large
Figure 6. Double Quadratic Potentials, $V(\phi_1,\phi_2) = \frac{1}{2} m_1^2 \phi_1^2 + \frac{1}{2} m_2^2 \phi_2^2$, \\ With Canonical Kinetic Terms
\large
\vskip 12 pt
Varying the Initial Condition, $\theta$, With the Lagrangian Parameter Fixed At $\frac{m_2}{m_1} = 5$
\normalsize
\vskip 12 pt
\includegraphics[height=213.5mm]{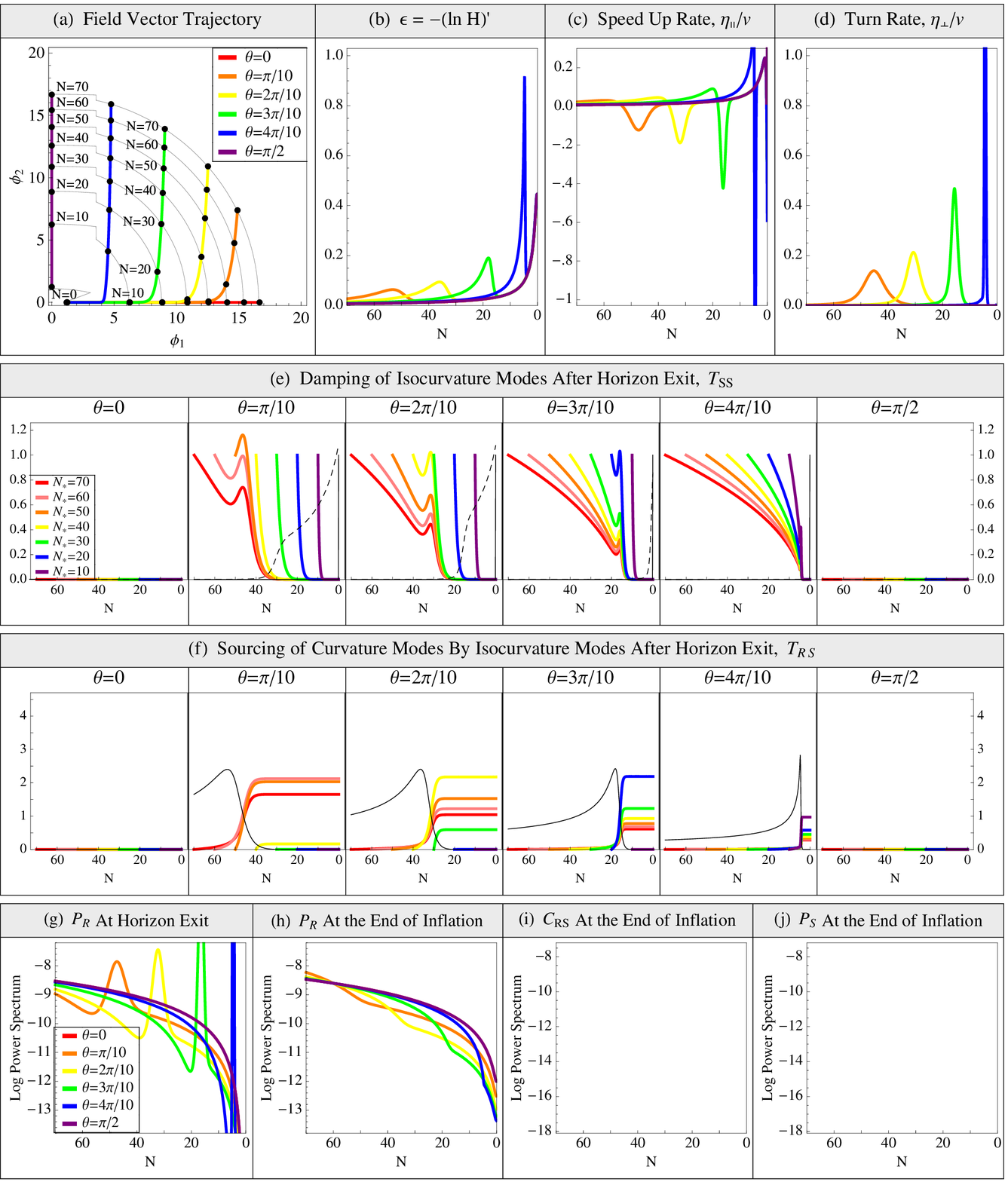}
\label{Quad VaryCoupling}
\end{minipage}
\end{figure*}

\clearpage

\noindent  trum, and appear roughly as if the equivalent single-field power spectrum from the second phase was essentially appended to the equivalent single-field power spectrum from the first phase, modulo a small transition region.   In all cases shown, since the large turn rate produced $T_{\mathcal{SS}} \lll 1$, the cross and isocurvature spectra are all vanishingly small, with amplitudes on the order of $10^{-20}$ or even much less.

\subsubsection{Varying the Lagrangian Parameter}

Now we vary the mass ratio while holding the initial condition fixed at $\theta = 45^o$.

The trajectories produced (Figure 7(a)) and the associated kinematical quantities (Figures 7(b)-(d)) depend strongly on the mass ratios of the fields.   For our fixed initial condition $\theta = 45^o$, trajectories with a mass ratio of around $1$ are approximately straight 45$^o$ lines in the flat field space, representing the fact that the two fields about equally contribute to the field dynamics.  For mass ratios $\frac{m_2}{m_1} \gtrsim 2$, the trajectories turn through roughly $90^o$ in field space, but the rate at which they complete this turn varies dramatically.  The greater the mass ratio, the sharper the turn is in field space, as before the phase transition, the more massive $\phi_2$ field initially decreases faster than the $\phi_1$ field.  For mass ratios in the range of about $1 \lesssim \frac{m_2}{m_1} \lesssim 3$, both fields significantly contribute to the energy density.  These scenarios exhibit typical slow-roll behavior, with small but still significant turn rates.   For mass ratios $\frac{m_2}{m_1} \gtrsim 3$, the turn in the trajectory is sharper and occurs largely within a couple to several $e$-folds of inflation, reflecting the two distinct phases of inflation we discussed earlier.  The larger the mass ratio, the sharper the turn and the larger and more abrupt the increase and subsequent decrease in $\epsilon$.  For very large mass ratios (not shown), inflation ends while the dynamics are still dominated by $\phi_2$.  Such scenarios are similar to those that we classified as being under the third subset of scenarios in the previous section.

The coefficients of the lowest-order effective mass matrix support our claims in this and the previous sections about the types of trajectories that can be produced by double quadratic potentials with canonical kinetic terms (analysis not shown).  The matrix coefficients vary by a few orders of magnitude over the region of field space shown, and they depend strongly on the combination of the initial conditions and the mass ratio of the fields.  For a large fraction of the different possible combinations of the initial conditions and mass ratio, the speed up rate ($\frac{\eta_{\parallel}}{v} \approx - \tilde{M}_{\parallel \parallel}$) will be larger than the turn rate ($\frac{\eta_{\perp}}{v} \approx - \tilde{M}_{\parallel \perp}$), and the speed up rate and turn rates are greatest near the axes.  However, for mass ratios typically greater than about 5, the SRST approximation breaks down close to the $\phi_1$ axis, and both $\tilde{M}_{\parallel \perp}$ and $\tilde{M}_{\perp \perp}$ typically increase by a few orders of magnitude and reach their largest values near the $\phi_1$ axis.  This corroborates both our earlier results and the results in this section that trajectories with larger mass ratios that approach close to the $\phi_1$ axis will turn quickly and have their isocurvature modes be rapidly suppressed thereafter.  

Indeed, we find that the isocurvature modes (Figure 7(e)) are initially steadily damped in the six scenarios depicted, but for those trajectories that later turn rapidly in field space, their isocurvature modes experience a transient boost from the drop in $\epsilon$, followed by a rapid suppression from the subsequent larger turn rate.  By the end of inflation, all isocurvature modes have been strongly suppressed.  The net effect of the turn rate's behavior and the suppression of isocurvature modes on $T_{\mathcal{RS}}$ is shown in Figure 7(f).  For small values of the mass ratio, $T_{\mathcal{RS}}$ is small and largely scale-independent.  However, as the mass ratio increases, $T_{\mathcal{RS}}$ significantly increases in value and becomes more scale-dependent.  As in the previous section, $T_{\mathcal{RS}}$ increases dramatically when the turn rate is large, and then levels off.   The value of $T_{\mathcal{RS}}$ at the end of inflation increases as the mass ratio increases and is largest for those modes that exit the horizon around the transition from $\phi_2$ to $\phi_1$ domination. 

The resultant curvature power spectra are shown in Figures 7(g)-(h).  Like in the previous section, the curvature power spectrum exhibits strongly scale-dependent features for those trajectories that correspond to two distinct phases of inflation, but is nearly scale-invariant for trajectories that exhibit very little turning in field space.  Also, like in the previous section, the scale-dependence at horizon exit is largely blunted by sourcing effects, so the final curvature spectrum has a much weaker scale dependence, and the two phases of inflation are distinctly evident in the final power spectrum.  Lastly, the cross spectrum (Figure 7(i)) is small and only marginally significant for those trajectories with small but non-zero turn rates.  Similarly, the isocurvature spectrum (Figure 6(j)) is only marginally significant for the trajectories with zero or very small turn rates, but still is a few orders of magnitude smaller than the curvature spectrum; for all other trajectories, it is negligibly small.

\subsubsection{Spectral Observables}

Now we vary both the initial conditions and the mass ratio, and we determine how these variations affect our spectral observables.  Using this information, we determine which scenarios are consistent with observational constraints.

First, we consider the three quantities that depend on the relative amplitudes of the power spectra.  The tensor-to-scalar ratio is $r_T \approx 0.13$, regardless of the initial conditions and the mass ratio, as shown in Figure 8(a).  By comparison, the cross-correlation ratio (Figure 8(b)) varies dramatically and can be large, particularly as $\theta$ decreases and $\frac{m_2}{m_1}$ increases, but is vanishingly small when one field dominates the dynamics during the last 60 $e$-folds of inflation.  When the cross-correlation ratio is large, the isocurvature fraction (Figure 8(c)) is negligibly 

\begin{figure*}[t]
\begin{minipage}[t]{1.0\linewidth}
\centering
\Large
Figure 7. Double Quadratic Potentials, $V(\phi_1,\phi_2) = \frac{1}{2} m_1^2 \phi_1^2 + \frac{1}{2} m_2^2 \phi_2^2$, \\ With Canonical Kinetic Terms
\large
\vskip 12 pt
Varying the Lagrangian Parameter, $\frac{m_2}{m_1}$, With the Initial Condition Fixed At $\theta = 45^o$
\normalsize
\vskip 12 pt
\includegraphics[height=213.5mm]{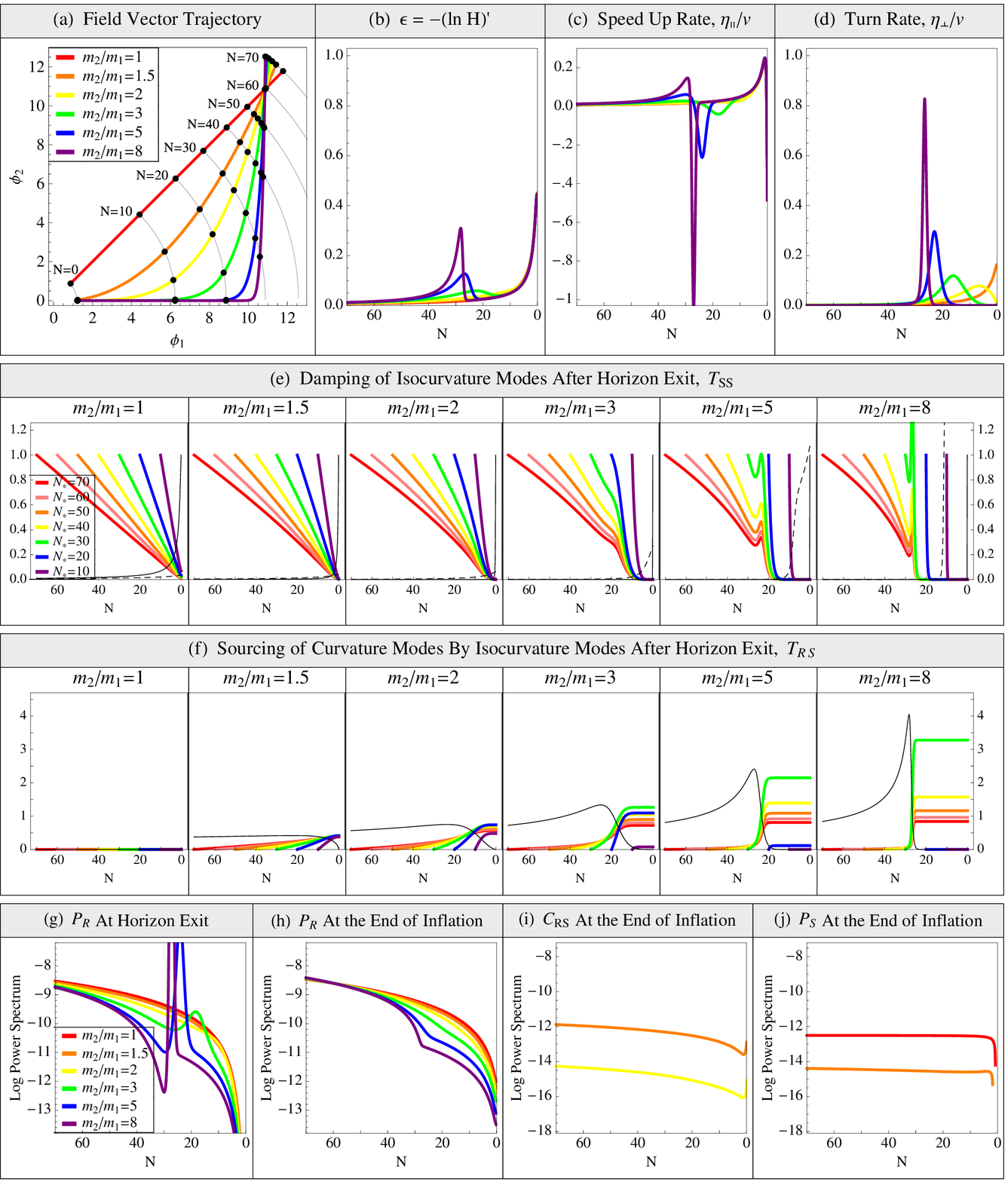}
\label{Quad VaryInitConds}
\end{minipage}
\end{figure*}

\begin{figure*}[t]
\begin{minipage}[t]{1.0\linewidth}
\centering
\Large
Figure 8. Double Quadratic Potentials, $V(\phi_1,\phi_2) = \frac{1}{2} m_1^2 \phi_1^2 + \frac{1}{2} m_2^2 \phi_2^2$, \\ With Canonical Kinetic Terms
\large
\vskip 12 pt
Varying Both the Initial Condition, $\theta$, and the Lagrangian Parameter, $\frac{m_2}{m_1}$
\normalsize
\vskip 12 pt
\includegraphics[height=210mm]{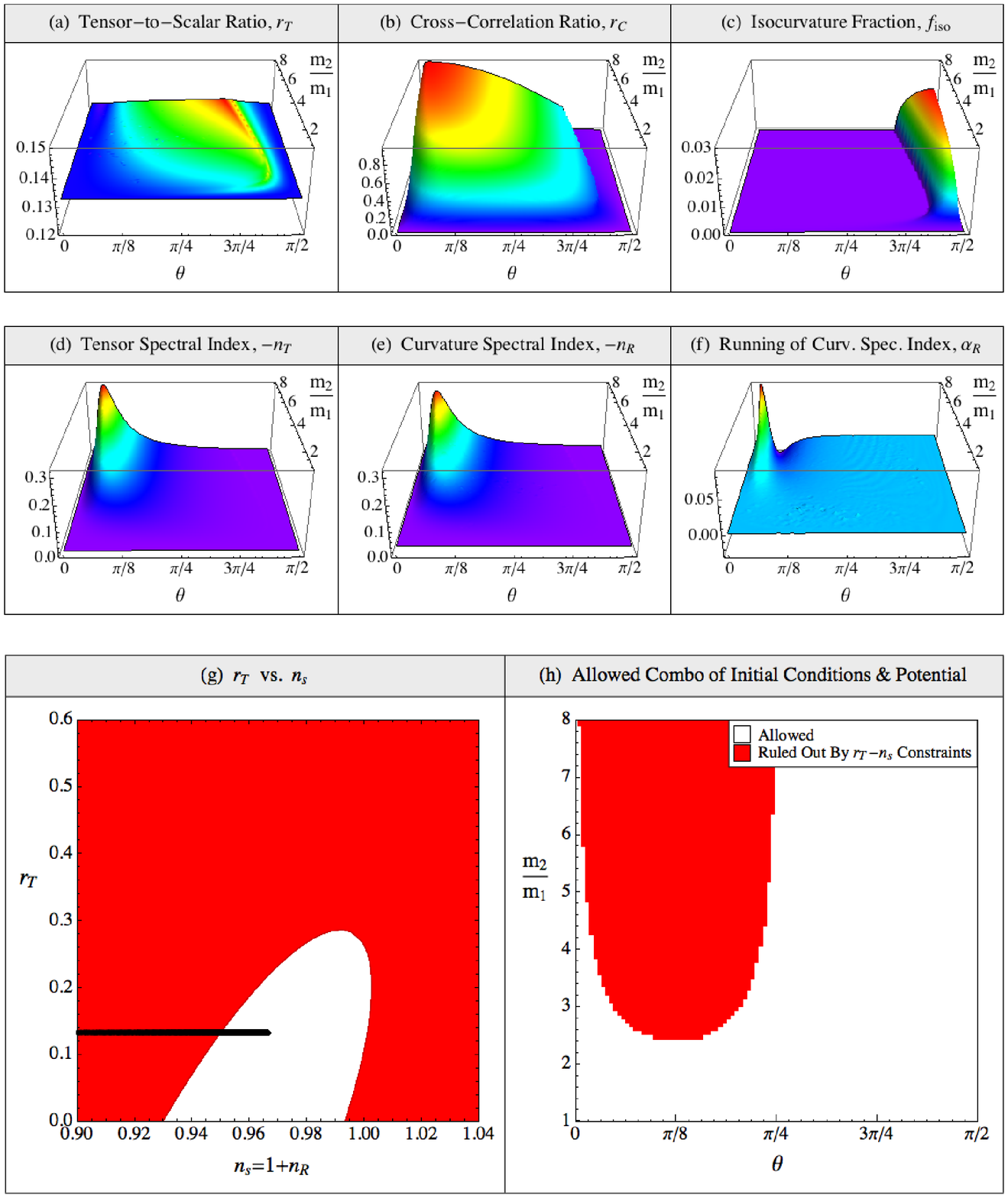}
\label{Quad VaryBoth}
\end{minipage}
\end{figure*}

\clearpage

\noindent small, but when the cross-correlation ratio is small, the isocurvature fraction becomes significant, reaching values of up to about 0.015.
 
Like for the multiplicative double polynomial potentials, the plots for both $-n_T$ (Figure 8(a)) and $-n_{\mathcal{R}}$ (Figure 8(b)) are quite similar to each other, illustrating that the main contribution to the spectral indices is from $\epsilon$.  Interestingly, the deviations from scale invariance stemming from the speed up and turn rates appear to largely cancel each other out.  Here, though, the spectral indices depend strongly on the initial conditions and the Lagrangian parameter.  This is particularly true for mass ratios above $\frac{m_2}{m_1} \approx 3$, for which the greatest deviations from scale invariance occur for small initial angles. This is because these trajectories correspond to scenarios where the turn rate is large around the pivot scale.  The significant scale dependence of $n_{\mathcal{R}}$ is reflected in the large magnitude of $\alpha_{\mathcal{R}}$ for certain combinations of initial conditions and mass ratios, as shown in Figure 8(d).

Finally, we examine which combinations of the initial conditions and the mass ratios can be ruled out based on observations.  Single-field $V \propto \phi^2$ models are viable under constraints on the combination of $r_T$ and $n_s$, so it is interesting to investigate the effect of adding a second inflaton.  Figure 8(f) shows that adding a second inflaton can indeed change the viability of the quadratic potential.  Scenarios with mass ratios roughly above 3 and initial conditions $\theta \lesssim 45^o$ are ruled out by the 95\% confidence region in the $n_{s}-r_T$ plane, largely due to the smaller values of the scalar spectral index.  Models with mass ratios below about 2.5 are viable for all initial conditions.

Therefore, for this particular class of models, the power spectra and associated observables are primarily determined by the mass ratio and whether one field dominates the dynamics or not.   Here, the role of the initial conditions for mass ratios sufficiently larger than 1 is to determine whether inflation consists of two separate phases, each dominated by a different field, or consists of a single phase where the two fields are co-dominant.  When the field trajectory corresponds to two distinct phases, these two phases will be reflected in the resultant curvature spectrum, which somewhat surprisingly is much less scale-dependent than the corresponding spectrum at horizon exit, due to the fact that the scale-dependences of the speed up and turn rates largely cancel each other out.   By contrast, for mass ratios close to 1, the inflationary dynamics are much less sensitive to the initial conditions and the curvature power spectrum is close to scale-invariant.   In general, for these models, the isocurvature and cross spectra are very small, if not completely negligible, reflecting the larger turn rate and effective entropy mass.

\subsection{Double Quartic Potential With Canonical Kinetic Terms}

A somewhat similar class of potentials to the double quadratic potential is the double quartic potential, which includes all potentials of the form
\begin{eqnarray}
\label{Quar Potl}
V = \frac{1}{4} \lambda_1^4 \phi_1^4 + \frac{1}{2} \lambda_2^4 \phi_2^4.
\end{eqnarray}
Since the addition of a second field to single-field $V \propto \phi^2$ models changes their viability in certain cases, it is interesting to consider whether the addition of a second field to the otherwise ruled out single-field $V \propto \phi^4$ models makes them viable for any particular combinations of the initial conditions and the ratio of coupling constants.  Here, we again take the kinetic terms to be canonical.  We take the adjustable Lagrangian parameter to be the ratio of the coupling constants, $\frac{\lambda_2}{\lambda_1}$, where $\frac{\lambda_2}{\lambda_1} \ge 1$.

As it turns out, the results for the double quartic potential are similar in most ways to those for the double quadratic potential, so we keep this discussion short.  Separately varying the initial conditions and the ratio of the coupling constants, we find very similar results for the trajectories, kinematics, transfer functions, and power spectra (plots not shown) as for the double quadratic models.  However, there are a few important differences worth mentioning.  For those trajectories that turn significantly in field space, they turn a bit further out from the $\phi_1$ axis, representing the fact that the $\phi_2$ field is still appreciably evolving during the $\phi_1$-dominated phase.  As a result, the turn rate is a bit smaller for these scenarios and the isocurvature modes are less strongly damped in the second phase of inflation, and so the isocurvature and cross spectra tend to be orders of magnitude larger.  Also for these trajectories, the speed up rate tends to be a bit larger in magnitude, so $\epsilon$ peaks at a higher value before $\phi_1$ domination begins in earnest.  This also results in a higher boost in the amplitude of isocurvature modes during the transition between $\phi_2$ and $\phi_1$ domination.  Lastly, the curvature power spectra produced by double quartic potentials naturally have larger spectral indices, just like the single-field quartic potential.

In Figure 9, we illustrate the ranges of values that the double quartic models can produce for the spectral observables. Since the kinematics are quite similar to those for the double quadratic models, it is no surprise that these observables share many of the same features.  The main differences are that the tensor-to-scalar ratio is about twice as large with a value of $r_T \approx 0.26$; the spectral indices $n_T$ and $n_{\mathcal{R}}$ are larger in magnitude, with their peak values being roughly three times higher and the differences between the two being proportionally greater; and the amplitude of the running of the curvature spectral index is significantly larger.   The net result of both $r_T$ and $n_{\mathcal{R}}$ being larger in magnitude (and hence $n_s$ being smaller) is that the double quartic model does not produce any scenarios falling in the allowed region in the $n_s-r_T$ plane.  Therefore, adding a second inflaton does 

\begin{figure*}[t]
\begin{minipage}[t]{1.0\linewidth}
\centering
\Large
Figure 9. Double Quartic Potentials, $V(\phi_1,\phi_2) = \frac{1}{4} \lambda_1^4 \phi_1^4 + \frac{1}{4} \lambda_2^4 \phi_2^4$, \\ With Canonical Kinetic Terms
\large
\vskip 12 pt
Varying Both the Initial Condition, $\theta$, and the Lagrangian Parameter, $\frac{\lambda_2}{\lambda_1}$
\normalsize
\vskip 12 pt
\includegraphics[height=210mm]{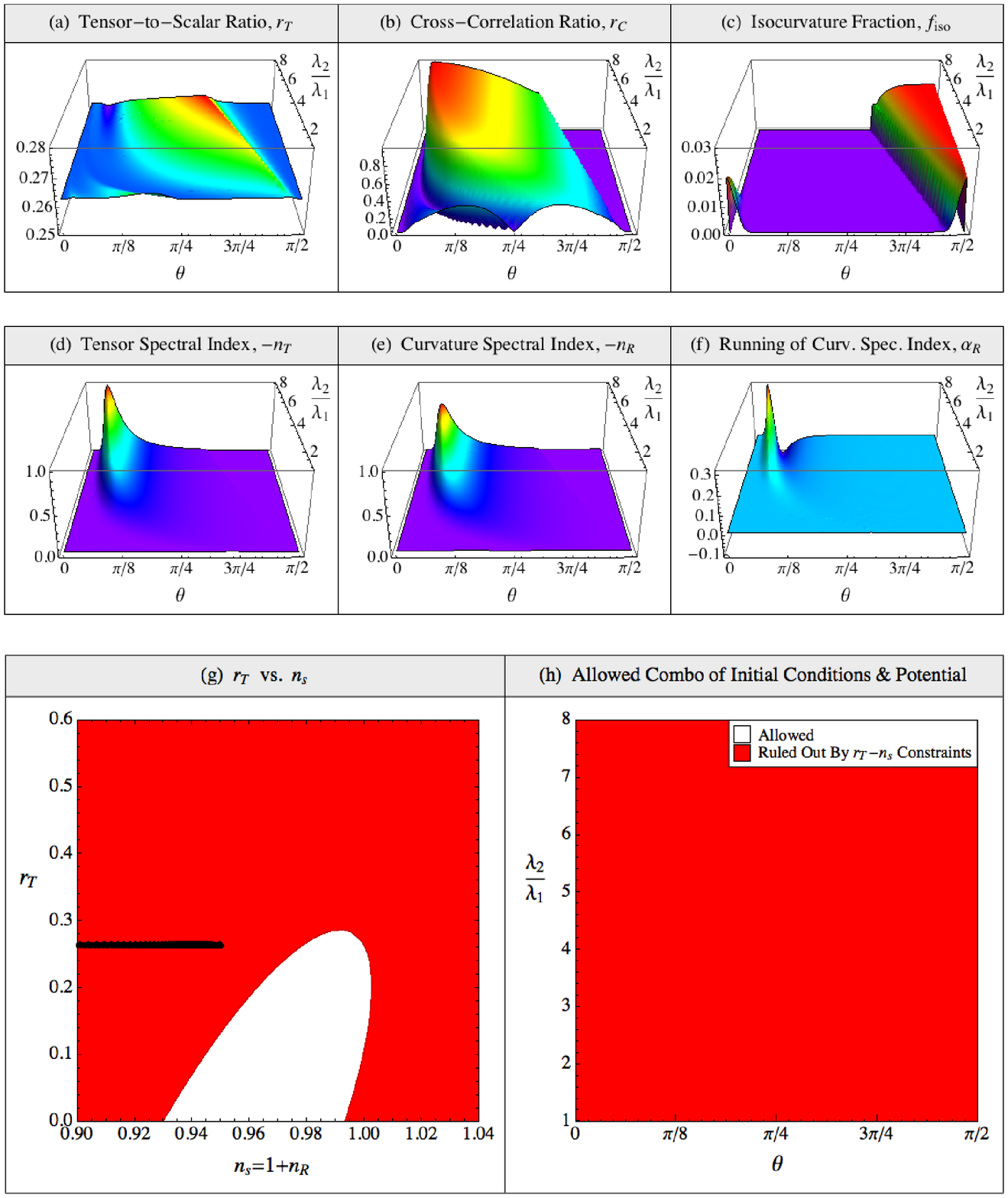}
\label{Quar VaryBoth}
\end{minipage}
\end{figure*}
 
\clearpage
 
\noindent not make these models viable.  In fact, it make things worse by decreasing the value of $n_s$ and hence shifting the scenario far away from the allowed region in $n_s-r_T$ plane.

\subsection{Double Quadratic Potential With Non-Canonical Kinetic Terms}

In the previous sections, we chose to investigate simple inflationary models that represent the most common types of kinematical behaviors.  In this section, we consider the addition of non-canonical kinetic terms, and in doing so, we also investigate the remaining unexplored kinematical limit: $\frac{\eta_{\perp}}{v} \gg \frac{\eta_{\parallel}}{v}$.  

The last class of models we consider is the double quadratic potential in equation (\ref{Quad Potl}) but with the addition of non-canonical kinetic terms in the form
\begin{align}
\label{kinetic terms form for example model}
\mathbf{G} = \left(\begin{array}{cc} 1 & 0 \\ 0 & e^{2 b \phi_1} \end{array}\right),
\end{align}
where $b$ is a constant.  The form of the field metric in equation (\ref{kinetic terms form for example model}) is the same as that which appears in Brans-Dicke theories after transformation to the Einstein frame.  The Ricci scalar for this field metric is 
\begin{align}
R = - 2 b^2,
\end{align}
so in keeping with our assumption that $|R| \lesssim 1$, we require that $|b| \lesssim \frac{1}{\sqrt{2}} \approx 0.71$.  For these models, we vary the Lagrangian parameter $b$, while holding the mass ratio fixed at $\frac{m_2}{m_1} = 2$.  This will allow us to see whether the addition of non-canonical kinetic terms to a model that is otherwise allowed by observations can affect the model's viability.

\subsubsection{Varying the Lagrangian Parameter}

Since the dependence of the kinematics, transfer functions, and power spectra on the initial conditions can be inferred from Section \ref{vary initial conds for double quadratic}, we skip directly to the varying the Lagrangian parameter, $b$, while holding the initial condition fixed at $\theta = 45^o$.

In Figure 10(a), we plot the resultant trajectories relative to the $\phi_1-\phi_2$ coordinate plane, which we call the field space for short, and we remind the reader that the field manifold no longer coincides with field space when the kinetic terms are non-trivial.  For our fixed initial condition $\theta = 45^o$, varying $b$ leads to trajectories starting and ending in very different regions of field space, and hence they correspond to quite different starting and ending values for the potential.  Also, the trajectories produced have different curvatures from each other, regardless of whether they are considered with respect to field space or the field manifold.  When $b=0$, both the fields contribute to the inflationary dynamics during most of the last 60 $e$-folds of inflation.  For negative values of $b$, the more negative $b$ is, the more strongly the field vector is initially driven towards the $\phi_1$ axis, and the more inflation tends to consist of two distinct phases, with the first phase dominated by $\phi_2$ and the second dominated by $\phi_1$.  Conversely, the more positive $b$ is, the more the field vector is initially driven towards the $\phi_2$ axis.  If $b$ is sufficiently large and positive, then two distinct inflationary stages emerge: initially $\phi_1$ evolves more much than $\phi_2$, but eventually $\phi_2$ speeds up substantially.  But here, unlike in previous examples, the evolution of the less massive field dominates the first inflationary stage, as if the large positive value of $b$ effectively increases the mass of the $\phi_1$ field.  Also, unlike previously, although one field initially evolves much faster than the other, both fields significantly contribute to the potential energy density. 

Depicted in Figures 10(b)-(d) are the associated inflationary kinematics, which in many ways can be inferred from our previous discussions, so we just make three new points.  First, the larger the magnitude of $b$, the greater the maximum value of the turn rate tends to be.  Second, the two trajectories corresponding to large negative values of $b$ have similar kinematics to double quadratic models with large mass ratios and canonical kinetic terms; for this reason, we can think of negative values of $b$ as effectively increasing the mass of the $\phi_2$ field.  Third, we point out that the trajectory corresponding to $b=0.5$ represents the interesting but less common case where $\frac{\eta_{\perp}}{v} \gg \frac{\eta_{\parallel}}{v}$ for many $e$-folds of inflation, which corresponds to the sourcing effects dominating the growth of adiabatic modes.  This last special scenario rounds out our coverage of the various possible kinematical behaviors.

Figures 10(e)-(f) show the transfer functions for these scenarios.  Relative to the $b=0$ trajectory, the larger $b$ is in magnitude, the greater the damping of isocurvature modes.  Sufficiently large and negative values of $b$ affect the transfer functions $T_{\mathcal{SS}}$ and $T_{\mathcal{RS}}$ in a similar manner to large mass ratios.  That is, the isocurvature modes are initially boosted by the drop in $\epsilon$ during the transition between phases and are then suppressed by the subsequent larger turn rates and effective entropy masses.  Also, when the turn rates are large, the function $T_{\mathcal{RS}}$ increases dramatically in value but then levels off.  At the end of inflation, $T_{\mathcal{RS}}$ is quite scale-dependent for these values of $b$, being largest for those modes that exit the horizon while the turn rate is significant.  For trajectories with positive values of $b$, the background kinematics are less scale-dependent, so the isocurvature modes are suppressed more gradually and evenly, and the function $T_{\mathcal{RS}}$ is less scale-dependent and smaller, with its value reflecting the size of the turn rate.  As before, an analysis of the coefficients of the effective mass matrix is in agreement with these findings, and also shows that the Ricci scalar curvature term only slightly decreases the damping of isocurvature modes.

Figures 10(g)-(j) show the resultant power spectra.  Many of the same results as for the double quadratic potential with canonical kinetic terms carry over here, in that the transfer function $T_{\mathcal{RS}}$ blunts much of the scale-

\begin{figure*}[t]
\begin{minipage}[t]{1.0\linewidth}
\centering
\Large
Figure 10. Double Quadratic Potentials, $V(\phi_1,\phi_2) = \frac{1}{2} m_1^2 \phi_1^2 + \frac{1}{2} m_2^2 \phi_2^2$, \\ With the Non-Canonical Field Metric $\mathbf{G} = \mathrm{diag}(1,e^{b\phi_1})$
\large
\vskip 12 pt
Varying the Lagrangian Parameter, $b$, With the Initial Condition Fixed At $\theta = 45^o$
\normalsize
\vskip 12 pt
\includegraphics[height=213.5mm]{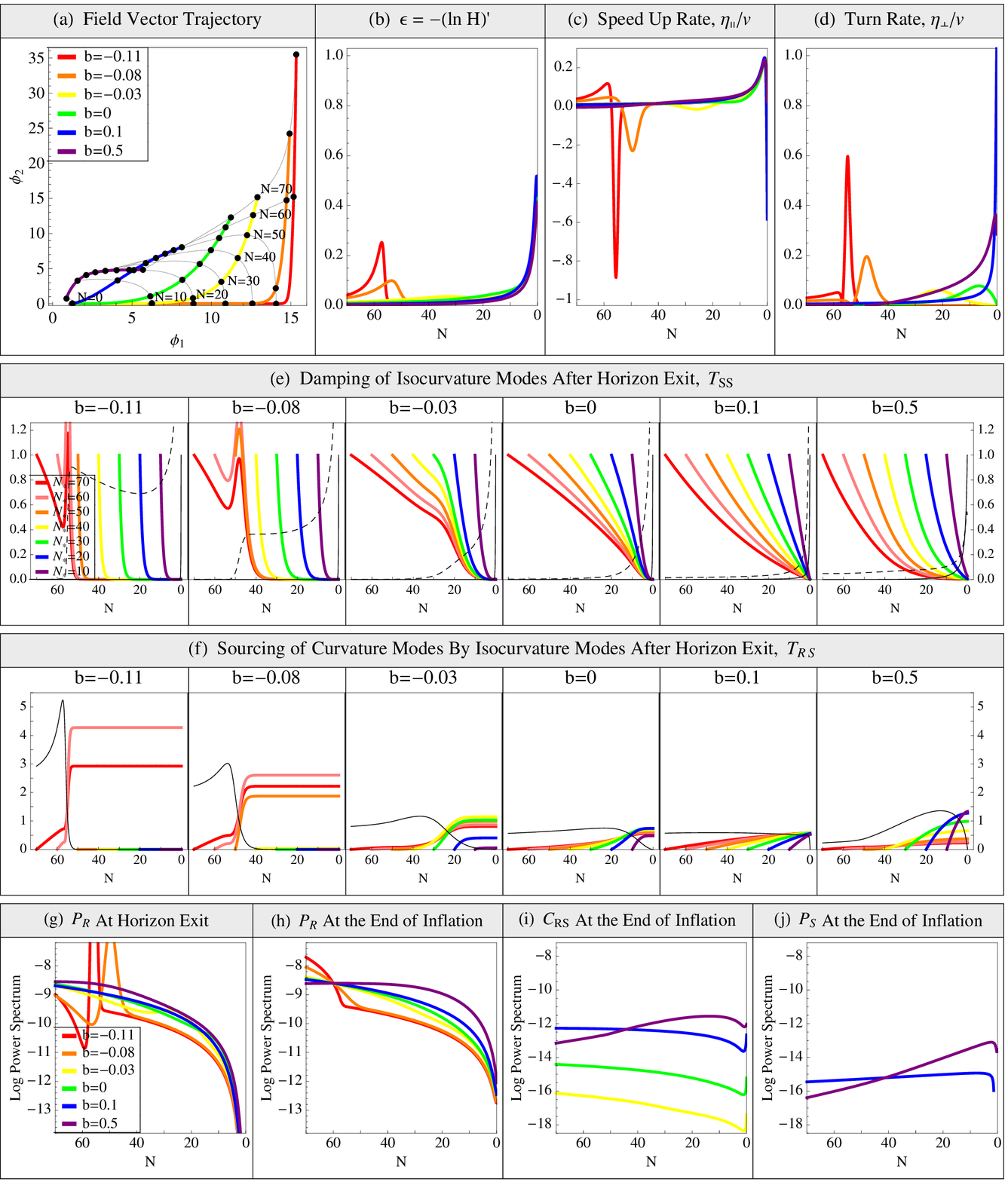}
\label{Quad VaryInitConds}
\end{minipage}
\end{figure*}

\begin{figure*}[t]
\begin{minipage}[t]{1.0\linewidth}
\centering
\Large
Figure 11. Double Quadratic Potentials, $V(\phi_1,\phi_2) = \frac{1}{2} m_1^2 \phi_1^2 + \frac{1}{2} m_2^2 \phi_2^2$, \\ With the Non-Canonical Field Metric $\mathbf{G} = \mathrm{diag}(1,e^{b\phi_1})$
\large
\vskip 12 pt
Varying Both the Initial Condition, $\theta$, and the Lagrangian Parameter, $b$
\normalsize
\vskip 12 pt
\includegraphics[height=210mm]{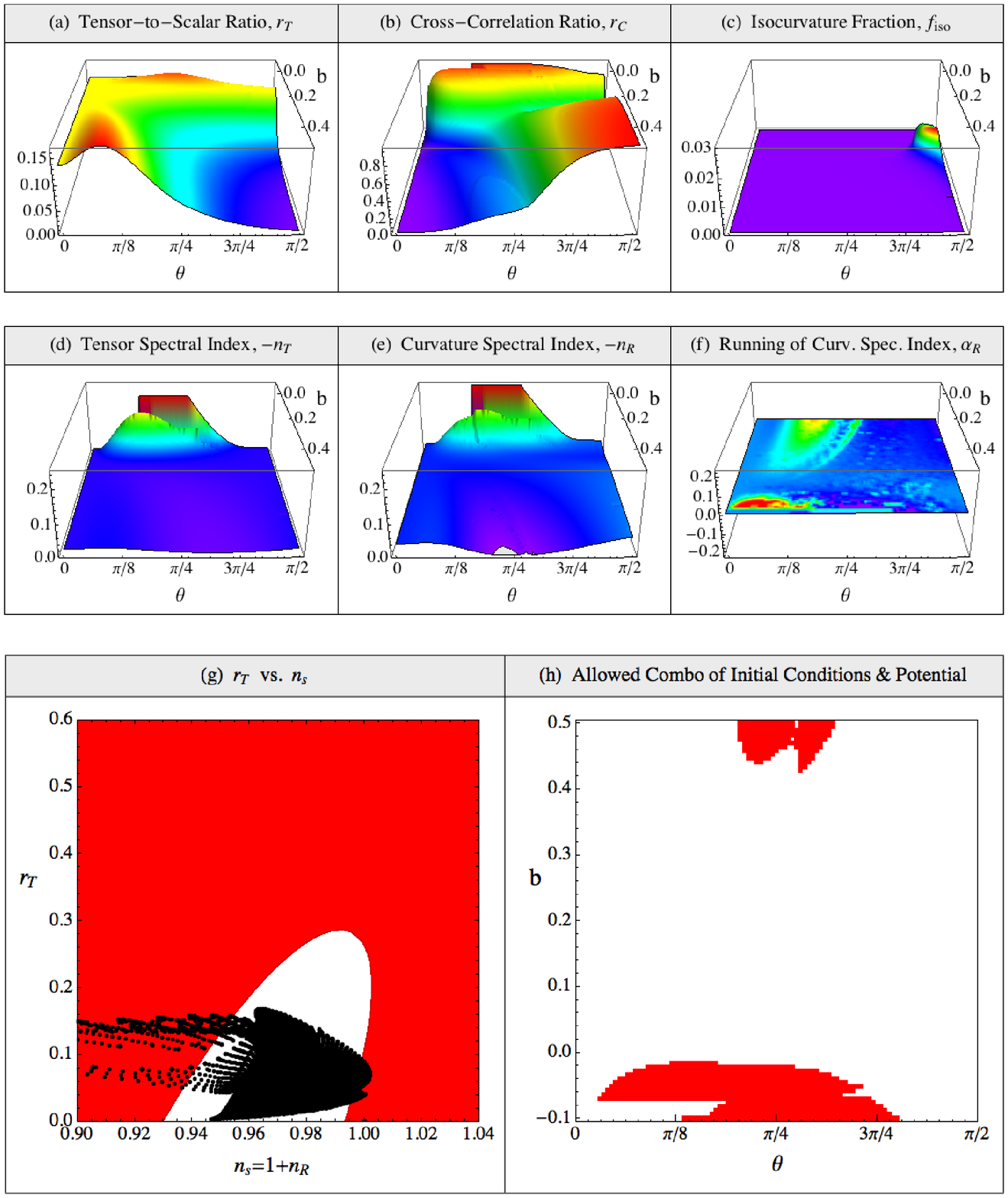}
\label{Quar VaryBoth}
\end{minipage}
\end{figure*}

\clearpage 
 
\noindent  dependence of the curvature spectrum at horizon exit.   Interestingly, positive values of $b$ make the curvature power spectrum at the end of inflation more scale-invariant than the corresponding spectrum at horizon exit \textit{and} than the final curvature spectrum for the $b=0$ scenario.  Indeed, the flatness of the curvature spectrum for the $b=0.5$ scenario is remarkable, suggesting that a significant and sustained turn rate accompanied by a significantly smaller roll rate (an example of the limit $\frac{\eta_{\perp}}{v} \gg \frac{\eta_{\parallel}}{v}$) provides a way of achieving a very highly scale-invariant curvature spectrum.   We also point out that the smaller $b$ is, the larger the cross and the isocurvature spectra tend to be, though both are at least a few orders of magnitude smaller than the curvature spectrum at the end of inflation.  Interestingly, the  the isocurvature spectrum for the $b=0.5$ scenario shows significantly greater power on smaller scales.

\subsubsection{Spectral Observables}

Now we vary both the initial conditions and the Lagrangian parameter $b$ to determine their effects on the spectral observables.  In Figures 11(a)-(e), we plot the values of the parameter $b$ in decreasing order so that more of the two-dimensional surfaces representing these observables are visible.  

Figure 11(a) shows that the parameter $b$ can dramatically affect the tensor-to-scalar ratio.  When $b=0$, $r_T \approx 0.13$ regardless of the initial conditions.   Negative values of $b$ do not effect this result much.  However, for positive values of $b$, the larger initial angle, the smaller $r_T$ is, reflecting the fact that the turn rate is larger, while $\epsilon$ and the speed up rate are still small.  The cross-correlation ratio, shown in Figure 11(b), has a complex dependence on the initial conditions and $b$.  For $b \approx 0$, the correlation is small.  For negative values of $b$, the more negative $b$ is, the larger the correlation ratio is, and this is largely independent of the initial conditions.   For positive values of $b$, the larger $b$ is, the larger the correlation is for large angles, but the smaller it is for small angles.  These complicated dependences largely mirror the turn rate profile.   Interestingly, the isocurvature fraction is minuscule for these scenarios, peaking around $f_{iso} \approx 0.006$ for $b \approx -0.1$ and $\theta \approx 90^o$ (Figure 11(c)).

Now we examine the key spectral indices.  As for all the other models we have considered, the profiles for   $-n_T$ (Figure 11(d)) and $-n_{\mathcal{R}}$ (Figure 11(e)) are strikingly similar.  Here, the spectral indices are very small, except for more negative values of $b$, for which the spectral indices are moderately large for a significant fraction of initial angles.   The greatest differences between the two spectral indices occurs for models with large turn rates, and hence their curvature spectral index gets a boost from the turn rate and effective entropy mass.  For the running of the curvature spectral index  (Figure 11(f)), surprisingly, it is small in all cases.

Finally, we examine which combinations of the initial conditions and the Lagrangian parameter $b$ can be ruled out based on observations.  The double quadratic model with a mass ratio of $\frac{m_2}{m_1} = 2$ and canonical kinetic terms is consistent with observations.  Adding the non-canonical kinetic terms in equation (\ref{kinetic terms form for example model}) makes the landscape of kinematical behaviors and spectral observables far richer and more complex.   Combinations of large negative or positive values of $b$ and moderate initial angles are ruled by the 95\% confidence region in the $n_{s}-r_T$ plane.  The former set corresponds to large turn rates accompanied by large speed up rates, with the result that this causes the curvature spectrum to be too scale-dependent to match observations.  The latter set corresponds to scenarios in which the kinematical behavior $\frac{\eta_{\perp}}{v} \gg \frac{\eta_{\parallel}}{v}$ causes the curvature spectrum to be too scale-independent, causing an inflationary model that is otherwise consistent with observations to be ruled out.   Therefore, the addition of non-canonical kinetic terms provides a powerful way to fine-tune a model, to achieve the right amount of scale-dependence in the density spectrum and to achieve a tensor-to-scalar ratio that is consistent with observations. 

Ultimately, this analysis, along with other analyses we have performed in Section \ref{Applications}, shows that having the right balance of the field speed, the speed up rate, and the turn rate (along with reasonable values for the effective entropy mass) is critical for ensuring that the curvature power spectrum is neither too scale-dependent nor scale-independent and for achieving a tensor-to-scalar ratio that is consistent with observations.

\section{Conclusion}
\label{Conclusions}

\subsection{Summary}

In this paper, we have constructed a complete covariant framework for understanding two-field models of inflation with an arbitrary potential and with arbitrary non-canonical kinetic terms.  We have derived the power spectra to second-order in the combined slow-roll and slow-turn (SRST) approximation, provided new insight into how the spectra can be inferred from the background kinematics and the field manifold, and we have illustrated how to rigorously test and constrain two-field models of inflation using observational data. 

We started by considering the background dynamics in Section \ref{unperturbed equations}.  After simplifying the background equations of motion using covariant vector notation and the number of $e$-folds as the time variable, we introduced a set of three covariant quantities to understand the kinematics of the background field vector: the field speed ($v = \sqrt{2\epsilon}$), the speed up rate ($\frac{\eta_{\parallel}}{v}$), and the turn rate ($\frac{\eta_{\perp}}{v}$), where the third quantity is unique to multi-field inflation and hence can be viewed as the marker of multi-field effects.  We used this kinematical framework to generalize the standard slow-roll approximation to two-field inflation, dividing it into two separate parts: a slow-roll approximation to represent limits on single-field-type behavior and a slow-turn approximation to represent limits on multi-field behavior.   We then derived first- and second-order expressions for the background equations, and we introduced the mass matrix, $\mathbf{M} =  \boldsymbol{\nabla}^\dag \boldsymbol{\nabla} \ln V$, whose coefficients estimate the speed up and turn rates.  

In Section \ref{perturbed equations}, we considered the perturbed equations of motion.  Working in terms of gauge-invariant quantities, we simplified the equation of motion for the field perturbations and showed that their evolution is determined primarily by the effective mass matrix, $\mathbf{\tilde{M}} =  \boldsymbol{\nabla}^\dag \boldsymbol{\nabla} \ln V + \frac{\epsilon R}{(3-\epsilon)}$, where $R$ is the Ricci scalar of the field manifold induced by the field metric.  Thereafter, we rotated to the kinematical basis and derived exact expressions for the evolution of adiabatic and entropy modes.  Working in the super-horizon limit, we found a simple exact expression for the evolution of adiabatic modes, revealing that the growth of adiabatic modes is determined by the speed up and turn rates, with the turn rate determining the degree of sourcing by the entropy modes.   Further, we found that the ratio $\frac{\eta_{\perp}}{v} / \frac{\eta_{\parallel}}{v}$, which controls the relative contribution of mode sourcing to the growth of adiabatic modes, can be viewed as indicating the relative importance of multi-field effects.  This also has an aesthetic appeal, since the same ratio provides a measure of the curvature of the background trajectory at a given time.  We also found approximate expressions for the super-horizon evolution of entropy modes, which is controlled by the effective entropy mass ($\mu_\perp$).  From analyzing the behavior of the effective entropy mass, we argued that assuming the effective entropy mass and slow-roll parameters are constant in order to estimate the power spectra \cite{BartoloEtAl-2001a,TsujikawaEtAl-2002,DiMarcoAndFinelli-2005,LalakEtAl-2007} generally leads to large inaccuracies in estimating the amplitude of modes, even in the SRST approximation.  Finally, we used these results to find semi-analytic expressions for the super-horizon amplitude of the related curvature and isocurvature perturbations, and we discussed how the general features of their evolution can largely be inferred from the background kinematics and the field manifold.   For example, a large turn rate rate produces strong sourcing of curvature modes by isocurvature modes, leading to a large boost in the amplitude of curvature modes at the expense of a dramatic suppression of isocurvature modes.  

Thereafter, in Section \ref{perturbed equations}, we calculated and interpreted the power spectra.  To do so, we quantized the field perturbations, solved the field perturbation equation in the three standard regimes of interest, and matched the solutions across the boundaries.   After rotation to the kinematical basis, we calculated the power spectra at horizon exit, and then used the transfer matrix formalism \cite{AmendolaEtAl-2001,WandsEtAl-2002} to derive compact expressions for the curvature, isocurvature, and cross spectra to second-order in the SRST limit.  We found that to lowest order, the spectra depend on just four functions ($H$, $\epsilon$, $T_{\mathcal{SS}}$, and $T_{\mathcal{RS}}$), which means that all features of the power spectra can be traced back to five fundamental kinematical parameters: $H$, $\epsilon$, $\frac{\eta_{\parallel}}{v}$, $\frac{\eta_{\perp}}{v}$, and $\mu_{\perp}$, which together are simply related to the value, gradient, and Hessian of $V$ (along with corrections from any trivial geometry of the field manifold.  We discussed how exactly these background kinematics and the curvature of the field manifold are reflected in the general features and relative sizes of the spectra.  For example, for models with large turn rates, the curvature power spectrum will be boosted and the cross and isocurvature spectra will be several orders of magnitude smaller, with profiles that reflect the scale-dependence of the turn rate, among other quantities.  We also provided conditions for when a two-field inflationary scenario effectively behaves like a single-field inflationary scenario.  Thereafter, we presented compact expressions for the tensor-to-scalar ratio, spectral indices, and runnings of the spectral indices, and a consistency relation among them, and we showed how can work backwards to reconstruct the background kinematics from these spectral observables.  

Finally, we illustrated for the first time how to rigorously test two-field inflationary models against observational data by incorporating initial conditions.  We tested four classes of inflationary models, varying both their initial conditions and a characteristic Lagrangian parameter in order to test tens of thousands of possible scenarios to determine the types of kinematical behaviors and power spectra they produce.  The four classes of models we considered covered all three limits for the relative ratio of the turn rate to the speed up rate, both distributed and abrupt rolling and turning behavior, and non-canonical kinetic terms.  For our three classes of models with canonical kinetic terms, we found that certain combinations of the initial conditions and Lagrangian parameter for the multiplicative double polynomial and double quadratic potentials are consistent with observational constraints on $r_T$ and $n_s$, but that double quartic models are completely ruled out, as is the corresponding single-field potential.  The double multiplicative models mostly produce scenarios in which one of the fields dominates the dynamics and hence primarily determines the observables $r_T$ and $n_s$, so the power to which the dominant field is raised determines whether the scenario is consistent with observations.  The double quadratic and double quartic models produce mostly scenarios in which either the fields co-dominate or one field dominates the dynamics, then followed by the second field dominating.  The latter scenarios possess large turn rates around the transition between phases, so if the transition occurs around the pivot scale, these scenarios tend to be ruled out.  The double quartic models, however, are ruled out regardless of the initial conditions and coupling constants, due to their larger values for $r_T$ and smaller values for $n_s$.  Interestingly, the addition of non-canonical kinetic terms to an otherwise viable double quadratic model creates a far richer landscape of potential background kinematics and power spectra, resulting in a wide range of values for the spectral observables, some of which are consistent with observations and some are not.  Of note, certain choices of the field metric produce distributed turning with $\frac{\eta_{\perp}}{v} \gg \frac{\eta_{\parallel}}{v}$, making the curvature power spectrum more scale-independent than in the equivalent case with canonical kinetic terms.   This means that the addition of the non-canonical kinetic terms may provide a mechanism for fine-tuning models to match observations.  Lastly, in the vast majority of scenarios for all models tested, the isocurvature and cross spectra were at least an order of magnitude or two smaller than the curvature spectrum.

\subsection{Outlook}

As mentioned in the introduction, there are compelling theoretical reasons to consider multi-field inflation.  This paper provides a complete theoretical framework and set of tools for parsing and analyzing two-field models of inflation, making it easier to calculate the power spectra and to understand what features a two-field model needs to possess in order to be consistent with observational constraints.  

However, there are many outstanding issues that merit further consideration.   Particularly important is the issue of initial conditions.  How should we weigh the initial conditions?  Since many inflationary models have attractor-like solutions, specifying initial conditions 60 $e$-folds before the end of inflation rather than at the beginning of inflation may correspond to a radically different measure on the space of trajectories and hence predicted spectra.  Also important are the nature of the end of inflation and reheating.  Some models predict further post-inflationary suppression of isocurvature perturbations, while others do not, complicating the use of observational isocurvature bounds for constraining inflationary models.  It is therefore important to make further progress in these areas in order to place tighter constraints on inflationary models.

As for the observational constraints themselves, we have seen that the combination of scalar and tensor power spectra alone have the potential to constrain or rule out large classes of multi-field inflation models.   Since the rapid progress in cosmic microwave background measurement precision is likely to continue for some time---with perhaps measurements such as the isocurvature fraction or the correlation angle placing useful constraints on multi-field models in the near future---it is therefore timely to investigate multi-field phenomenology in greater detail.

\acknowledgements{The authors wish to thank Matias Zaldarriaga and Mark Hertzberg for helpful discussions, and Andrew Liddle, Jonny Frazer, and the referee who made insightful suggestions for improving this paper.  This work was supported by an NSF Graduate Research Fellowship, NSF grants AST-0071213 \& AST-0134999, NASA grants NAG5-9194 \& NAG5-11099, a grant from the John Templeton Foundation, a fellowship from the David and Lucile Packard Foundation, and a Cottrell Scholarship from Research Corporation.}

% ../MathematicaFilesAndImages/ImagesAndPlots/

\end{document}